\documentclass[11pt]{article}
\pdfoutput = 1
\usepackage[utf8]{inputenc}
\usepackage{color,graphicx}
\usepackage{epsfig}
\usepackage{amsmath}
\usepackage{verbatim}
\usepackage{amssymb}
\usepackage{mathabx}
\usepackage{stmaryrd}
\usepackage{empheq}
\usepackage{simpler-wick}
\usepackage{esint}
\usepackage{physics} 
\usepackage{amsfonts}
\usepackage{array}
\usepackage{setspace}
\usepackage{xcolor}
\usepackage{braket}
\usepackage{float}
\usepackage{url}
\usepackage{mathtools}
\usepackage{tensor}
\usepackage{bbold}
\usepackage{slashed}
\usepackage{tikz}
\usetikzlibrary{decorations.markings}
\usepackage{tikz,pgfplots}
\usetikzlibrary{tikzmark,calc}
\usepackage{epstopdf}
\usepackage{subfigure}
\usepackage[labelfont=bf]{caption}
\usepackage{pgfplots}
\usepackage{mathrsfs}
\usepackage{fancybox}
\usepackage{eurosym}
\usepackage{tcolorbox}
\usepackage{tensor}
\allowdisplaybreaks
\usepackage{changepage}
\usepackage[mathcal]{eucal}

\usepackage[margin = 2.2cm]{geometry}
\usepackage[ragged]{footmisc}
\usepackage[maxbibnames=99,
    sorting=none,
    backend=biber,
    style=numeric,
]{biblatex}
\usepackage[bookmarks=true,bookmarksnumbered=false,hyperindex=true,bookmarksopen=true,hyperfigures=true,colorlinks=true,linkcolor=webblue,citecolor=webgreen,urlcolor=webblue,breaklinks]{hyperref}
\definecolor{webgreen}{rgb}{0, 0.5, 0}
\definecolor{webblue}{rgb}{0, 0, 0.5}
\definecolor{webred}{rgb}{0.5, 0, 0}
\definecolor{darkgreen}{rgb}{0,0.5,0}

\def\ben{\begin{equation}}
\def\een{\end{equation}}

 \let\b=\beta   \let\e=\varepsilon
 \let\h=\eta  
   \let\x=\xi  \let\r=v
   \let\c=\chi

\def\be{\begin{equation}}
\def\ee{\end{equation}}
\def\ba{\begin{array}}
\def\ea{\end{array}}

\def\dalemb#1#2{{\vbox{\hrule height .#2pt
       \hbox{\vrule width.#2pt height#1pt \kern#1pt
               \vrule width.#2pt}
       \hrule height.#2pt}}}

\newcommand{\bea}{\begin{eqnarray}}
\newcommand{\eea}{\end{eqnarray}}

\let\tilde=\widetilde

\numberwithin{equation}{section}

\def\is{\!&\! = \! & \!}

\newcommand{\uncrossedB}{
  \mathrel{
    \tikz[baseline=-0.8ex]{
      \draw[line width=1pt] (0,0) circle (1.5ex);
      \draw[line width=1pt] (135:1.5ex) to[out=330, in=30] (225:1.5ex);
      \draw[line width=1pt] (45:1.5ex) to[out=210, in=150] (315:1.5ex);
      }}}
\newcommand{\crossedB}{
  \mathrel{
    \tikz[baseline=-0.8ex]{
      \draw[line width=1pt] (0,0) circle (1.5ex);
      \draw[line width=1pt] (135:1.5ex) to[out=315, in=135] (315:1.5ex);
      \draw[line width=1pt] (45:1.5ex) to[out=225, in=45] (225:1.5ex);
      }}}
\newcommand{\uncrossedLLB}{
  \mathrel{
    \tikz[baseline=-0.8ex]{
      \draw[line width=1pt] (0,0) circle (1.5ex);
      \draw[line width=1pt] (130:1.5ex) to[out=0, in=0] (230:1.5ex);
      \draw[line width=1pt] (150:1.5ex) to[out=0, in=0] (210:1.5ex);
      }}}
\newcommand{\uncrossedLLRinG}{
  \mathrel{
    \tikz[baseline=-0.8ex]{
      \draw[line width=1pt] (0,0) circle (1.5ex);
      \draw[green!60!black, line width=1pt] (130:1.5ex) to[out=0, in=0] (230:1.5ex);
      \draw[red, line width=1pt] (150:1.5ex) to[out=0, in=0] (210:1.5ex);
      }}}
\newcommand{\uncrossedLLGinR}{
  \mathrel{
    \tikz[baseline=-0.8ex]{
      \draw[line width=1pt] (0,0) circle (1.5ex);
      \draw[red, line width=1pt] (130:1.5ex) to[out=0, in=0] (230:1.5ex);
      \draw[green!60!black, line width=1pt] (150:1.5ex) to[out=0, in=0] (210:1.5ex);
      }}}
\newcommand{\uncrossedLRB}{
  \mathrel{
    \tikz[baseline=-0.8ex]{
      \draw[line width=1pt] (0,0) circle (1.5ex);
      \draw[line width=1pt] (30:1.5ex) to[out=180, in=180] (330:1.5ex);
      \draw[line width=1pt] (150:1.5ex) to[out=0, in=0] (210:1.5ex);
      }}}

\newcommand{\marknode}[2]{\tikzmarknode{#1}{#2}}

\newcommand{\contractaa}[5][]{%
\begin{tikzpicture}[overlay, remember picture]
\coordinate (L) at ($(pic cs:#2)+(0,2ex)$);
\coordinate (R) at ($(pic cs:#3)+(0,2ex)$);
\coordinate (Lt) at ($(L)+(0,#4)$);
\coordinate (Rt) at ($(R)+(0,#4)$);
\coordinate (#5) at ($(Lt)!0.5!(Rt)$);
\draw[#1]
  (L) -- (Lt) -- (Rt) -- (R);
\end{tikzpicture}%
}

\newcommand{\arctoR}[4][]{%
\tikz[overlay, remember picture]{
\coordinate (S) at (#2);
\coordinate (T) at ($(pic cs:#3)+(0,2ex)$);
\coordinate (M) at ($(S)!0.5!(T)$);
\coordinate (A) at ($(M)+(0,#4)$);
\draw[#1]
  (S) to[out=90,in=180] (A);
\draw[#1]
  (T) to[out=90,in=0] (A);}%
}
\newcommand{\arctoL}[4][]{%
\tikz[overlay, remember picture]{
\coordinate (S) at (#2);
\coordinate (T) at ($(pic cs:#3)+(0,2ex)$);
\coordinate (M) at ($(S)!0.5!(T)$);
\coordinate (A) at ($(M)+(0,#4)$);
\draw[#1]
  (S) to[out=90,in=0] (A);
\draw[#1]
  (T) to[out=90,in=180] (A);}%
}

\newcommand{\upnode}[3][]{%
\tikz[overlay, remember picture]{
\draw[#1,->]
  ($(pic cs:#2)+(0,2ex)$)
  -- ++(0,#3);
}%
}

\def\spc{\hspace{1pt}}

\begin{document}

\thispagestyle{empty}
\begin{adjustwidth}{-1cm}{-1cm}
\begin{center}
    ~\vspace{9mm}
    
     {\LARGE \bf 
   Baby Universe in a Coupled SYK Model
   }
   \end{center}
    \end{adjustwidth}
    \begin{center}
   \vspace{0.4in}
    
    {\bf Andrew Sontag, Herman Verlinde }

    \vspace{0.4in}
    {
    Department of Physics, Princeton University, Princeton, NJ 08544, USA
    }
    \vspace{0.1in}
    
    {\tt sontaga@princeton.edu, verlinde@princeton.edu}
\end{center}

\vspace{0.4in}

\begin{abstract}
We analyze three saddle points of the path integral computing the partition function of the SYK model with a Maldacena-Qi coupling \cite{Maldacena:2018lmt} in the double scaling limit. The three saddle points are holographically dual to three topologically different spacetimes: a pair of Euclidean black holes (two thermal disks), a thermal AdS$_2$ (a cylinder), and a thermal AdS$_2$ with a baby universe (a cylinder with a handle). We develop explicit chord rules that span and probe these three bulk geometries. We derive the rules by expanding the effective $G,\Sigma$ action in powers of the coupling $\mathcal{J}$ and writing the partition function as a weighted sum of chord diagrams. By slicing the diagrams open, we generate a Hilbert space description on a spatial slice for each saddle point. The Hartle-Hawking chord state for the third saddle point has genuine entanglement between the baby universe and the external spacetimes, providing evidence that a closed universe can support a nontrivial Hilbert space.

\end{abstract}

\def\nspc{\hspace{-1pt}}
\pagebreak
\setcounter{page}{1}
\tableofcontents

\pagebreak

\section{Introduction}\label{sect:intro}

 Finding a microscopic realization of a closed universe cosmology is one of the central challenges in the study of quantum gravity. Besides the fact that we may well live inside one, the analogy between closed universe quantum gravity and black hole interiors \cite{Stanford:2022fdt,Blommaert:2024ftn,Iliesiu:2024cnh,Akers:2022qdl,Engelhardt:2025azi,Akers:2025ahe,Abdalla:2025gzn} provides another motivation for studying this problem. While a direct microscopic description of quantum gravity in a (readily closed) de Sitter space is still in the very early stages of development \cite{Narovlansky:2025tpb,Collier:2025lux,Tietto:2025oxn,marini20263dneardesittergravity}, some interesting recent progress has been made by situating closed baby universes into asymptotically Anti-de Sitter spacetimes \cite{Antonini:2023hdh,Antonini:2024mci,Sasieta:2025vck}, enabling the use of the AdS/CFT holographic dictionary \cite{Maldacena:1997,Witten:1998qj,Gubser:1998bc} to study their quantum properties \cite{Gesteau:2025obm,Engelhardt:2025vsp,Kudler-Flam:2025cki,Liu:2025cml,liu2025filteringcftslargen}. It remains an open question, however, to what extent a well-behaved CFT (or ensemble of such) can encode the semiclassical physics in the interior of a closed universe \cite{Marolf:2020xie,McNamara:2020uza,dong2024nullstatestimeevolution,Antonini:2025ioh} or even if this interior has a nontrivial Hilbert space \cite{usatyuk2025closeduniversesfactorizationensemble,Usatyuk_2024,Harlow:2025pvj}.

This motivates us to study the possibility of creating a closed baby universe in the context of a soluble model of low dimensional holography, the double scaled SYK (DSSYK) model. DSSYK is a fermionic quantum many-body model most commonly studied as a candidate dual to quantum gravity in  AdS$_2$ \cite{Berkooz:2018qkz,Berkooz:2018jqr,Goel:2023svz,Lin:2023trc}, while for high temperature, it exhibits features that hint at a duality with de~Sitter gravity \cite{Susskind:2022bia,Rahman:2022jsf,Narovlansky:2023lfz,Verlinde:2024znh,Verlinde:2024zrh,Blommaert:2024ymv,Blommaert:2025eps}. The DSSYK correlation functions can be written as sums over chord diagrams, which can in turn be reformulated as matrix elements in an auxiliary ``chord'' Hilbert space \cite{Berkooz:2024lgq}. The chord diagrams and Hilbert space furnish a practical description for studying the bulk properties of the gravity dual of DSSYK \cite{Lin:2023trc,Lin_2022} and as such provide a logical framework for studying potential quantum realizations of a  closed baby universe and shed light on recent puzzles \cite{Sasieta:2025vck,Engelhardt:2025vsp,Liu:2025cml,Antonini:2025ioh}. Developing the DSSYK chord rule technology in the regime where the bulk dual has non-trivial topology is a first necessary step. This is the goal of our project.

We study three topologically inequivalent phases of the doubled scaled SYK model \cite{Berkooz:2018jqr,Kitaev:2017awl,Maldacena:2016hyu} with Maldacena-Qi coupling \cite{Maldacena:2018lmt} and derive an explicit set of effective chord rules in each of them.  The three phases are dual to three topologically different spacetimes: a pair of Euclidean black holes, a thermal AdS$_2$, and an AS$^2$ geometry \cite{Antonini:2023hdh}. We develop explicit chord rules that span and probe these three bulk geometries.
The chord rules allow us to slice these geometries into chord states and assign chord Hilbert spaces to disconnected components of that slice. We construct the chord Hartle-Hawking states for each of these phases and analyze their entanglement structures. In the state dual to the AS$^2$ geometry, which contains a baby universe, we find nontrivial entanglement between the external AdS regions and the baby universe.

We begin with a review of the AS$^2$ geometry and AR puzzle in section \ref{subsect:introASSR} and the SYK model in section \ref{subsect:introSYK}. In section \ref{sect:MQPhases} we motivate the use of the coupled SYK model and discuss its topological phases and Hawking-Page transition. In section \ref{sect:ChordRules} we derive the chord rules for these phases and show that they can be phrased in a simple form. In section \ref{sect:ChordStates} we use these chord rules to define a microscopic chord Hilbert space, Hamiltonian, and Hartle-Hawking state for each phase. In section \ref{sect:ASSRinSYK} we construct the AS$^2$ geometry within the SYK model and repeat the above analysis to endow it with a microscopic chord description, showing nontrivial entanglement with the baby universe. Appendices \ref{app:A}-\ref{app:D} contain the explicit computations supporting the above.

\subsection{Review of AS$^2$ and AR}\label{subsect:introASSR}

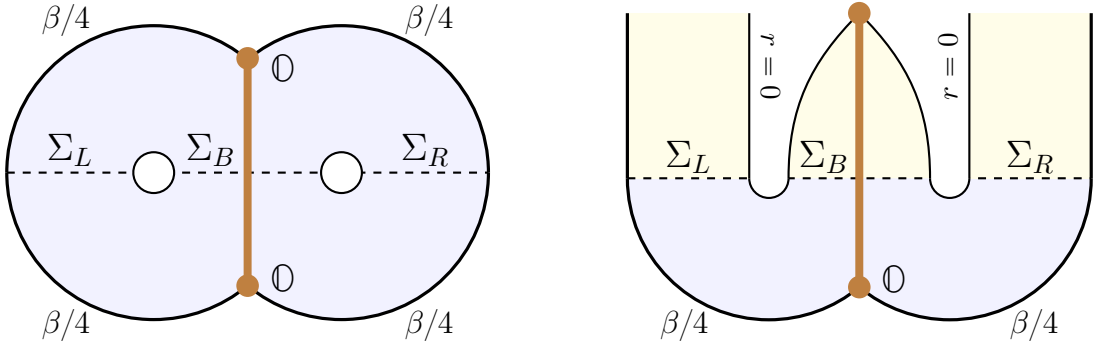
\begin{figure}[t]
\centering
\begin{tikzpicture}[scale=.54]
\def\rout{3.6}
\def\rin{0.5}
\def\br{1.8} 
\def\h{2.8}
\fill[blue!5]
(-\br-\rin-\rout,0)
arc[start angle=180,end angle=-180,x radius=\rout,y radius=\rout] -- cycle;
\fill[blue!5]
(\br+\rin+\rout,0)
arc[start angle=360,end angle=0,x radius=\rout,y radius=\rout] -- cycle;

\draw[thick, dashed] (-\br-\rin-\rout,0) -- (\br+\rin+\rout,0);

\fill[white]
(-\br-2*\rin,0)
arc[start angle=-180,end angle=180,x radius=\rin,y radius=\rin] -- cycle;
\fill[white]
(\br+2*\rin,0)
arc[start angle=0,end angle=360,x radius=\rin,y radius=\rin] -- cycle;

\draw[very thick] (-\br-\rin-\rout,0) arc[start angle=180,end angle={360-acos((\rin+\br)/\rout)},x radius=\rout,y radius=\rout];
\draw[very thick] (\br+\rin+\rout,0) arc[start angle=360,end angle={180+acos((\rin+\br)/\rout)},x radius=\rout,y radius=\rout];
\draw[very thick] (-\br-\rin-\rout,0) arc[start angle=180,end angle={0+acos((\rin+\br)/\rout)},x radius=\rout,y radius=\rout];
\draw[very thick] (\br+\rin+\rout,0) arc[start angle=0,end angle={180-acos((\rin+\br)/\rout)},x radius=\rout,y radius=\rout];
\draw[thick] (-\br-2*\rin,0) arc[start angle=-180,end angle=180,x radius=\rin,y radius=\rin];
\draw[thick] (\br+2*\rin,0) arc[start angle=360,end angle=0,x radius=\rin,y radius=\rin];

\draw[brown, line width=3pt]
    ({0}, {-1*sqrt(\rout^2 - (\br+\rin)^2)}) node[fill=brown, circle, inner sep=3pt] {}
    -- (0,\h) node[fill=brown, circle, inner sep=3pt] {};
\node at ({0.5}, {-1*sqrt(\rout^2 - (\br+\rin)^2)+0.2}) {\ \ \ \Large $\mathbb{O}$};
\node at ({0.5}, -{-1*sqrt(\rout^2 - (\br+\rin)^2)-0.2}) {\ \ \ \Large $\mathbb{O}$};
\node at (-\br-1.5*\rin-\rout/2, 0.55) {\Large $\Sigma_L$};
\node at (\br+1.5*\rin+\rout/2, 0.55) {\Large $\Sigma_R$};
\node at (-\br/2, 0.55) {\Large $\Sigma_B$};
\node at (-\br-\rin-0.5*\rout-0.5*0.7, -0.87*\rout-0.87*0.7) {\large $\beta/4$};
\node at (\br+\rin+0.5*\rout+0.5*0.7, -0.87*\rout-0.87*0.7) {\large $\beta/4$};
\node at (-\br-\rin-0.5*\rout-0.5*0.7, 0.87*\rout+0.87*0.7) {\large $\beta/4$};
\node at (\br+\rin+0.5*\rout+0.5*0.7, 0.87*\rout+0.87*0.7) {\large $\beta/4$};
\medskip
\end{tikzpicture}~~~~~~~~~~~~~
\begin{tikzpicture}[scale=.52]
\def\rout{3.6}
\def\rin{0.5}
\def\br{1.8} 
\def\h{4.2}
\fill[blue!5]
(-\br-\rin-\rout,0)
arc[start angle=180,end angle=360,x radius=\rout,y radius=\rout] -- cycle;
\fill[blue!5]
(\br+\rin+\rout,0)
arc[start angle=360,end angle=180,x radius=\rout,y radius=\rout] -- cycle;
\fill[yellow!10]
(-\br-\rin-\rout,0) -- (-\br-\rin-\rout,\h) -- (-\br-2*\rin,\h) -- (-\br-2*\rin,0) -- cycle;
\fill[yellow!10]
(\br+\rin+\rout,0) -- (\br+\rin+\rout,\h) -- (\br+2*\rin,\h) -- (\br+2*\rin,0) -- cycle;
\fill[yellow!10]
(-\br,0) to[out=90, in=225] (0,\h) to[out=315, in=90] (\br,0) -- cycle;

\draw[thick, dashed] (-\br-\rin-\rout,0) -- (\br+\rin+\rout,0);

\fill[white]
(-\br-2*\rin,0)
arc[start angle=-180,end angle=180,x radius=\rin,y radius=\rin] -- cycle;
\fill[white]
(\br+2*\rin,0)
arc[start angle=0,end angle=360,x radius=\rin,y radius=\rin] -- cycle;

\draw[very thick] (-\br-\rin-\rout,0) arc[start angle=180,end angle={360-acos((\rin+\br)/\rout)},x radius=\rout,y radius=\rout];
\draw[very thick] (\br+\rin+\rout,0) arc[start angle=360,end angle={180+acos((\rin+\br)/\rout)},x radius=\rout,y radius=\rout];
\draw[thick] (-\br-2*\rin,0) arc[start angle=180,end angle=360,x radius=\rin,y radius=\rin];
\draw[thick] (\br+2*\rin,0) arc[start angle=360,end angle=180,x radius=\rin,y radius=\rin];
\draw[very thick] (-\br-\rin-\rout,0) -- (-\br-\rin-\rout,\h);
\draw[very thick] (\br+\rin+\rout,0) -- (\br+\rin+\rout,\h);
\draw[thick] (-\br-2*\rin,0) -- (-\br-2*\rin,\h) node[rotate=-90, xshift=20, yshift=8] {$r=0$};
\draw[thick] (\br+2*\rin,0) -- (\br+2*\rin,\h) node[rotate=90, xshift=-20, yshift=8] {$r=0$};
\draw[thick] (-\br,0) to[out=90, in=225] (0,\h) to[out=315, in=90] (\br,0);

\draw[brown, line width=3pt]
    ({0}, {-1*sqrt(\rout^2 - (\br+\rin)^2)}) node[fill=brown, circle, inner sep=3pt] {}
    -- (0,\h) node[fill=brown, circle, inner sep=3pt] {};
\node at ({0.5}, {-1*sqrt(\rout^2 - (\br+\rin)^2)+0.2}) {\ \ \ \Large $\mathbb{O}$};
\node at (-\br-1.5*\rin-\rout/2, 0.55) {\Large $\Sigma_L$};
\node at (\br+1.5*\rin+\rout/2, 0.55) {\Large $\Sigma_R$};
\node at (-\br/2, 0.55) {\Large $\Sigma_B$};
\node at (-\br-\rin-0.5*\rout-0.5*0.7, -0.87*\rout-0.87*0.7) {\large $\beta/4$};
\node at (\br+\rin+0.5*\rout+0.5*0.7, -0.87*\rout-0.87*0.7) {\large $\beta/4$};
\medskip
\end{tikzpicture}
\caption{Schematic diagrams of the AS$^2$ geometry in full Euclidean space (left panel) and with the top half Wick rotated to Lorentzian spacetime (right panel). The bottom boundary in each panel represents the Euclidean preparation of the two-sided CFT state dual to the dashed Cauchy slice $\Sigma=\Sigma_L\cup\Sigma_B\cup\Sigma_R$ (the $\Sigma_B$ section continues past the brown line). This two-sided state is initialized by a heavy particle insertion $\mathbb{O}$, represented by the brown point, and then evolved by $\beta/4$ Euclidean time on each side. The bulk geometry is prepared by Euclidean gravitational path integral (blue shaded region) with half circle boundary and the same $\mathbb{O}$ insertion. For sufficiently large $\beta$ and sufficiently massive $\mathbb{O}$, it contains a baby universe $\Sigma_B$. The Lorentzian continuation (yellow shaded region) of $\Sigma$ contains two thermal AdS regions and a big bang / big crunch cosmology.}
\label{fig:ASSR}
\end{figure}

It was proposed in \cite{Antonini:2023hdh} that a closed big bang/big crunch universe entangled with a pair of thermal AdS spaces is dual to a partially entangled thermal state (PETS) \cite{Goel:2018ubv} of the CFT living on its shared Euclidean boundary. 
For a matter operator given in the CFT by $\mathbb{O}_{ab}$, this means the CFT state is
\begin{equation}\label{eq:intro_psi_CFT}
    |\Psi_\mathbb{O}\rangle=\frac{1}{\sqrt{Z}}\sum_{a,b}e^{-(\beta/4)(E_a+E_b)}\mathbb{O}_{ab}|E_a\rangle_L|E_b\rangle_R
\end{equation}
where $Z$ ensures that $\langle\Psi_\mathbb{O}|\Psi_\mathbb{O}\rangle=1$. 
At very low temperature, with $\beta$ large enough so that each one-sided system undergoes a Hawking-Page transition, and for a matter operator ${\mathbb{O}_{ab}}$ with large enough conformal dimension, this leads to the AS$^2$ geometry depicted in Figure~\ref{fig:ASSR}. The bottom boundary of each panel represents the Euclidean preparation of this two-sided CFT state \eqref{eq:intro_psi_CFT} dual to the dashed Cauchy slice $\Sigma=\Sigma_L\cup\Sigma_B\cup\Sigma_R$. The brown point represents $\mathbb{O}_{ab}$, and the black lines indicate the Euclidean time evolution.

The bulk state on $\Sigma$ dual to $|\Psi_\mathbb{O}\rangle$ is prepared by a Euclidean gravitational path integral with half circle boundary and a heavy particle insertion $\mathbb{O}$, whose worldline is the brown line in the figure. If both sides of the system are
below the Hawking-Page temperature, the dominant saddle is two copies of thermal AdS stitched together along the heavy worldline. This geometry is shown in Figure~\ref{fig:ASSR}. The Euclidean portion (left panel) notably has two separate ``holes'' where the areal radius vanishes. Cutting the spacetime along a spatial slice through both holes yields a tri-partitioned Cauchy slice with a portion $\Sigma_B$ disconnected from the two boundary portions. If continued to Lorentzian signature (top half of right panel), this generates a closed big bang/big crunch cosmology (the ``baby universe'') disconnected from the two thermal AdS regions. This closed universe has negative curvature everywhere except for a kink at the worldline created by the heavy operator.

From the bulk path integral perspective, it is clear that the dominant Lorentzian saddle point contains the baby universe, and that the state of this universe shares non-trivial entanglement with the thermal AdS states on left and right. From the CFT perspective, however, this is far less obvious. The CFT state $|\Psi_\mathbb{O}\rangle$ is pure on the combined Hilbert space of left and right and, as argued in \cite{Antonini:2024mci}, equally well describes two copies of thermal AdS with small entangled matter excitations and no additional geometry. The idea that this same CFT state could be dual to two drastically different bulk geometries has come to be known as the AR puzzle.

One proposed solution to the puzzle is that the quantum gravity Hilbert space of a closed universe is trivial: there is only one state. If this is true, there's nothing to entangle with, and the puzzle is solved. But this does make it difficult to explain the experience of a semiclassical observer in such a closed universe. Another solution proposed in \cite{Liu:2025cml} is that the state $|\Psi_\mathbb{O}\rangle$ may not have a well-defined large $N$ limit due to highly oscillatory phases in the matrix elements $\mathbb{O}_{ab}$. This leaves the question of how to define averaged or filtered \cite{liu2025filteringcftslargen} semiclassical bulk observables. A third option is to axiomatically include a classical observer in the path integral with macroscopic entropy \cite{Engelhardt:2025azi,Akers:2025ahe,Abdalla:2025gzn,Harlow:2025pvj,Chen:2025fwp}.

To gain more insight, it would be valuable to have a well-controlled microscopic theory that can directly probe the topology of the bulk spacetime and identify states on topologically disconnected segments of the bulk spatial slice. Having such a description would allow us to compute $|\Psi_\mathbb{O}\rangle$ and see which bulk saddle point it is dual to, and thereby test the validity or limitations of each of the above arguments. To address this, we turn to the SYK model.

\subsection{Review of SYK}\label{subsect:introSYK}

The SYK model consists of $N$ Majorana fermions $\{\psi_i,\psi_j\}=2\delta_{ij}$ interacting via a $p$-body Hamiltonian with Gaussian random couplings
\begin{eqnarray}\label{eq:SYKHamiltonian}
   & & H_{\text{SYK}}=i^{p/2}\sum_{1\leq i_1<...<i_p\leq N}J_{i_1...i_p}\psi_{i_1}...\psi_{i_p}\\[3mm]
\label{eq:SYK_J_distribution}
  & & \ \ \ \ \ \   \langle J_{i_1...i_p} J_{i_1...i_p}\rangle\, =\, \frac{2^{p-1} N }{p^2 \spc \Bigl(  {{N}\atop p}  \Bigr)}\,
  {\mathcal{J}}^2
\end{eqnarray}
for $\mathcal{O}(1)$ constant $\mathcal{J}$ with units of energy. The double scaling limit consists of taking $N$ and $p$ both to infinity while holding fixed the ratio $\lambda=2p^2/N$. In this limit, the model's partition function can be expressed as a path integral over the bilocal collective field
\begin{equation}
    G(\tau,\tau')=\frac{1}{N}\sum_{i=1}^N \psi_i(\tau)\psi_i(\tau')
\end{equation}
and its Lagrange multiplier $\Sigma(\tau,\tau')$ \cite{Maldacena:2016hyu}. 

By translation invariance, the saddle point of this path integral is of the form $G(\tau-\tau')$. It can be computed at any $\mathcal{J}$ from the equations of motion
\begin{equation}
    \frac{1}{G(\omega)}=-i\omega-\Sigma(\omega),\quad\Sigma(\tau-\tau')=\frac{\mathcal{J}^2}{p}\big(2G(\tau-\tau')\big)^{p-1}
\end{equation}
where the former is in frequency space and the latter position space. Defining
\begin{equation}
    G(\tau,\tau')=\frac{1}{2}\text{sgn}(\tau-\tau')\Big(1+\frac{g(\tau-\tau')}{p}\Big)
\end{equation}
and solving these equations to leading order in $1/p$ yields the single differential equation\footnote{Specializing here to even $p$. We will eventually be taking $p\equiv0\text{ mod }4$.}
\begin{equation}
    \partial_\tau^2\big[\text{sgn}(\tau)g(t)\big]=2\mathcal{J}^2 e^{g(\tau)}
\end{equation}
whose solution given fermionic boundary conditions is
\begin{equation}
    e^{g(\tau)}=\Bigg(\frac{\cos\frac{\pi v}{2}}{\cos\Big(\pi v\big(\frac{1}{2}-\frac{|\tau|}{\beta}\big)\Big)}\Bigg)^2,\quad\beta\mathcal{J}=\frac{\pi v}{\cos\frac{\pi v}{2}}.
\end{equation}

In the low energy limit $\beta\mathcal{J}\rightarrow\infty$, $-g(\tau)$ becomes\footnote{Up to a difference in renormalization scheme, i.e. an added constant.} the geodesic distance $\ell(\tau)$ between $\tau$-separated points on the boundary of the static Euclidean AdS$_2$ disk of the same inverse temperature $\beta$ \cite{harlow2019factorizationproblemjackiwteitelboimgravity}. This motivates the holographic dictionary
\begin{equation}\label{eq:intro_ell}
\Big\langle\big(2G(\tau)\big)^{p}\Big\rangle\,\longleftrightarrow\,\Big\langle e^{-\ell(\tau)}\Big\rangle
\end{equation}
where the left hand side is a thermal expectation value in DSSYK and the right hand side is the exponential of a path-integrated bulk geodesic length in $D=2$ gravity. Another way to reach this same conclusion is to compute thermal matter correlators
\begin{equation}
    \big\langle\mathbb{O}(\tau)\mathbb{O}(0)\big\rangle=\frac{1}{Z_{\text{SYK}}(\beta)}\Tr\Big[e^{-(\beta-\tau)H_{\text{SYK}}}\,\mathbb{O}^\dagger\;e^{-\tau H_{\text{SYK}}}\,\mathbb{O}\Big]
\end{equation}
with
\begin{equation}
    \mathbb{O}=i^{p'/2}\sum_{1\leq i_1<...<i_{p'}\leq N}K_{i_1...i_{p'}}\psi_{i_1}...\psi_{i_{p'}}\,,\quad\;\;\frac{p'}{p}=\Delta
\end{equation}
where the $K_{i_1...i_{p'}}$ are distributed identically to, but independently of, the $J_{i_1...i_p}$. This correlator can be broken down into a sum over terms of the form $\Tr(H_\text{SYK}^m)$ and $\Tr(H_\text{SYK}^m\mathbb{O}^\dagger H_\text{SYK}^n\mathbb{O})$, each of which permits a chord diagram expansion due to the self-averaging property of the $J$ and $K$ disorder. See \cite{Berkooz:2024lgq} for a more detailed introduction. The upshot is that $\Tr(H_\text{SYK}^m)$ is equal to a weighted sum over chord diagrams where each diagram is a circle with $m$ nodes on it, chords connecting each node to one other node, and weight given by $e^{-\lambda\#}$, where $\#$ is the total number of chord-crossings in the diagram. See the leftmost disk of Figure~\ref{fig:Disks_Chord_Diagram} for an example diagram contributing to $\Tr(H_\text{SYK}^{12})$ with weight $e^{-5\lambda}$. The entire sum is also multiplied by $(\mathcal{J}/\sqrt{\lambda})^m$, see Appendix \ref{app:B}.

The $\Tr(H_\text{SYK}^m\mathbb{O}^\dagger H_\text{SYK}^n\mathbb{O})$ terms are identical, but with a pair of matter nodes added to the circle such that $m$ of the Hamiltonian nodes are between them on one side and $n$ on the other side. A matter chord connects the two matter nodes, and the diagram's penalty is multiplied by $e^{-\lambda\Delta\#'}$ where $\#'$ is the number of Hamiltonian chords crossing the matter chord. In this sense the two-point matter correlator is really measuring the thermal expectation value of $e^{-\lambda\Delta\#'}$. Taking $\ell=\lambda \#'$ shows that this is really the same information contained in our bilocal collective field
\begin{equation}
    \big\langle\mathbb{O}(\tau)\mathbb{O}(0)\big\rangle=\Big\langle\big(2G(\tau)\big)^{p\Delta}\Big\rangle\longleftrightarrow \Big\langle e^{-\Delta\ell(\tau)}\Big\rangle.
\end{equation} 
This reinforces our holographic dictionary, since we expect massive boundary-to-boundary correlators in AdS$_2$ to look like $e^{-\Delta\ell}$. These two methods are interchangeable: we can derive the $G,\Sigma$ equations of motion with chord diagram reasoning, and (crucially for this project) we can derive the chord rules by solving the free $G,\Sigma$ theory exactly and then expanding the partition function in powers of $\mathcal{J}$.

The chord diagrams are particularly useful because they can be ``cut open'' to yield a Hilbert space of chord states. We define the chord state $|n\rangle$ as in \cite{Lin_2022} to represent any configuration of any number of chords anchored on a half circle boundary with precisely $n$ free, unconnected ends (which have not crossed each other in the past). The chord space Hamiltonian $H_\text{chord}$ acts by inserting a new chord node to the left of all existing ones. The chord coming from this node can either become another unconnected end or connect to one of the previously unconnected ends, changing $n$ by $\pm 1$. We define \cite{Berkooz:2018jqr,Lin:2023trc}
\begin{equation}\label{eq:H_chord_SYK}
    \mathfrak{a}^\dagger |n\rangle=|n+1\rangle,\quad\alpha|n\rangle=|n-1\rangle,\quad H_\text{chord}=- \frac{\mathcal{J}}{\sqrt{\lambda}}\Big(\mathfrak{a}^\dagger+\alpha[n]_q\Big)
\end{equation}
where $[n]_q$ is the $q$-deformed integer $n$, given by
\begin{equation}\label{eq:q-deformed_ints}
    [n]_q=\frac{1-q^n}{1-q},\quad q=e^{-\lambda}.
\end{equation}
The norm of one of these chord states is given by the $q$-deformed factorial
\begin{equation}\label{eq:chord_inner_product}
    \langle m | n\rangle = \delta_{mn} [n]_q!=\delta_{mn}\prod_{j=1}^n[j]_q.
\end{equation}
This is the inner product that makes $H_\text{chord}$ Hermitian. These definitions let us write
\begin{align}
    \Tr\Big((H_\text{SYK})^m\Big)&=\langle 0|(H_\text{chord})^m|0\rangle\\ \Tr\Big((H_\text{SYK})^{m_1}\,\mathbb{O}^\dagger\,(H_\text{SYK})^{m_2}\,\mathbb{O}\Big)&=\langle 0|(H_\text{chord})^{m_1}\,e^{-\lambda\Delta n}\,(H_\text{chord})^{m_2}|0\rangle.
\end{align}
The partition function and thermal matter correlators in DSSYK can then all be written purely in terms of chord space variables. This has proved to be a powerful tool for understanding the microscopic origin of complexity, bulk length discretization, and gravitational algebras in DSSYK \cite{Lin_2022,rabinovici2023bulkmanifestationkrylovcomplexity}.

\section{Phases in SYK with MQ coupling}\label{sect:MQPhases}

In order to study the analog of the AS$^2$ geometry in a controlled environment like DSSYK, we need to develop a theory of the latter in which spacetime topology change is possible. The most straightforward example of such a change is the Hawking-Page transition, in which a spacetime changes from one containing a black hole (with a contractible thermal circle) to one containing no black hole (without a contractible thermal circle). In $D$ spacetime dimensions, the topology of a Euclidean spacetime containing a black hole is $B^2\times S^{D-2}$, where the boundary of the disk (the $2$-ball $B^2$) is the thermal circle. The topology without a black hole is that of an $S^1\times B^{D-1}$, where the $S^1$ is the thermal circle. Specializing to $D=2$ then tells us that the black hole geometry should be isomorphic to a pair of disks (since $S^0$ is a pair of points) and the no-black-hole geometry should be isomorphic to a tube (since a $1$-ball $B^1$ is an interval).

 In $D>2$, spacelike boundary slices are isomorphic to $S^{D-2}$ and so are connected. We should expect, then, that generic locally-coupled holographic theories living on these slices should have the fields at every point coupled to their immediate neighbors. In $D=2$, where $S^0 = \{{\rm two \ points}\}$ is disconnected, this local coupling is not manifest. This motivates us to consider a holographic theory composed of a pair of SYK models (one for every point on $S^0$) coupled by a simple interaction. One example of such a theory is the Maldacena-Qi model \cite{Maldacena:2018lmt}. This model consists of two sets of $N$ Majorana fermions $\{\psi_i^{(0,1)}\}$ with Hamiltonian
\begin{equation}\label{eq:H_MQ}
    H_{MQ}=H_{SYK}^{(0)}+H_{SYK}^{(1)}+i\frac{\mu}{p}\sum_{i=1}^N \psi_i^{(0)}\psi_i^{(1)}
\end{equation}
where
$H_{SYK}^{(a)}$ is a copy of \eqref{eq:SYKHamiltonian} labeled by $a\in\{0,1\}$. The couplings $J^{(a)}_{i_1...i_p}$ could potentially also depend nontrivially on $(a)$, but for simplicity we will take them to be equal\footnote{More intricate options where $J_I^{(0)}$ and $J_I^{(1)}$ are non-maximally correlated were explored in \cite{Maldacena:2018lmt}.}
\begin{equation}\label{eq:J^LRstats}
    J_{i_1...i_p}^{(1)}= J_{i_1...i_p}^{(0)}
\end{equation}
with $J_{i_1...i_p}^{(0)}$ distributed as in \eqref{eq:SYK_J_distribution}. We will also take $p\equiv 0 \text{ mod }4$ for the remainder of this paper.

\subsection{Saddles}\label{subsect:saddles}

We want to find the saddle point geometries that dominate the partition function
\begin{equation}\label{eq:ZofBeta}
    Z_{MQ}[\beta]=\Tr\Big(e^{-\beta H_{MQ}}\Big).
\end{equation}
at large $N$, to leading order in large $p$. We can do this as before via path integral with a set of four collective fields\footnote{Similar to those defined in \cite{Maldacena:2018lmt}, but with a shift that will make later calculations easier.}
\begin{equation}\label{eq:defineGs}
    G_{ab}(\tau,\tau')=\frac{(-i)^{(a+b)^2}}{N}\sum_{i=1}^N \psi_i^{(a)}(\tau)\psi_i^{(b)}(\tau')
\end{equation}
and their Lagrange multipliers $\Sigma_{ab}(\tau,\tau')$, where $a$ and $b$ are in $\{0,1\}$. The phase in the definition of the off-diagonal $G_{01}$ and $G_{10}$ ensures that all of the $G_{ab}$ are real. We restrict these fields to satisfy the fermion anticommutation relation $G_{ab}(\tau,\tau')=-G_{ba}(\tau',\tau)$ and antiperiodicity $G_{ab}(\tau+\beta,\tau')=G_{ab}(\tau,\tau'+\beta)=-G_{ab}(\tau,\tau')$. The same restrictions apply to $\Sigma_{ab}$.\footnote{The pieces of $\Sigma_{ab}$ that do not satisfy these restrictions are zero-modes of the action, so we remove them.} 

Integrating out the fermions $\psi_j^{(a)}$ yields the effective action
\begin{align}\label{eq:SeffGSig}
    -\frac{S_{\text{eff}}[G,\Sigma]}{N}&=\frac{1}{2}\log \det \begin{pmatrix} \partial_\tau - \Sigma_{00} & i\frac{\mu}{p}+i\,\Sigma_{01} \\ -i\frac{\mu}{p}+i\,\Sigma_{10} & \partial_\tau - \Sigma_{11}\end{pmatrix} \nonumber\\[2mm]
    &-\frac{1}{2}\iint\limits_{\;00}^{\,\beta\beta}d\tau\, d\tau'\,\sum_{a,b} \Big(\Sigma_{ab}(\tau,\tau')G_{ab}(\tau,\tau')-\frac{\mathcal{J}^2}{2p^2}(2G_{ab}(\tau,\tau'))^p\Big).
\end{align}
On-shell, we know that all of the $G,\Sigma\,(\tau,\tau')$ fields will be time-translation invariant and so can be diagonalized in a single Fourier basis as $\tilde{G},\tilde{\Sigma}\,(\omega)$. Restricted to this subspace, the 4 blocks of the matrix in the kinetic term of \eqref{eq:SeffGSig} commute with each other. We also know that $H_{MQ}$ is unchanged by the antiunitary operator
\begin{equation}\label{eq:antiunitary}
    U\,\Big(z\,\psi_i^{(a)}\Big)=\bar{z}\,(-1)^a\,\psi_i^{(a)}
\end{equation}
which tells us that on-shell
\begin{equation}\label{eq:GXYonshell}
    G_{01}(\tau,\tau')=G_{01}(\tau',\tau)=-G_{10}(\tau,\tau').
\end{equation}
The derivation of the equations of motion for the on-shell $G_{ab}$ can be found in \cite{Maldacena:2018lmt}. We reproduce the ``free'' ($\mathcal{J}=0$, but $\mu\neq 0$) solutions for reference and analysis here. 

We define $\nu=\frac{\mu}{p}$ for brevity, still considering $\mu=\mathcal{O}(p^0)$. The $\mathcal{J}=0$ saddle point is
\begin{align}
    &G_{00}(\tau,\tau')\Big|_{\mathcal{J}=0}=G_{11}(\tau,\tau')\Big|_{\mathcal{J}=0}=\text{sgn}(\tau-\tau')\frac{\cosh\big(\nu(\frac{\beta}{2}-|\tau-\tau'|)\big)}{2\cosh\big(\frac{\nu\beta}{2}\big)}\equiv \overline{G_{00}}(\tau,\tau')\label{eq:G0c}\\[2mm]
    &G_{01}(\tau,\tau')\Big|_{\mathcal{J}=0}=-G_{10}(\tau,\tau')\Big|_{\mathcal{J}=0}=\frac{\sinh\big(\nu(\frac{\beta}{2}-|\tau-\tau'|)\big)}{2\cosh\big(\frac{\nu\beta}{2}\big)}\equiv \overline{G_{01}}(\tau,\tau')\label{eq:G0s}
\end{align}
for $\tau,\tau'\in[0,\beta)$. As in \eqref{eq:intro_ell}, our holographic dictionary tells us that the geodesic distance\footnote{As measured by an isolated test particle. This distinction will be important for the AS$^2$ saddle.} between boundary points $\tau$ and $\tau'$ on circles $a$ and $b$ through the quantum gravity bulk should be given by
\begin{eqnarray}\label{eq:ellab}
   \boxed{\ \  \big\langle e^{-\ell_{ab}(\tau,\tau')}\big\rangle\  \longleftrightarrow \ \Big\langle\big(2G_{ab}(\tau,\tau')\big)^p\Big\rangle\large{}^{\strut}_{\strut} \ }
\end{eqnarray}
We can use this to diagnose the different phases of the model. 

\pagebreak

\begin{figure}[t]
\centering
\begin{tikzpicture}[xscale=1.2,yscale=-1.2]

\draw[ultra thick, ->] (0,0) -- (10,0);
\draw[thick, ->] (9.6,.6) -- (10.3,.6) node[right]{$+\,\mu\beta$};

\foreach \x/\lab in {0/$\mu\beta=0$, 3/$\mu\beta\sim\sqrt{p}$, 6/$\mu\beta=p\log p$} {
    \node[fill=black, circle, inner sep=2pt] at (\x,0) {};
    \node[yshift=-20] at (\x,0) {\lab};
}
\draw[ultra thick, green!60!black] (0,-0.2) -- (3,-0.2);
\node[yshift=25, green!60!black] at (1.4,0) {\large Disks};
\draw[ultra thick, blue] (5.5,-0.2) -- (6.5,-0.2);
\node[yshift=25, blue] at (6,0) {\large Tube};
\draw[ultra thick, purple, ->] (6.5,-0.2) -- (10,-0.2);
\node[yshift=25, purple] at (8.25,0) {\large Ribbon};

\fill[white]
(0,-0.3) -- (-0.3, -0.3) -- (-0.3, 0.3) -- (0, 0.3) -- cycle;
\draw[ultra thick] (0,-0.3) -- (0,0.3);
\end{tikzpicture}
\caption{Axis diagram representing which regions of inverse temperature $\beta$ correspond to which geometric bulk phases of the coupled SYK model. Infinite temperature ($\beta=0$) is on the left edge.}
\label{fig:beta_number_line}
\end{figure}
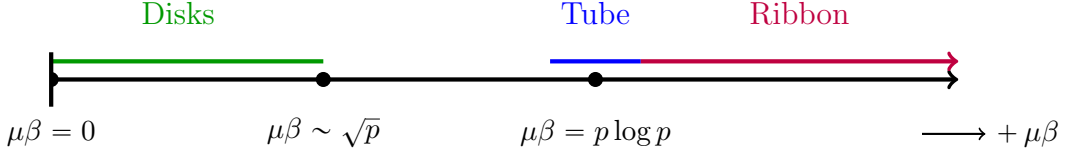

When $\beta=\mathcal{O}(p^0)$, we have $\nu\beta=\mu\beta/p=\mathcal{O}(1/p)$, so at $\mathcal{J}=0$, $\ell_{00}$ and $\ell_{11}$ are both equal to their $\mu=0$ values plus subleading corrections in $p$. The cross-site lengths $\ell_{01}$ and $\ell_{10}$ are both $\gg\mathcal{O}(p)$. The standard DSSYK expansion in $\mathcal{J}$ then yields two uncoupled (and undeformed) disks. This is solidly in the ``disks'' phase up to $\mu\beta\lesssim \sqrt{p}$, represented by the green region in Figure~\ref{fig:beta_number_line}. As the inverse temperature increases to beyond $\mathcal{O}(\sqrt{p})$, these disks start deforming into shallow cigars. This deformation indicates the beginning of the transition to the low temperature phase in which the $\mu$-chords created by the MQ coupling start to condense. 

\begin{figure}[t]
\centering
\includegraphics[width = .9\textwidth]{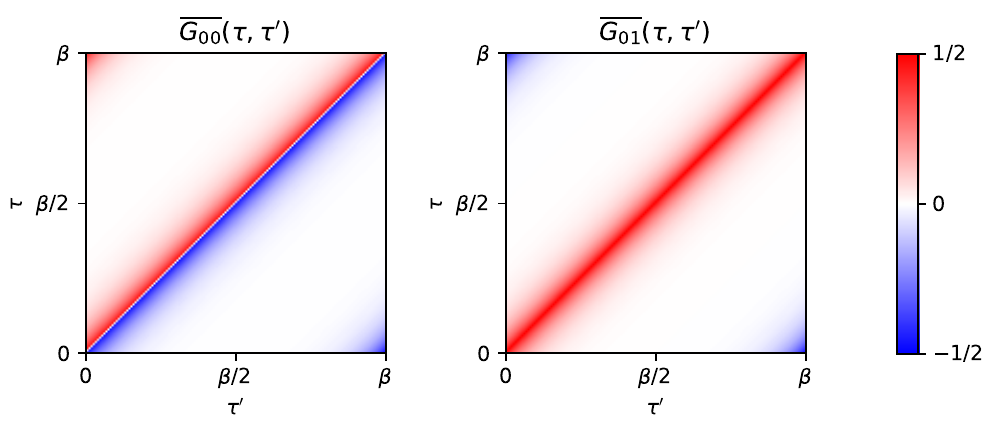}
\caption{Plots of the free cylinder saddle point solution in the DSSYK-MQ model with $\nu\beta=15$, for comparison to Figures \ref{fig:ASSR_2matter_Plots} and \ref{fig:ASSR_1matter_Plots} and also to Appendix \ref{app:C} Figures \ref{fig:ASSR_2matter_Realcolor_Plots} and \ref{fig:ASSR_1matter_Realcolor_Plots}. \label{fig:ASSR_0matter_Plots}}
\end{figure}

In the regime where
\begin{equation}\label{eq:beta_tube}
    \beta=\frac{\alpha}{\mu}\,p\log p\,,\quad \alpha>1
\end{equation}
the $\mu$-chords proliferate. This proliferation manifests itself in the behavior of the geodesic distance $\ell_{ab}$ defined in equation \eqref{eq:ellab}. The transition is already apparent at zero SYK coupling $\mathcal{J}$ where
\bea
    \ell_{ab}(\tau,\tau')\Big|_{\mathcal{J}=0}=\mu\,\min\big(|\tau-\tau'|,\,\beta-|\tau-\tau'|\big)\label{eq:elltubeab}
\eea
for all four combinations of $a,b$. The fact that this length variable grows linearly with $\mu$ times the distance, as well as the presence of the ``min'' function and the nonzero $01$-distance, tells us that the $\mu$-chords have condensed and in effect suppress the ability of other chords to travel finite distances larger than ${\cal  O}(\mu^{-1})$. This is the characteristic property of the ``ribbon'' phase, represented by the purple region in Figure~\ref{fig:beta_number_line}. Expanding in $\mathcal{J}$ gives this ribbon some finite thickness and flares it out near the boundaries, but does not change its topology. For later reference we present a plot of the two-point function in this phase in Figure \ref{fig:ASSR_0matter_Plots}. 

It will be helpful for later discussion to examine more closely the leading order difference between the free $\ell_{00}$ and $\ell_{01}$ very near the transition point $\alpha=1$.  We will parameterize this fine-tuned temperature range by
\bea
\label{eq:beta_transition}
    \beta=\frac{1}{\mu}\,p\log\Big(\frac{2p}{\sigma}\Big)\,,\qquad\sigma=\mathcal{O}(p^0).
\eea
We will call this regime the ``tube'' phase, represented by the blue region in Figure~\ref{fig:beta_number_line}. With this scaling, we find that equation \eqref{eq:elltubeab} is unchanged for $\ell_{00} \Big|_{\mathcal{J}=0}$ and $\ell_{11}\Big|_{\mathcal{J}=0}$, but that now
\bea
    \ell_{01}(\tau,\tau')\Big|_{\mathcal{J}=0}
    \is \mu\,\min\big(|\tau-\tau'|,\,\beta-|\tau-\tau'|\big)\,+\,\sigma\label{eq:elltube_sigma}
\eea
so the tube has an inherent (i.e. $\mathcal{J}=0$) thickness given by $\sigma$. As $\sigma\rightarrow0$, the tube phase smoothly connects to the ribbon phase. Moving forward, we will use the term ``cylinder phase'' for the union of the tube phase and the ribbon phase. Between the tube phase and the disks phase lies a continuum of geometries interpolating between a progressively more spaghettified tube and a pair of progressively more cigar-ish disks. For further discussion see section (5.4) in \cite{Maldacena:2018lmt}.

\subsection{Topological Action}\label{subsect:entropies}

The actions of these saddle points can be readily computed \cite{Maldacena:2018lmt,Maldacena:2016hyu} and include a temperature-independent term which we should think of as a topological contribution. This contribution can be examined without loss of generality at $\mathcal{J}=0$. Here $\Sigma_{ab}=0$ on shell, so our action is purely
\begin{equation}\label{eq:SeffGSigJ0}
    -S_{\text{eff}}^{\mathcal{J}=0}=\frac{N}{2}\log \det \begin{pmatrix} \partial_\tau  & i\nu \\ -i\nu & \partial_\tau \end{pmatrix}=\frac{N}{2}\Tr \log \begin{pmatrix} -i\omega  & i\nu \\ -i\nu & -i\omega \end{pmatrix}
\end{equation}
where we restrict $\omega$ to be $n\pi/\beta$ for odd integer $n$ to satisfy antiperiodicity. Following the literature \cite{Maldacena:2018lmt,Maldacena:2016hyu} and renormalizing properly yields 
\begin{eqnarray}\label{eq:SeffGSigJ0sum}
    -S_{\text{eff}}^{\mathcal{J}=0}=N\log 2 + N \log \cosh \frac{\nu\beta}{2}.
\end{eqnarray}
We see that in the disks phase, this is the standard total microscopic SYK$^2$ entropy $N\log 2$,\footnote{If we actually restrict to this phase by manually setting $G_{01}=G_{10}=0$ in the action, we recover this $N \log 2$ for any $\beta$.} while in the cylinder phase, the factor of $1/2$ in $\cosh$ cancels against this to yield only a term linear in $\beta$ (i.e. zero topological contribution). This makes sense, because the expected topological term in the action should be of the form $S_0\,\chi$ where $e^{S_0}$ is the total dimension of the (single site) SYK Hilbert space, namely $2^{N/2}$, and $\chi$ is the Euler characteristic ($\chi^\text{disks}=2$, $\chi^\text{cylinder}=0$). 


\section{Chord Rules}\label{sect:ChordRules}

As discussed above, much of our understanding of the microscopic behavior of DSSYK comes from our ability to express the partition function as a sum over chord diagrams, with a simple set of rules computing each diagram's weight. 
We can find the DSSYK chord rules by Taylor expanding the collective field action in powers of $\mathcal{J}$, as in Appendix H of \cite{Lin:2023trc}. With a bit more effort, we can reproduce this calculation for the coupled DSSYK partition function, and write down the new set of chord rules at any temperature. In general, these are complicated and not particularly conducive to a clean microscopic picture. However, when the system is in the disks phase, the chord rules are approximately those for two copies of the standard DSSYK disk. And when the system is in the cylinder phase, the chord rules become the analogous set for a cylinder background. We give a bare bones derivation here, but leave the majority of the heavy lifting to Appendices \ref{app:A} and \ref{app:B}.


\subsection{Generic Temperature}\label{subsect:CRGeneric}

We start with the effective $G,\Sigma$ action \eqref{eq:SeffGSig} and set $\mathcal{J}=0$. We then integrate out $\Sigma_{ab}$ to get
\begin{equation}\label{eq:SeffGonly}
    -\frac{S_{\text{eff}}^{\mathcal{J}=0}[G]}{N}=-\frac{1}{2}\Tr \log \begin{pmatrix} G_{00} & iG_{01} \\ iG_{10} & G_{11}\end{pmatrix}+\frac{1}{2}\Tr\Bigg[ \begin{pmatrix} \partial_\tau & i\nu \\ -i\nu & \partial_\tau\end{pmatrix}\begin{pmatrix} G_{00} & iG_{01} \\ iG_{10} & G_{11}\end{pmatrix}-I\Bigg].
\end{equation}
From here we let 
\begin{align}
    &G_{aa}(\tau,\tau')=\overline{G_{00}}(\tau,\tau')+\frac{\text{sgn}(\tau-\tau')}{2p}\delta G_{aa}(\tau,\tau')\label{eq:define_dGLR}\\
    &G_{01}(\tau,\tau')=\overline{G_{01}}(\tau,\tau')+\frac{1}{2p}\delta G_{01}(\tau,\tau')\label{eq:define_dGX}\\
    &G_{10}(\tau,\tau')=-\overline{G_{01}}(\tau,\tau')-\frac{1}{2p}\delta G_{10}(\tau,\tau')\label{eq:define_dGY}
\end{align}
and then expand to leading order in $p$ to find
\begin{equation}\label{eq:SeffdG}
    -S_{\text{eff}}^{\mathcal{J}=0}[G]=\frac{N}{2}\log \det \begin{pmatrix} \partial_\tau & i\nu \\ -i\nu & \partial_\tau\end{pmatrix}+\frac{1}{8\lambda}\int_0^\beta d\tau \int_0^{\tau} d\tau' \;\delta G_{ab}(\tau,\tau') M_{abcd}\, \delta G_{cd}(\tau,\tau')
\end{equation}
where $M_{abcd}$ is a differential operator written explicitly in \eqref{eq:A.defOpm}-\eqref{eq:A.SeffFinal}. Inverting this operator with the appropriate boundary conditions on $\delta G$ allows us to compute $\langle \delta G_{ab}(\tau,\tau')\delta G_{cd}(\sigma,\sigma')\rangle_0$, where by $\langle\cdot\rangle_0$ we mean the expectation value in the $\mathcal{J}=0$ theory.\footnote{Normalized such that $\langle1\rangle_0=1$.} Naming the $\delta G$-independent piece $S_0[\beta]$, we then restore $\mathcal{J}$ to find
\begin{equation}\label{eq:<expJ>}
    Z_{MQ}[\beta]=e^{S_0[\beta]} \;\Bigg\langle \exp\bigg( \frac{\mathcal{J}^2}{\lambda}\int_0^\beta d\tau \int_0^{\tau} d\tau' \sum_{a,b} \Big(2G_{ab}(\tau,\tau')\Big)^p \bigg) \Bigg\rangle_0.
\end{equation}
Taylor expanding in powers of $\mathcal{J}$ and rearranging integrals yields
\begin{align}
    Z_{MQ}[\beta]=&e^{S_0[\beta]}\sum_{\substack{n_{0},n_{1}=0 \\ n_{0}+n_{1} \text{ even}}}^\infty\frac{1}{n_{0}!\;n_{1}!}\Big(\frac{\beta\mathcal{J}}{\sqrt{\lambda}}\Big)^{n_{0}+n_{1}}\sum_{\substack{c=0 \\ n_{0}+c \text{ even}}}^{\min(n_{0},n_{1})}\nonumber\\*[2mm]
    &\qquad \int_D \!\! dV \;\exp\bigg(p\sum_{j} \log|2\overline{G_j}|+\sum_{j<k}\frac{\langle\delta G_j \delta G_{k}\rangle_0}{(2\overline{G_j})(2\overline{G_{k}})}\bigg)\label{eq:Z_sum_diagrams}
\end{align}
where $n_0$ represents the total number of nodes on circle $0$, $n_1$ represents the total number of nodes on circle $1$, and $c$ represents the total number of chords that cross from circle $0$ to circle $1$. The region $D$ describes the locations $\{\tau_j,\tau'_j\}$ of these nodes. Details are given in Appendix \ref{app:B}. The upshot is that this is an integral over chord diagrams with known weights. This integral would become the standard sum over chord diagrams if the weight depended only on the ordering of chord nodes on the boundary circles as opposed to their actual $\tau$ locations. This does not happen at generic temperatures, but approximately does happen (with some qualifications) in each of the distinct phases. The formulae for the term containing $\langle\delta G_j \delta G_{k}\rangle_0$ at generic $\beta$ are given in \eqref{eq:A.GpGp}-\eqref{eq:A.HmHm}, and we summarize them here specialized to the disks and cylinder phases.


\subsection{Disks Phase}\label{subsect:CRDisks}

\begin{figure}[H]
\centering
\begin{tikzpicture}[scale=1.05]

\def\r{1.5}

\draw[very thick, fill=blue!5] (0,0) circle (\r);

\foreach \theone/\thetwo in {30/300, 120/180, 0/90, 240/270, 60/210, 150/330}{
\node[fill=green!60!black, circle, inner sep=2pt](A) at ({\r*cos(\theone)}, {\r*sin(\theone)}) {};
\node[fill=green!60!black, circle, inner sep=2pt](B) at ({\r*cos(\thetwo)}, {\r*sin(\thetwo)}) {};
\draw[green!60!black, thick] (A) to[out=\theone+180, in=\thetwo+180] (B);}

\begin{scope}[shift={(4,0)}]

\draw[very thick, fill=blue!5] (0,0) circle (\r);
\foreach \theone/\thetwo in {0/216, 288/324, 108/252, 36/180, 72/144}{
\node[fill=green!60!black, circle, inner sep=2pt](A) at ({\r*cos(\theone)}, {\r*sin(\theone)}) {};
\node[fill=green!60!black, circle, inner sep=2pt](B) at ({\r*cos(\thetwo)}, {\r*sin(\thetwo)}) {};
\draw[green!60!black, thick] (A) to[out=\theone+180, in=\thetwo+180] (B);}

\node at (0:\r*4) {\Large $\displaystyle=\;\;\Bigg[\bigg(\frac{\mathcal{J}}{\sqrt{\lambda}}\bigg)^{12}e^{-5\lambda}\Bigg]\;\Bigg[\bigg(\frac{\mathcal{J}}{\sqrt{\lambda}}\bigg)^{10}e^{-3\lambda}\Bigg]$};

\end{scope}

\end{tikzpicture}
\caption{One chord diagram contributing to the coupled SYK partition function in the disks phase. The green arcs on the left  represent $00$-chords and those on the right represent $11$-chords. The diagram's weight is computed by counting the numbers of chord nodes and intersections.}
\label{fig:Disks_Chord_Diagram}
\end{figure}
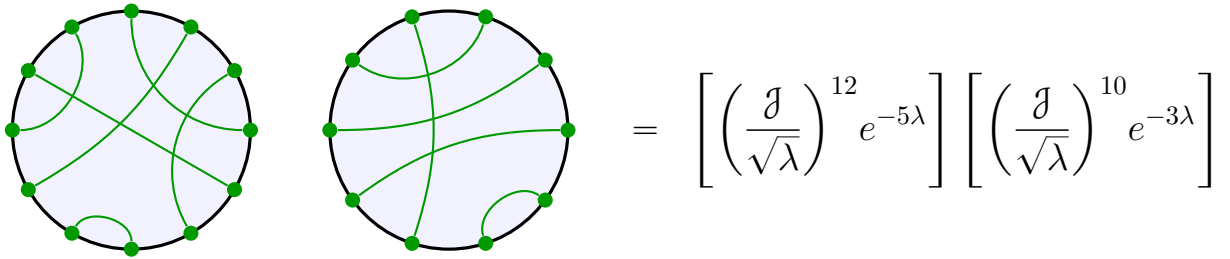

In the disks phase ($\beta=\mathcal{O}(p^0)$) we recover the same chord rules as in two copies of the usual DSSYK model. The first term in the exponential in \eqref{eq:Z_sum_diagrams} puts a large penalty on having any $01$-chords or $10$-chords, $\mathcal{O}(|2\overline{G_{01}}|^p)=\mathcal{O}(p^{-p})$.\footnote{We need to take a little bit of care in this limit, since $\overline{G_{01}}=\mathcal{O}(1/p)$ is the same order as its variation. See Appendix \ref{app:B}.} Any diagrams with these chords in them can be safely ignored (i.e. the sum over $c$ in \eqref{eq:Z_sum_diagrams} is restricted to $c=0$). The analogous penalty for having any $00$-chords or $11$-chords is only $\mathcal{O}(|2\overline{G_{00}}|^p)=1+\mathcal{O}(1/p)$, and we take $1/p\ll\lambda$, so these chords contribute as if we had $\mu=0$. The second term in the exponential in \eqref{eq:Z_sum_diagrams} reduces to the standard DSSYK rule, where $\langle\delta G_{00}(\tau_1,\tau_1')\delta G_{00}(\tau_2,\tau_2')\rangle_0$ gives $-\lambda$ if the chord on the disk connecting $\tau_1$ to $\tau_1'$ crosses that connecting $\tau_2$ to $\tau_2'$, and gives $0$ otherwise. The same rule applies to $\langle\delta G_{11}\delta G_{11}\rangle_0$. The mixed correlator $\langle\delta G_{00}\delta G_{11}\rangle_0$ vanishes. The outer sum in \eqref{eq:Z_sum_diagrams} then factorizes into two sums over single-disk chord diagrams, each with coupling $\mathcal{J}$. See Figure~\ref{fig:Disks_Chord_Diagram} for an example of one term in this  sum.

\medskip

If we push our temperature down to $\beta=\mathcal{O}(\sqrt{p})$, then we start seeing contributions from the MQ coupling in the chord rules, since now the chord penalty for $00$-chords and $11$-chords is $\mathcal{O}(1)$. There are still no contributing diagrams with $01$-chords or $10$-chords. This matches the analysis in section (5.4) of \cite{Maldacena:2018lmt}, where the saddle point at this temperature starts deforming into a shallow cigar.


\subsection{Ribbon Phase}\label{subsect:CRTube1}

At low enough temperature $\beta=\frac{\alpha}{\mu}p\log p$ for $\alpha>1$, the $\mu$-chords condense.  This is the ribbon phase. In this regime, we find that the first term in the exponential in \eqref{eq:Z_sum_diagrams} looks like a ``traveling'' penalty, with every chord being suppressed by 
\bea
\label{eq:travelingp}
\bigl(2\overline{G}(\tau,\tau')\bigr)^p=\, e^{-\ell(\tau,\tau')}\, =\, e^{-\mu\,\min\big((|\tau-\tau'|,\,\beta-|\tau-\tau'|\big)}
\eea
from \eqref{eq:elltubeab}.  This traveling penalty gives rise to an emergent geometry in the shape of a cylinder instead of a pair of disks. All four types of chords ($00$, $11$, $01$, $10$) can contribute to the sum over chord diagrams, but chords that stretch over more than an $\mathcal{O}(1)$ $\tau$-band are exponentially suppressed.\footnote{This statement follows from the extra restriction \eqref{eq:A.large_enough_N}.} This lets us unambiguously choose how each chord wraps the cylinder: there can never be two different wrappings of a single chord that both have $\mathcal{O}(1)$ traveling penalty.

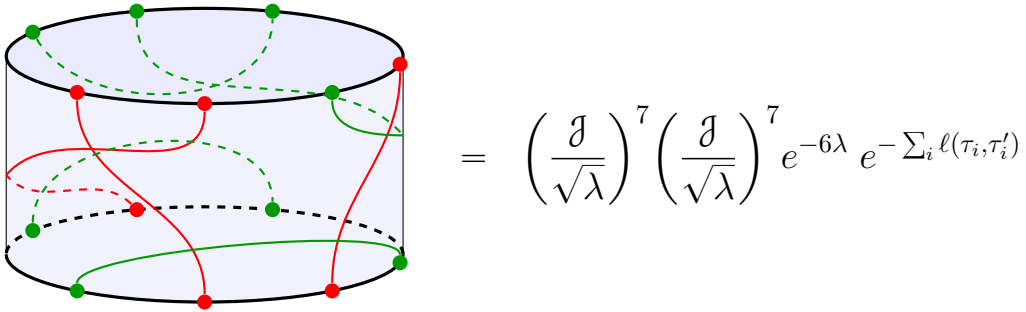
\begin{figure}[H]
\centering
\begin{tikzpicture}[scale=1.05]
\def\r{2.5}      
\def\h{2.5}      
\def\e{0.6}    

\fill[blue!5]
(-\r,0)
arc[start angle=180,end angle=360,x radius=\r,y radius=\e]
-- (\r,\h)
arc[start angle=360,end angle=180,x radius=\r,y radius=\e]
-- cycle;
\draw[very thick, fill=blue!8] (0,\h) ellipse [x radius=\r, y radius=\e];

\draw[very thick] (-\r,0) arc[start angle=180,end angle=360,x radius=\r,y radius=\e];
\draw[very thick, dashed] (\r,0) arc[start angle=0,end angle=180,x radius=\r,y radius=\e];

\draw (-\r,0) -- (-\r,\h);
\draw (\r,0) -- (\r,\h);

\node[fill=red, circle, inner sep=2pt](r1a) at ({\r*cos(310)}, {\e*sin(310)}) {};
\node[fill=red, circle, inner sep=2pt](r1b) at ({\r*cos(350)}, {\h + \e*sin(350)}) {};
\draw[red, thick] (r1a) to[out=90, in=270] (r1b);

\node[fill=red, circle, inner sep=2pt](r2a) at ({\r*cos(270)}, {\e*sin(270)}) {};
\node[fill=red, circle, inner sep=2pt](r2b) at ({\r*cos(230)}, {\h + \e*sin(230)}) {};
\draw[red, thick] (r2a) to[out=90, in=270] (r2b);

\node[fill=red, circle, inner sep=2pt](r3a) at ({\r*cos(110)}, {\e*sin(110)}) {};
\node[fill=red, circle, inner sep=2pt](r3b) at ({\r*cos(270)}, {\h + \e*sin(270)}) {};
\draw[red, thick, dashed] (r3a) to[out=135, in=310] ({-\r},{\h*0.4});
\draw[red, thick] ({-\r},{\h*0.4}) to[out=50, in=270] (r3b);

\node[fill=green!60!black, circle, inner sep=2pt](g1a) at ({\r*cos(350)}, {\e*sin(350)}) {};
\node[fill=green!60!black, circle, inner sep=2pt](g1b) at ({\r*cos(230)}, {\e*sin(230)}) {};
\draw[green!60!black, thick] (g1a) ..controls ({\r*cos(350)}, {\e*sin(350)+0.5}) and ({\r*cos(230)}, {\e*sin(230)+0.5}).. (g1b);

\node[fill=green!60!black, circle, inner sep=2pt](g2a) at ({\r*cos(150)}, {\e*sin(150)}) {};
\node[fill=green!60!black, circle, inner sep=2pt](g2b) at ({\r*cos(70)}, {\e*sin(70)}) {};
\draw[green!60!black, thick, dashed] (g2a) to[out=90, in=90] (g2b);


\node[fill=green!60!black, circle, inner sep=2pt](g4a) at ({\r*cos(150)}, {\h+\e*sin(150)}) {};
\node[fill=green!60!black, circle, inner sep=2pt](g4b) at ({\r*cos(70)}, {\h+\e*sin(70)}) {};
\draw[green!60!black, thick, dashed] (g4a) to[out=300, in=270] (g4b);

\node[fill=green!60!black, circle, inner sep=2pt](g5a) at ({\r*cos(110)}, {\h+\e*sin(110)}) {};
\node[fill=green!60!black, circle, inner sep=2pt](g5b) at ({\r*cos(310)}, {\h+\e*sin(310)}) {};
\draw[green!60!black, thick, dashed] (g5a) to[out=270, in=130] ({\r},{\h*0.6});
\draw[green!60!black, thick] ({\r},{\h*0.6}) to[out=180, in=270] (g5b);

\node at ({\r*2.7},{\h*0.5}) {\Large $\displaystyle=\;\;\bigg(\frac{\mathcal{J}}{\sqrt{\lambda}}\bigg)^{7}\bigg(\frac{\mathcal{J}}{\sqrt{\lambda}}\bigg)^{7}e^{-6\lambda}\;e^{-\sum_i \ell(\tau_i,\tau'_i)}$};

\end{tikzpicture}
\caption{One chord diagram contributing to the coupled SYK partition function in the ribbon phase. The green arcs anchored on the top  represent $00$-chords, and those anchored on the bottom represent $11$-chords. The red lines represent $01$-chords and $10$-chords. The diagram's weight is computed by counting the chord nodes on each boundary ($7$ each) as well as the total number of chord intersections on the cylinder ($6$). The last factor $e^{-\sum_i \ell(\tau_i,\tau'_i)}$  is the traveling penalty.}
\vspace{-2mm}
\label{fig:Tube_Chord_Diagram}
\end{figure}

The second term in the exponential in \eqref{eq:Z_sum_diagrams} gives us the crossing penalties for these chords. If all chords are drawn on a static cylinder, then this term gives $e^{-\lambda}$ for any two chords that cross,\footnote{As in DSSYK, this penalty only applies to chords that \emph{must} cross given their boundary nodes. This is never true of a $00$-chord and a $11$-chord, so $\langle\delta G_{00}\delta G_{11}\rangle_0$ still vanishes.} regardless of their type. For $\langle\delta G_{00}\delta G_{00}\rangle_0$ and $\langle\delta G_{11}\delta G_{11}\rangle_0$, this is actually the same rule as before. But now these chords can also cross $01$-chords and $10$-chords, and these $01$-chords and $10$-chords can also cross each other. With the above restriction in hand to always choose the ``short'' wrapping direction around the tube, these chord rules are unambiguous. See Figure~\ref{fig:Tube_Chord_Diagram} for an example of such a diagram.

\medskip

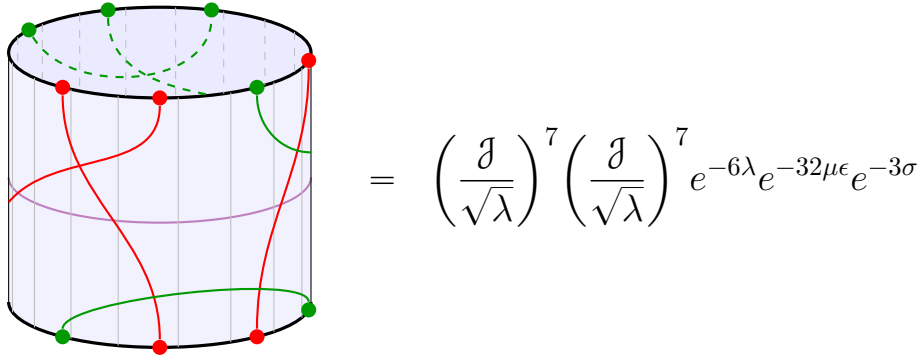
\begin{figure}[H]
\centering
\begin{tikzpicture}
\def\r{2}      
\def\h{3.3}      
\def\e{0.6}    

\fill[blue!5]
(-\r,0)
arc[start angle=180,end angle=360,x radius=\r,y radius=\e]
-- (\r,\h)
arc[start angle=360,end angle=180,x radius=\r,y radius=\e]
-- cycle;
\draw[very thick, fill=blue!8] (0,\h) ellipse [x radius=\r, y radius=\e];

\draw[violet!50, thick] (-\r,\h/2) arc[start angle=180,end angle=360,x radius=\r,y radius=\e];

\draw[very thick] (-\r,0) arc[start angle=180,end angle=360,x radius=\r,y radius=\e];

\draw (-\r,0) -- (-\r,\h);
\draw (\r,0) -- (\r,\h);

\foreach \thet in {193, 214, 234, 254, 277, 298, 319, 340}{
\draw[black!25] ({\r*cos(\thet)},{\e*sin(\thet)}) to ({\r*cos(\thet)},{\h+\e*sin(\thet)});
}
\foreach \thet in {200, 219, 238, 257, 276, 295, 314, 333}{
\draw[black!25, dashed] ({\r*cos(\thet)},{\h-\e*sin(\thet)}) to ({\r*cos(\thet)},{\h+\e*sin(\thet)});
}

\node[fill=red, circle, inner sep=2pt](r1a) at ({\r*cos(310)}, {\e*sin(310)}) {};
\node[fill=red, circle, inner sep=2pt](r1b) at ({\r*cos(350)}, {\h + \e*sin(350)}) {};
\draw[red, thick] (r1a) to[out=90, in=270] (r1b);

\node[fill=red, circle, inner sep=2pt](r2a) at ({\r*cos(270)}, {\e*sin(270)}) {};
\node[fill=red, circle, inner sep=2pt](r2b) at ({\r*cos(230)}, {\h + \e*sin(230)}) {};
\draw[red, thick] (r2a) to[out=90, in=270] (r2b);

\node[fill=red, circle, inner sep=2pt](r3b) at ({\r*cos(270)}, {\h + \e*sin(270)}) {};
\draw[red, thick] ({-\r},{\h*0.4}) to[out=50, in=270] (r3b);

\node[fill=green!60!black, circle, inner sep=2pt](g1a) at ({\r*cos(350)}, {\e*sin(350)}) {};
\node[fill=green!60!black, circle, inner sep=2pt](g1b) at ({\r*cos(230)}, {\e*sin(230)}) {};
\draw[green!60!black, thick] (g1a) ..controls ({\r*cos(350)}, {\e*sin(350)+0.5}) and ({\r*cos(230)}, {\e*sin(230)+0.5}).. (g1b);



\node[fill=green!60!black, circle, inner sep=2pt](g4a) at ({\r*cos(150)}, {\h+\e*sin(150)}) {};
\node[fill=green!60!black, circle, inner sep=2pt](g4b) at ({\r*cos(70)}, {\h+\e*sin(70)}) {};
\draw[green!60!black, thick, dashed] (g4a) to[out=300, in=270] (g4b);

\node[fill=green!60!black, circle, inner sep=2pt](g5a) at ({\r*cos(110)}, {\h+\e*sin(110)}) {};
\node[fill=green!60!black, circle, inner sep=2pt](g5b) at ({\r*cos(310)}, {\h+\e*sin(310)}) {};
\draw[green!60!black, thick, dashed] (g5a) to[out=270, in=170] ({\r*0.342},{\h-\e*0.94});
\draw[green!60!black, thick] ({\r},{\h*0.6}) to[out=180, in=270] (g5b);

\node at ({\r*3.2},{\h*0.5}) {\Large $\displaystyle=\;\;\bigg(\frac{\mathcal{J}}{\sqrt{\lambda}}\bigg)^{7}\bigg(\frac{\mathcal{J}}{\sqrt{\lambda}}\bigg)^{7}e^{-6\lambda}e^{-32\mu\epsilon}e^{-3\sigma}$};

\end{tikzpicture}
\caption{The same chord diagram as in Figure~\ref{fig:Tube_Chord_Diagram}, but now with a background of gray $\mu$-chords and a single purple $\sigma$-chord integrated in. The surface of the tube has been made opaque to reduce clutter. The weight is now updated with the total number of $\mu$-chord crossings (including those out of view) and $\sigma$-chord crossings.}
\vspace{-3mm}
\label{fig:Tube_Chord_Diagram_Background}
\end{figure}

The rules described above, while fully exact and practical, do not produce a topological sum of chord diagrams, since the traveling penalty \eqref{eq:travelingp}  depends on the locations of the chord nodes and not just the order of the chord nodes. One way to make the rules fully topological is to integrate in an additional set of $\mu$-chords with the following simple crossing rules. Choose an $\epsilon\ll1/\mu$, and then insert a background of evenly-distributed $\mu$-chord nodes on the top circle $0$ and bottom circle $1$ with the distance between adjacent nodes being $\epsilon$. The $\mu$-chords all go directly from the node on circle $0$ to the corresponding node on circle $1$ without crossing each other. Then the integral over chord diagrams with traveling penalty is represented by a sum over discrete chord diagrams with this $\mu$-chord background, where the penalty for any Hamiltonian chord crossing a $\mu$-chord is $e^{-\mu\epsilon}$. See Figure~\ref{fig:Tube_Chord_Diagram_Background} for an example of such a diagram with its weight.

\bigskip

\subsection{Tube Phase}\label{subsect:CRTube2}

As we discussed at the end of subsection \ref{subsect:saddles}, in the tube phase \eqref{eq:beta_transition} the $01$-chords and $10$-chords that traverse from the top circle to the bottom circle acquire an additional $e^{-\sigma}$ chord factor 
\bea
e^{-\ell_{01}(\tau,\tau')} \is e^{-\sigma}\, e^{-\ell_{00}(\tau,\tau')} 
\eea
from \eqref{eq:elltube_sigma}. In the ribbon phase $e^{-\sigma} = 1$, indicating that the width of the ribbon at zero $\mathcal{J}$-coupling is negligible. When $e^{-\sigma}<1$, on the other hand, the ribbon acquires a finite width and becomes more ``tube shaped." 

If we wished, we could directly incorporate the extra traversing penalty $e^{-\sigma}$ into the definition of the chord factors for the $01$-chords and $10$-chords. Alternatively, we can incorporate its effect by introducing a single background ``$\sigma$-chord'' that wraps the waist of the tube and comes with the chord intersection rule that every Hamiltonian chord crossing it receives a penalty of $e^{-\sigma}$. We depict this chord in purple in Figure~\ref{fig:Tube_Chord_Diagram_Background}. To transition smoothly to the ribbon phase, we simply set $\sigma=0$ and this $\sigma$-chord becomes transparent.



\section{Chord States}\label{sect:ChordStates}

With our chord rules in hand, we can now proceed to slice our diagrams in half at $\tau=0$ and $\tau=\beta/2$ to generate chord states. For the disks phase, this is very straightforward. Our chord Hilbert space is simply two independent copies of the DSSYK (0-particle) chord Hilbert space described in \cite{Lin_2022}, and its Hamiltonian is the sum of their Hamiltonians. We have
\bea
\mathcal{H}_\text{chord}^\text{disks}\is \Big(\mathcal{H}_\text{chord}^\text{0p}\Big)^2=\; \Bigl\{\, |m\rangle_{\textcolor{blue} {0}}\otimes|n\rangle_{\textcolor{blue} {1}},\;\;\;m,n\in\mathbb{Z}_{\geq 0}\, \Bigr\}\\[3.5mm]
&& \hspace{-3mm} H_\text{chord}^\text{disks}=H_\text{chord}^{(\textcolor{blue} {0})}+H_\text{chord}^{(\textcolor{blue} {1})}
\eea
where $|m\rangle_{\textcolor{blue} {0}}$ and $|n\rangle_{\textcolor{blue} {1}}$ are chord number eigenstates, such as (with $m=4$ and $n=3$) 
\begin{equation}
\label{bras}
| m\rangle_{\textcolor{blue} {0}}\otimes |n\rangle_{\textcolor{blue} {1}}  = ~\raisebox{-5mm}{
\begin{tikzpicture} 
 \begin{scope}[yscale=1,xscale=1]
 \tikzset{partial ellipse/.style args={#1:#2:#3}{
        insert path={+ (#1:#3) arc (#1:#2:#3)}
        }
}        
\draw[thick,fill=lightgray,opacity =.3] (0,0)  [partial ellipse=180:360: 8.5mm and 8.5mm];
\draw (0,.22) node {$m$};
\draw (-.85,.2) node {\textcolor{blue}{\scriptsize $0$}};
\draw (.85,.2) node {\textcolor{blue}{\scriptsize $0$}};
\draw[green!30!black,thin] (0,-.05) -- (.3,-.8);
\draw[green!30!black,thin] (.3,-.05) -- (-.3,-.8);
\draw[green!30!black,thin] (.4,-.05) -- (.6,-.6);
\draw[green!30!black,thin] (-.5,-.05) -- (-.6,-.6);
\end{scope}
\end{tikzpicture}} ~\otimes~  \raisebox{-5mm}{\begin{tikzpicture}\begin{scope}[yscale=1,xscale=1]
 \tikzset{partial ellipse/.style args={#1:#2:#3}{
        insert path={+ (#1:#3) arc (#1:#2:#3)}
        }
}       
\draw[green!40!black] (.6,-.6) arc [start angle=60, end angle=130, radius=1];
\draw (0,.22) node {$n$};
\draw[blue] (-.85,.2) node {\scriptsize $1$};
\draw[blue] (.85,.2) node {\scriptsize $1$};
\draw[green!30!black] (.4,-.05) -- (.7,-.5);
\draw[green!30!black] (0,-.05) -- (0,-.8);
\draw[green!30!black] (-.5,-.05) -- (-.6,-.6);
\draw[thick,fill=lightgray,opacity =.3] (0,0)  [partial ellipse=180:360: 8.5mm and 8.5mm];
\end{scope}
\end{tikzpicture}}
\end{equation}
and the $H_\text{chord}^{(a)}$, $a\in\{0,1\}$, are defined as in \eqref{eq:H_chord_SYK}. We will use $H_a$ as shorthand for $H_\text{chord}^{(a)}$. We can compute $\ell_{00}$ and $\ell_{11}$ in this Hilbert space as $\lambda$ times the expectation values of the chord number $n$ in sector $(0)$ and sector $(1)$, respectively. The Hartle-Hawking state for this geometry is unentangled between these sectors
\bea
\label{eq:disks_Hartle_Hawking}
    \big|\,\text{HH}\big\rangle_\text{disks}=e^{-\beta H_0/2}|0\rangle_0 \otimes  e^{-\beta H_1/2}|0\rangle_1.
\eea
In the tube or ribbon phase, we can define the analogous algebra by slicing the cylinder into two halves at $\tau=0$ and $\tau=\beta/2$. See Figure~\ref{fig:Tube_Chord_Diagram_Split} for an illustration. We will perform the analysis first for the tube phase \eqref{eq:beta_transition} and then take $\sigma\rightarrow 0$ to enter the ribbon phase.

\begin{figure}[t]
\centering
\begin{tikzpicture}[scale=1.07]
\def\r{2}      
\def\h{3}      
\def\e{0.6}    

\fill[blue!5]
(-\r,0)
arc[start angle=180,end angle=360,x radius=\r,y radius=\e]
-- (\r,\h)
arc[start angle=360,end angle=180,x radius=\r,y radius=\e]
-- cycle;
\draw[very thick, fill=blue!8] (0,\h) ellipse [x radius=\r, y radius=\e];

\draw[violet!50, thick] (-\r,\h/2) arc[start angle=180,end angle=360,x radius=\r,y radius=\e];
\draw[violet!50, thick, dashed] (\r,\h/2) arc[start angle=0,end angle=180,x radius=\r,y radius=\e];

\draw[very thick] (-\r,0) arc[start angle=180,end angle=360,x radius=\r,y radius=\e];
\draw[very thick, dashed] (\r,0) arc[start angle=0,end angle=180,x radius=\r,y radius=\e];

\draw (-\r,0) -- (-\r,\h);
\draw (\r,0) -- (\r,\h);

\foreach \thet in {193, 214, 235, 256, 277, 298, 319, 340}{
\draw[black!25] ({\r*cos(\thet)},{\e*sin(\thet)}) to ({\r*cos(\thet)},{\h+\e*sin(\thet)});
}
\foreach \thet in {203, 224, 245, 266, 287, 308, 329}{
\draw[black!25, dashed] ({\r*cos(\thet)},{\h-\e*sin(\thet)}) to ({\r*cos(\thet)},{-\e*sin(\thet)});
}

\node[fill=red, circle, inner sep=2pt](r1a) at ({\r*cos(310)}, {\e*sin(310)}) {};
\node[fill=red, circle, inner sep=2pt](r1b) at ({\r*cos(350)}, {\h + \e*sin(350)}) {};
\draw[red, thick] (r1a) to[out=90, in=270] (r1b);

\node[fill=red, circle, inner sep=2pt](r2a) at ({\r*cos(270)}, {\e*sin(270)}) {};
\node[fill=red, circle, inner sep=2pt](r2b) at ({\r*cos(230)}, {\h + \e*sin(230)}) {};
\draw[red, thick] (r2a) to[out=90, in=270] (r2b);

\node[fill=red, circle, inner sep=2pt](r3a) at ({\r*cos(110)}, {\e*sin(110)}) {};
\node[fill=red, circle, inner sep=2pt](r3b) at ({\r*cos(270)}, {\h + \e*sin(270)}) {};
\draw[red, thick, dashed] (r3a) to[out=135, in=310] ({-\r},{\h*0.4});
\draw[red, thick] ({-\r},{\h*0.4}) to[out=50, in=270] (r3b);

\node[fill=green!60!black, circle, inner sep=2pt](g1a) at ({\r*cos(350)}, {\e*sin(350)}) {};
\node[fill=green!60!black, circle, inner sep=2pt](g1b) at ({\r*cos(230)}, {\e*sin(230)}) {};
\draw[green!60!black, thick] (g1a) ..controls ({\r*cos(350)}, {\e*sin(350)+0.5}) and ({\r*cos(230)}, {\e*sin(230)+0.5}).. (g1b);

\node[fill=green!60!black, circle, inner sep=2pt](g2a) at ({\r*cos(150)}, {\e*sin(150)}) {};
\node[fill=green!60!black, circle, inner sep=2pt](g2b) at ({\r*cos(70)}, {\e*sin(70)}) {};
\draw[green!60!black, thick, dashed] (g2a) to[out=90, in=90] (g2b);


\node[fill=green!60!black, circle, inner sep=2pt](g4a) at ({\r*cos(150)}, {\h+\e*sin(150)}) {};
\node[fill=green!60!black, circle, inner sep=2pt](g4b) at ({\r*cos(70)}, {\h+\e*sin(70)}) {};
\draw[green!60!black, thick, dashed] (g4a) to[out=300, in=270] (g4b);

\node[fill=green!60!black, circle, inner sep=2pt](g5a) at ({\r*cos(110)}, {\h+\e*sin(110)}) {};
\node[fill=green!60!black, circle, inner sep=2pt](g5b) at ({\r*cos(310)}, {\h+\e*sin(310)}) {};
\draw[green!60!black, thick, dashed] (g5a) to[out=270, in=130] ({\r},{\h*0.6});
\draw[green!60!black, thick] ({\r},{\h*0.6}) to[out=180, in=270] (g5b);

\node at ({\r*1.5},{\h*0.45}) {\Large $\displaystyle=$};

\begin{scope}[shift={(6,\h*0.5)}]

\draw[very thick, fill=blue!5] (0,0) circle (\r);
\draw[very thick, fill=white] (0,0) circle (\r*0.15);
\draw[violet!50, thick] (0,0) circle (\r*0.5);

\foreach \thet in {0, 180}{
\draw[dashed][very thick] ({\r*cos(\thet)*0.15},{\r*sin(\thet)*0.15}) to ({\r*cos(\thet)},{\r*sin(\thet)});
}

\foreach \thet in {0, 1, 2, 3, 4, 5, 6, 7, 8, 9, 10, 11, 12, 13, 14, 15}{
\draw[black!25] ({\r*cos(11.25+22.5*\thet)},{\r*sin(11.25+22.5*\thet)}) to ({0.15*\r*cos(11.25+22.5*\thet)},{0.15*\r*sin(11.25+22.5*\thet)});
}

\foreach \theone/\thetwo in {310/350, 270/230}{
\node[fill=red, circle, inner sep=1.5pt](A) at ({\r*cos(\theone)*0.15}, {\r*sin(\theone)*0.15}) {};
\node[fill=red, circle, inner sep=2pt](B) at ({\r*cos(\thetwo)}, {\r*sin(\thetwo)}) {};
\draw[red, thick] (A) to[out=\theone, in=\thetwo+180] (B);}

\node[fill=red, circle, inner sep=1.5pt](r3a) at ({\r*cos(110)*0.15}, {\r*sin(110)*0.15}) {};
\node[fill=red, circle, inner sep=2pt](r3b) at ({\r*cos(270)}, {\r*sin(270)}) {};
\draw[red, thick] (r3a) to[out=110, in=90] ({-\r*0.7},{0});
\draw[red, thick] ({-\r*0.7},{0}) to[out=270, in=110] (r3b);

\node[fill=green!60!black, circle, inner sep=2pt](g1a) at ({\r*cos(70)}, {\r*sin(70)}) {};
\node[fill=green!60!black, circle, inner sep=2pt](g1b) at ({\r*cos(150)}, {\r*sin(150)}) {};
\draw[green!60!black, thick] (g1a) to[out=70+180, in=20] ({-\r*0.239},{\r*0.658});
\draw[green!60!black, thick] ({-\r*0.239},{\r*0.658}) to[out=200, in=150+180] (g1b);

\node[fill=green!60!black, circle, inner sep=2pt](g2a) at ({\r*cos(110)}, {\r*sin(110)}) {};
\node[fill=green!60!black, circle, inner sep=2pt](g2b) at ({\r*cos(310)}, {\r*sin(310)}) {};
\draw[green!60!black, thick] (g2a) to[out=290, in=120] ({\r*0.606},{\r*0.35});
\draw[green!60!black, thick] ({\r*0.606}, {\r*0.35}) to[out=300, in=310+180] (g2b);

\node[fill=green!60!black, circle, inner sep=1.5pt](g3a) at ({\r*cos(70)*0.15}, {\r*sin(70)*0.15}) {};
\node[fill=green!60!black, circle, inner sep=1.5pt](g3b) at ({\r*cos(150)*0.15}, {\r*sin(150)*0.15}) {};
\draw[green!60!black, thick] (g3a) to[out=70, in=20] ({-\r*0.103},{\r*0.282});
\draw[green!60!black, thick] ({-\r*0.103},{\r*0.282}) to[out=200, in=150] (g3b);

\node[fill=green!60!black, circle, inner sep=1.5pt](g4a) at ({\r*cos(220)*0.15}, {\r*sin(220)*0.15}) {};
\node[fill=green!60!black, circle, inner sep=1.5pt](g4b) at ({\r*cos(340)*0.15}, {\r*sin(340)*0.15}) {};
\draw[green!60!black, thick] (g4a) to[out=230, in=200] ({\r*0.103},{-\r*0.282});
\draw[green!60!black, thick] ({\r*0.103},{-\r*0.282}) to[out=20, in=350] (g4b);

\end{scope}

\medskip

\end{tikzpicture}
\caption{Left: the same diagram as Figure~\ref{fig:Tube_Chord_Diagram_Background}, but not opaque. Right: the same diagram again but viewed from above, as an annulus. The black dashed line cuts this annulus in half at $\tau=0$ and $\tau=\beta/2$. The bra (top) and ket (bottom) are each a two-sided state, with support on $L$ (left) and $R$ (right) sectors.}
\label{fig:Tube_Chord_Diagram_Split}
\end{figure}
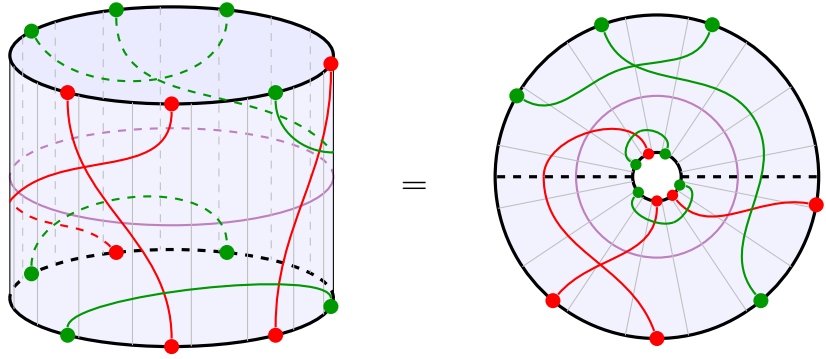

If we ignore the $\mu$-chords, then the ket state on the lower half-annulus looks like a two-sided 1-particle state from standard DSSYK, where the $\sigma$-chord acts as the ``particle'' worldline, with ``mass'' $\Delta_\sigma=\sigma/\lambda$.\footnote{We do not think this ``$\sigma$-particle'' is in any way a real object, just a mathematical trick for keeping track of the number of $01$-chords and $10$-chords.} Calling the left side $L$ and the right side $R$, ket states live in the Hilbert space
\bea
\label{eq:trans_chord_Hilbert_space}
   & &  \mathcal{H}^\text{tube}_\text{chord}=\Big(\mathcal{H}^\text{1p}_\text{chord}\Big)^2=\,    \Bigl\{\,|m_0,m_1\rangle_L\otimes|n_0,n_1\rangle_R,\;\;m_i,n_i \in\mathbb{Z}_{\geq 0}\, \Bigr\}\qquad \\[2mm]
& & \qquad\ \ \raisebox{6mm}{$|m_0,m_1\rangle_L\otimes|n_0,n_1\rangle_R\; =$} \; {\begin{tikzpicture} 
 \begin{scope}[yscale=1,xscale=1]
 \tikzset{partial ellipse/.style args={#1:#2:#3}{
        insert path={+ (#1:#3) arc (#1:#2:#3)}
        }
}        

\draw[thick,fill=lightgray,opacity =.3] (0,0)  [partial ellipse=180:360: 11mm and 11mm];
\draw[thick,fill=lightgray,opacity =.4] (.9,0)  [partial ellipse=180:360: 11mm and 11mm];
\draw[white,fill=white] (0.9,0)  [partial ellipse=180:244: 10.7mm and 10.7mm];
\draw[white,fill=white] (-0.18,0) -- (.45,-.98) -- (1.07,0) -- cycle;
\draw[white,fill=lightgray, opacity = 0.35]  (-.1,-1.13) -- (1,-1.13) -- (.45,-1) -- cycle;
\draw[white,fill=white] (0,0)  [partial ellipse=296:360: 10.7mm and 10.7mm];
\draw[thick, dotted, white,opacity =1] (.9,0)  [partial ellipse=244:267: 11mm and 11mm];
\draw[thick,opacity =.6] (0,0)  [partial ellipse=180:360: 11mm and 11mm];
\draw (-1.3,-.22) node {\small $L$};
\draw (2.2,-.22) node {\small $R$};
\draw (-.9,.22) node {\small $m_0$};
\draw (-.3,.22) node {\small $m_1$};
\draw (1.3,.22) node {\small $n_0$};
\draw (1.87,.22) node {\small $n_1$};
\draw[gray] (-.05,-1.105) -- (.9,-1.105);
\draw[violet, opacity=.6] (.45,0)  [partial ellipse=262:360: 11.2mm and 11.2mm];
\draw[violet,opacity=.4] (.45,0)  [partial ellipse=180:256: 11mm and 11mm];
\draw[thin,green!30!black,opacity=.3] (-.88,0)  [partial ellipse=258:360: 5.7mm and 4.7mm];
\draw[thin,green!30!black,opacity=.3] (1.2,0)  [partial ellipse=250:360: 5.7mm and 4.7mm];
\draw[thin,green!30!black,opacity=.3] (2,0)  [partial ellipse=223:180: 5.7mm and 12.7mm];
\draw[thin,green!30!black,opacity=.3] (0.3,0)  [partial ellipse=220:180: 8mm and 16mm];
\draw[thin,green!30!black,opacity=.3] (1.4,0)  [partial ellipse=230:290: 8mm and 8mm];
\draw[thin,green!30!black,opacity=.3] (-.45,-.17)  [partial ellipse=266:313: 8mm and 8mm];
\draw[green!30!black,thin,opacity=.3] (-.9,-.0) -- (-.7,-.87);
\end{scope}
\end{tikzpicture}}
\eea
The darker gray semicircle is the $(0)$ boundary, and the lighter gray semicircle is the $(1)$ boundary. We can evolve each side of such a state using the transfer matrices
\begin{align}\label{eq:trans_transfer_matrices}
    T^{\rm tube}_A&=T_{A,0}+T_{A,1}\nonumber\\[2mm]
    T_{A,a}&=\mathfrak{a}_a^\dagger+\alpha_a[n_a]_q+q^{n_a}e^{-\sigma}\alpha_{\overline{a}}[n_{\overline{a}}]_q
\end{align}
for $A\in\{L,R\}$ and $a\in\{0,1\}$, operator conventions as in \eqref{eq:H_chord_SYK}-\eqref{eq:q-deformed_ints}. 

To enter the ribbon phase, we send $\sigma\rightarrow 0$ and see our 1-particle states $|n_0,n_1\rangle$ become 0-particle states $|n_0+n_1\rangle$, so
\bea\label{eq:trans_chord_Hilbert_space2}
\mathcal{H}^\text{ribbon}_\text{chord}=\Big(\mathcal{H}^\text{0p}_\text{chord}\Big)^2 \is 
\Bigl\{|m\rangle_L\otimes|n\rangle_R,\;\;\;m,n\in\mathbb{Z}_{\geq 0}\Bigr\}.
\eea
The transfer matrix becomes simply 
\bea\label{eq:tube_transfer_matrices}
T_A^\text{ribbon}\!\!\is 2\big(\mathfrak{a}^\dagger+\alpha[n]_q\big) 
\eea
where the factor of 2 is to be expected, since placing a new chord node on the $(0)$ or the $(1)$ boundary generates two different chord diagrams but the same chord state on $L,R$.

To restore the $\mu$ traveling penalty (anywhere in the cylinder phase), we assign each chord diagram an additional weight of
\begin{align}\label{eq:tube_mu_weight}
    W=e^{-\mu\,\tau_\text{total}}=\exp\Big(-\mu\sum_{\text{chords }j} |\tau_j-\tau'_j|\Big).
\end{align}
This amounts to adding a chemical potential to the chord Hamiltonian that keeps count of the chord number 
\bea
\label{eq:tube_chord_Hamiltonian}
    H^\text{cylinder}_A=-\frac{\mathcal{J}}{\sqrt{\lambda}}\, T^\text{cylinder}_A+\mu\,N_A\equiv H_A
\eea
where $T^\text{cylinder}_A$ is either \eqref{eq:trans_transfer_matrices} or \eqref{eq:tube_transfer_matrices} and $N_A$ is the total chord number\footnote{Not including the $\sigma$-chord} on side $A \in \{L,R\}$.

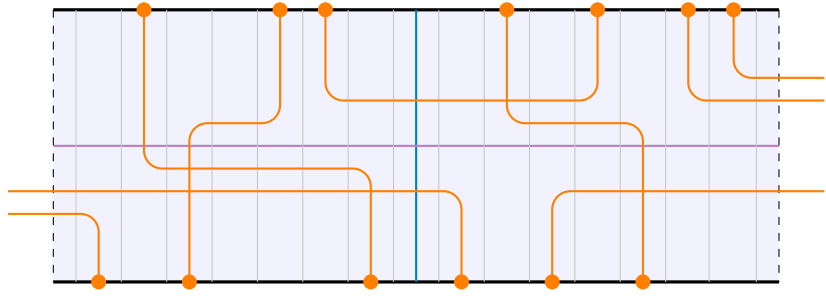
\begin{figure} [t] 
\centering
\begin{tikzpicture}[scale=1.2]

\def\h{3}
\def\r{0.2}

\fill[blue!5] (-4,0) -- (4,0) -- (4,\h) -- (-4,\h) -- cycle;
\draw[very thick] (-4,\h) -- (4,\h);
\draw[very thick] (-4,0) -- (4,0);
\draw[cyan!60!blue, thick] (0,\h) -- (0,0);
\draw[violet!50, thick] (-4,\h/2) -- (4,\h/2);
\draw[dashed] (4,\h) -- (4,0);
\draw[dashed] (-4,\h) -- (-4,0);

\foreach \thet in {-3.75, -3.25, -2.75, -2.25, -1.75, -1.25, -0.75, -0.25, 0.25, 0.75, 1.25, 1.75, 2.25, 2.75, 3.23, 3.75}{
\draw[black!22] (\thet, 0) -- (\thet, \h);
}

\draw[orange, thick]
    (-1,\h) node[fill=orange, circle, inner sep=2pt] {}
    -- (-1,2+\r) arc(180:270:\r) -- (2-\r,2) arc(270:360:\r) -- (2,\h)
    node[fill=orange, circle, inner sep=2pt] {};

\draw[orange, thick]
    (-2.5,0) node[fill=orange, circle, inner sep=2pt] {}
    -- (-2.5,1.75-\r) arc(180:90:\r) -- (-1.5-\r,1.75) arc(270:360:\r) -- (-1.5,\h)
    node[fill=orange, circle, inner sep=2pt] {};

\draw[orange, thick]
    (2.5,0) node[fill=orange, circle, inner sep=2pt] {}
    -- (2.5,1.75-\r) arc(0:90:\r) -- (1+\r,1.75) arc(270:180:\r) -- (1,\h)
    node[fill=orange, circle, inner sep=2pt] {};

\draw[orange, thick]
    (-0.5,0) node[fill=orange, circle, inner sep=2pt] {}
    -- (-0.5,1.25-\r) arc(0:90:\r) -- (-3+\r,1.25) arc(270:180:\r) -- (-3,\h)
    node[fill=orange, circle, inner sep=2pt] {};

\draw[orange, thick]
    (3,\h) node[fill=orange, circle, inner sep=2pt] {}
    -- (3,2+\r) arc(180:270:\r) -- (4.5, 2);

\draw[orange, thick]
    (3.5,\h) node[fill=orange, circle, inner sep=2pt] {}
    -- (3.5,2.25+\r) arc(180:270:\r) -- (4.5, 2.25);

\draw[orange, thick]
    (1.5,0) node[fill=orange, circle, inner sep=2pt] {}
    -- (1.5,1-\r) arc(180:90:\r) -- (4.5, 1);

\draw[orange, thick]
    (-3.5,0) node[fill=orange, circle, inner sep=2pt] {}
    -- (-3.5,0.75-\r) arc(0:90:\r) -- (-4.5, 0.75);

\draw[orange, thick]
    (0.5,0) node[fill=orange, circle, inner sep=2pt] {}
    -- (0.5,1-\r) arc(0:90:\r) -- (-4.5, 1);

\end{tikzpicture}
\medskip

\caption{One chord diagram contributing to the Hartle-Hawking state in the cylinder phase of the DSSYK-MQ model. The top line is half of the $(0)$ boundary, and the the bottom line is half of the $(1)$ boundary. The gray lines are the $\mu$-chords, the purple line is the $\sigma$-chord, and the blue line is the maximally entangled state $|\mathbb{1}\rangle$. The orange lines are Hamiltonian chords, dropping the green/red designation from before since the endpoints of open chords are left undetermined. This diagram comes from the term in \eqref{eq:tube_Hartle_Hawking} with $(H_L)^6(H_R)^7$, since there are 6 nodes on the left side and 7 nodes on the right side of the diagram, and from the term in $|\mathbb{1}\rangle$ corresponding to total chord number 2, since there are 2 orange chords crossing the blue line.}
\label{fig:Tube_HH_Term}
\end{figure}

Putting all of this together gives us the chord representation of the Hartle-Hawking state
\bea\label{eq:tube_Hartle_Hawking}
    \big|\,\text{HH}\big\rangle_\text{cylinder}=e^{-(\beta/4)(H_L+H_R)}|\mathbb{1}\rangle
\eea
where $|\mathbb{1}\rangle$ is the maximally entangled state on $L,R$. By expanding the exponent and using the above chord algebraic rules \cite{Lin:2023trc}, one generates the complete set of chord states. See Figure~\ref{fig:Tube_HH_Term} for a visualization of one term in this exponential.


There are a few important things to notice about this Hartle-Hawking state $|\,\text{HH}\,\rangle_\text{cylinder}$. Since the SYK chord states are not orthonormal, the maximally entangled state $|\mathbb{1}\rangle$ is not simply the sum over $n_i$ of $|n_i\rangle_L|n_i\rangle_R$. We must first compute the Gram matrix on this basis $G(m_i,n_i)=\langle m_i|n_i\rangle$ and then invert it to get
\bea
\label{eq:max_entangled_tube}
    |\mathbb{1}\rangle \is \sum_{m_i,n_i}G^{-1}(m_i,n_i)\;|m_i\rangle_L\;|n_i\rangle_R.
\eea
In the ribbon phase, $G^{-1}(m,n)$ is simply $\delta_{m,n}/\big([n]_q!\big)$. In the tube phase it is more complicated, since 1-particle states with the same total $N_A$ have nontrivial overlap with each other.

 With this definition, we can see that the Hartle-Hawking state could equivalently be expressed solely with $e^{-(\beta/2)H_L}$ or $e^{-(\beta/2)H_R}$, as we would hope.\footnote{ One might also have worried about this preparation including nonphysical or redundant diagrams, such as those where a pair of chords cross both on the left and right sides of the blue line in Figure~\ref{fig:Tube_HH_Term}. If we had not included the inverse Gram matrix in \eqref{eq:max_entangled_tube}, this would indeed be an issue. But with $G^{-1}$ included, we see that $\langle j_i|_L\langle k_i|_R \; |\mathbb{1}\rangle=\langle j_i|k_i\rangle$, and the latter does not include any such diagrams, so our chord basis decomposition of the state $ |\mathbb{1}\rangle$ is correct.}
 As expected, we see that the Hartle-Hawking state $\big|\,\text{HH}\, \rangle_\text{cylinder}$ is an entangled thermofield double state between the $L$ and $R$ sectors. The entanglement entropy between the two sectors is equal to the thermal entropy with temperature $\beta$. We leave the computation of the spectral density in this tube phase (and thus the explicit temperature dependence of this thermal entropy) for future study.


\section{AS$^2$ in SYK}\label{sect:ASSRinSYK}

We will now outline our proposed realization of the AS$^2$ baby universe geometry within the coupled DSSYK-MQ model. A priori, it is not entirely obvious that this can work. The fact that the model with the MQ coupling exhibits a Hawking-Page transition at low temperature is an encouraging sign. Intuitively, one can think of this transition as a BCS-like phase transition in which the $\psi^{(0)}\psi^{(1)}$ pairs condense, thus creating an effective mass for the chords that travel a distance along the boundary. This encourages the Hamiltonian chords to go parallel to the $\mu$-chords, thereby creating an emergent cylindrical geometry that connects the two SYK sites. 

Now imagine that we add an extra very heavy matter chord of weight $\Delta$ that traverses from, say, $\tau=\beta/4$ to $\tau=3\beta/4$. This matter chord needs to find the most economical route to make it from its beginning to end-point. Suppose the system stays in the cylinder phase. The heavy matter chord would then have to travel along the full half circle to the other side, as shown in Figure~\ref{fig:as-squared-one} below. This trajectory crosses all of the $\mu$-chords in that half-circle and thus comes with a traveling penalty of $e^{-\Delta\,  \mu \beta/2}$. Following the AS$^2$ intuition, one would expect that the coupled DSSYK-MQ system will be able to find a more efficient route, at the cost of creating a wormhole geometry with a shorter length than $\beta/2$. An explicit derivation of the bulk saddle point indeed bears out this intuition.

\bigskip

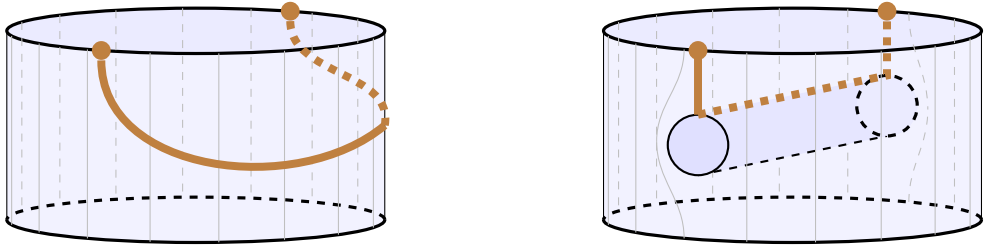
\begin{figure}[H]
\centering
\begin{tikzpicture}[scale=1.0]
\def\r{2.5}      
\def\h{2.5}      
\def\e{0.3}    

\fill[blue!5]
(-\r,0)
arc[start angle=180,end angle=360,x radius=\r,y radius=\e]
-- (\r,\h)
arc[start angle=360,end angle=180,x radius=\r,y radius=\e]
-- cycle;
\draw[very thick, fill=blue!8] (0,\h) ellipse [x radius=\r, y radius=\e];

\draw[very thick] (-\r,0) arc[start angle=180,end angle=360,x radius=\r,y radius=\e];
\draw[very thick, dashed] (\r,0) arc[start angle=0,end angle=180,x radius=\r,y radius=\e];

\draw (-\r,0) -- (-\r,\h);
\draw (\r,0) -- (\r,\h);
\foreach \thet in {203, 224, 245, 266, 287, 308, 329}{
\draw[black!25, dashed] ({\r*cos(\thet)},{\h-\e*sin(\thet)}) to ({\r*cos(\thet)},{-\e*sin(\thet)});
}
\foreach \thet in {193, 214, 235, 256, 277, 298, 319, 340}{
\draw[black!25] ({\r*cos(\thet)},{\e*sin(\thet)}) to ({\r*cos(\thet)},{\h+\e*sin(\thet)});
}

\node[fill=brown, circle, inner sep=2.5pt](g5a) at ({\r*cos(60)}, {\h+\e*sin(60)}) {};
\node[fill=brown, circle, inner sep=2.5pt](g5b) at ({\r*cos(240)}, {\h+\e*sin(240)}) {};
\draw[brown, line width=3pt, dashed] (g5a) to[out=270, in=80] ({\r},{\h*0.5});
\draw[brown,  line width=3pt] ({\r},{\h*0.5}) to[out=220, in=270] (g5b);

\end{tikzpicture}~~~~~~~~~~~~~~~~~~~~~\begin{tikzpicture}
\def\r{2.5}      
\def\h{2.5}      
\def\e{0.3}    
\def\hr{0.4}   

\fill[blue!5]
(-\r,0)
arc[start angle=180,end angle=360,x radius=\r,y radius=\e]
-- (\r,\h)
arc[start angle=360,end angle=180,x radius=\r,y radius=\e]
-- cycle;
\draw[very thick, fill=blue!8] (0,\h) ellipse [x radius=\r, y radius=\e];

\draw[very thick] (-\r,0) arc[start angle=180,end angle=360,x radius=\r,y radius=\e];
\draw[very thick, dashed] (\r,0) arc[start angle=0,end angle=180,x radius=\r,y radius=\e];

\draw (-\r,0) -- (-\r,\h);
\draw (\r,0) -- (\r,\h);

\foreach \thet in {203, 224, 245, 266, 287, 329}{
\draw[black!25, dashed] ({\r*cos(\thet)},{\h-\e*sin(\thet)}) to ({\r*cos(\thet)},{-\e*sin(\thet)});
}
\draw[black!25, dashed] ({\r*cos(308)},{\h-\e*sin(308)}) to[out=270, in=90] ({0.716*\r},{\h/2 +0.695*\e});
\draw[black!25, dashed] ({\r*cos(308)},{-\e*sin(308)}) to[out=90, in=270] ({0.716*\r},{\h/2 +0.695*\e});

\fill[blue!10]
(-\r/2,\h/2-0.866*\e+\hr) -- (\r/2,\h/2+0.866*\e+\hr)
arc[start angle=450,end angle=270,x radius=\hr,y radius=\hr]
-- (-\r/2,\h/2-0.866*\e-\hr);
\draw[thick, dashed] (-\r/2,\h/2-0.866*\e-\hr) -- (\r/2,\h/2+0.866*\e-\hr);

\draw[thick, fill=blue!12] (-\r/2,\h/2-0.866*\e) ellipse [x radius=\hr, y radius=\hr];
\draw[very thick, dashed, fill=blue!8] (\r/2,\h/2+0.866*\e) ellipse [x radius=\hr, y radius=\hr];

\foreach \thet in {193, 214, 256, 277, 298, 319, 340}{
\draw[black!25] ({\r*cos(\thet)},{\e*sin(\thet)}) to ({\r*cos(\thet)},{\h+\e*sin(\thet)});
}
\draw[black!25] ({\r*cos(235)},{\e*sin(235)}) to[out=90, in=270] ({-0.719*\r},{\h/2 -0.695*\e});
\draw[black!25] ({\r*cos(235)},{\h+\e*sin(235)}) to[out=270, in=90] ({-0.719*\r},{\h/2 -0.695*\e});

\draw[brown, line width=3pt]
    (-\r/2,\h-\e*0.866) node[fill=brown, circle, inner sep=2.5pt] {}
    -- (-\r/2,\h/2-0.866*\e+\hr);
\draw[brown, line width=3pt, dashed]
    (-\r/2,\h/2-0.866*\e+\hr) -- (\r/2,\h/2+0.866*\e+\hr)
    -- (\r/2,\h+\e*0.866) node[fill=brown, circle, inner sep=2.5pt] {};


\end{tikzpicture}
\bigskip

\caption{A very heavy matter particle's creation and annihilation operators are placed along the top boundary of the cylinder geometry at opposite sides of the thermal circle. The heavy matter chord traveling along the cylinder (left) accumulates an exponential chord penalty of $e^{-\Delta \,\mu \beta/2}$. For large enough $\Delta$, a wormhole geometry forms (right) allowing the heavy matter chord to follow a more efficient route.}
\label{fig:as-squared-one}
\medskip
\end{figure}

\subsection{AS$^2$ with two heavy operators}

The standard AS$^2$ protocol for creating a baby universe geometry \cite{Antonini:2023hdh} starts by taking a thermal AdS spacetime with topology $S^1\times B^{D-1}$, smearing a heavy matter creation operator $\mathbb{O}$ over the $S^{D-2}$ boundary of the spatial slice at Euclidean time $\tau$, and smearing its Hermitian conjugate over that at Euclidean time $\tau+\beta/2$. The smearing procedure has the advantage that it preserves the rotation symmetry of the thermal AdS spacetime.
Applying the analogous procedure to $D=2$, this means we ``smear'' the heavy operator over a pair of boundary points $S^0$: we insert $\mathbb{O}^{(0)}$ on the $(0)$ boundary at $\tau=\beta/4$ and $\tau=3\beta/4$, and insert $\mathbb{O}^{(1)}$ on the $(1)$ boundary at $\tau=\beta/4$ and $\tau=3\beta/4$. The two operators $\mathbb{O}^{(0)}$ and  $\mathbb{O}^{(1)}$ are assumed to have the same number of Majorana oscillators, but are microscopically distinct, meaning they do not have any Wick contractions with each other. 

Following this scenario in the coupled SYK model amounts to analyzing the saddle point of the path integral computing the thermal expectation value 
\bea\label{eq:ASS_Z_2matter}
    Z_{{\rm AS}^2}[\beta] = \Tr\Big[e^{-\frac{\beta}{2}H_{MQ}}\Big(\mathbb{O}^{(0)}\mathbb{O}^{(1)}\Big)^\dagger e^{-\frac{\beta}{2}H_{MQ}}\Big(\mathbb{O}^{(0)}\mathbb{O}^{(1)}\Big)\Big].
\eea
In the high temperature disks phase, the backreaction due the heavy operator squeezes each disk and leads to a bulk geometry containing two Euclidean black holes with expanded interiors, each with two horizons separated by a domain wall created by the corresponding heavy worldline. We are instead interested in the low temperature regime  $\beta\gtrsim\mathcal{O}(p\log p)$ corresponding to the cylindrical geometry. The mass $\Delta$ of the operators $\mathbb{O}^{(a)}$ should be $\mathcal{O}(1/\lambda)$ and parametrically large enough to create a baby universe. So we will work in the regime where\footnote{The first equation should be compared with the critical temperature \eqref{eq:beta_tube} marking the transition between the disks and cylinder phases. The $\beta$ we need in this scenario is indeed twice what we needed for the standard cylinder phase, matching the AS$^2$ story from section \ref{subsect:introASSR}.}
\bea\label{eq:ASS_regime}
    & & \beta=\frac{\alpha}{\mu}\,p\log p\,,\quad\alpha>2,\,\qquad \quad
1\, \ll \, \lambda \Delta\, \ll \, p.
\eea
Finding the ``free'' saddle point at $\mathcal{J}=0$ is easiest in the $G,\Sigma$ formalism, where we can analyze 
\begin{equation}
    \int\mathcal{D}G\,\mathcal{D}\Sigma\,\Big(2G_{00}(3\beta/4,\beta/4)\cdot 2G_{11}(3\beta/4,\beta/4)\Big)^{p\Delta}\,e^{-S^{\mathcal{J}=0}_\text{eff}[G,\Sigma]}.
\end{equation}
We then consolidate everything into a new effective action $S^{\mathcal{J}=0}_{\text{eff},\Delta}[G,\Sigma]$.
We leave the technical details to Appendix \ref{app:C} and discuss the solutions here.

\begin{figure}[t]
\centering
\hspace{-1cm}
\includegraphics[width = .5\textwidth]{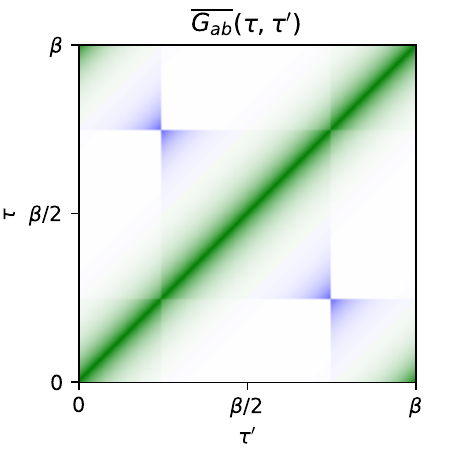}
\caption{Plot of the  chord factor $|\overline{G_{ab}}|$ for the AS$^2$ saddle point with two heavy matter insertions, colored to match the terms in equations \eqref{eq:ASS_2matter_Aab_body} and \eqref{eq:ASS_2matter_Bab_body}. The green diagonal ridge describes the short chords that connect nearby points on the (same or opposite) edge of the cylinder geometry. The blue peaks reveal the presence of a wormhole connecting the diametrically opposite points $\beta/4$ and $3\beta/4$.  The sharp-edged transitions at $\tau$ or $\tau'$ equal to $\beta/4$ and $3\beta/4$ indicate the presence of two heavy matter chords, as explained in Appendix \ref{app:D}. This solution is computed with $\nu\beta=15$ and $\lambda\Delta=p/5$. \label{fig:ASSR_2matter_Plots}}
\end{figure}

The saddle point solution for the two-point function $G_{ab}(\tau,\tau')$ of \eqref{eq:ASS_Z_2matter} is computed in Appendix~\ref{app:C}, written analytically in equations \eqref{eq:C.ASS_2matter_first}-\eqref{eq:C.ASS_2matter_z} and shown graphically in Figure~\ref{fig:ASSR_2matter_Realcolor_Plots}. 
Keeping only the leading order terms at large $p$, we find that this saddle point takes on a simple form. It will be convenient to split the single fermion two-point function into two segments according to whether $\tau$ and $\tau'$ are on the same side of the cylinder (as separated by the pair of heavy matter insertions) or on different sides of the cylinder. This is the same as specifying whether $\cos(2\pi\tau/\beta)$ and $\cos(2\pi\tau'/\beta)$ have the same or different sign. So we define:
\bea
{G_{ab}}(\tau,\tau') &=\begin{cases} A_{ab}(\tau,\tau'):&\cos(2\pi\tau/\beta)\cos(2\pi\tau'/\beta)>0\\[1mm]B_{ab}(\tau,\tau'):&\cos(2\pi\tau/\beta)\cos(2\pi\tau'/\beta)<0\end{cases}\label{eq:ASS_2matter_Gab_body}
\eea
The Hamiltonian chord factors are given by the $p$-th power of $2G_{ab}$. From \eqref{eq:C.ASS_2matter_first}-\eqref{eq:C.ASS_2matter_z} we find that to leading order in large $p$, these chord factors reduce to
\begin{align}
      \bigl(2A_{ab}(\tau,\tau')\bigr)^p&=\exp\left(-\min\begin{Bmatrix}\textcolor{green!30!black}{\mu|\tau-\tau'|}\\\textcolor{blue}{\ell_\text{wh}+\mu(\beta/2-|\tau-\tau'|)}\end{Bmatrix}\right)
    \label{eq:ASS_2matter_Aab_body}\\*[3mm]
    \bigl(2B_{ab}(\tau,\tau')\bigr)^p&=\exp\left(-\lambda\Delta-\min\begin{Bmatrix}\textcolor{green!30!black}{\mu|\tau-\tau'|}\\\textcolor{green!30!black}{\mu(\beta-|\tau-\tau'|)}\end{Bmatrix}\right).
    \label{eq:ASS_2matter_Bab_body}
\end{align}
Here each argument of the min functions in \eqref{eq:ASS_2matter_Aab_body} and \eqref{eq:ASS_2matter_Bab_body} represents a sharp peak that dominates at its peak location. The green terms are those we would expect from the usual cylinder geometry and are represented by the diagonal green ridge in Figure~\ref{fig:ASSR_2matter_Plots}. We can see that the location of the green ridge matches the colored portion of the  two-point function in the cylinder phase shown in Figure~\ref{fig:ASSR_0matter_Plots}. 

The blue term in \eqref{eq:ASS_2matter_Aab_body} is new relative to the standard cylinder phase without the heavy operators. Its presence generates a non-negligible amplitude for a Hamiltonian chord to jump across the thermal circle from $\beta/4$ to $3\beta/4$, despite the fact that the distance between these points along the cylinder geometry goes to infinity at large $p$. This chord has found a short-cut: it is traversing the wormhole! This wormhole term is represented by the two blue sharp-edged peaks in Figure~\ref{fig:ASSR_2matter_Plots} centered at  $(\beta/4,3\beta/4)$ and $(3\beta/4,\beta/4)$.

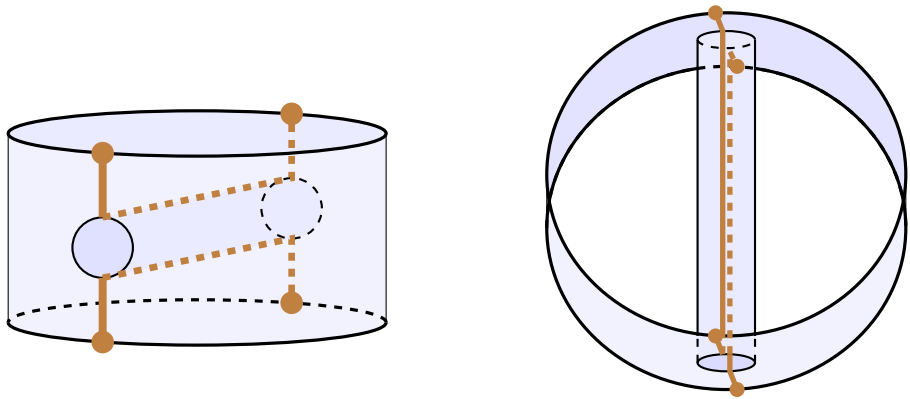
\begin{figure}[t]
\centering
\begin{tikzpicture}
\def\r{2.5}      
\def\h{2.5}      
\def\e{0.3}    
\def\hr{0.4}   

\fill[blue!5]
(-\r,0)
arc[start angle=180,end angle=360,x radius=\r,y radius=\e]
-- (\r,\h)
arc[start angle=360,end angle=180,x radius=\r,y radius=\e]
-- cycle;
\draw[very thick, fill=blue!8] (0,\h) ellipse [x radius=\r, y radius=\e];

\draw[very thick] (-\r,0) arc[start angle=180,end angle=360,x radius=\r,y radius=\e];
\draw[very thick, dashed] (\r,0) arc[start angle=0,end angle=180,x radius=\r,y radius=\e];

\draw (-\r,0) -- (-\r,\h);
\draw (\r,0) -- (\r,\h);

\fill[blue!8]
(-\r/2,\h/2-0.866*\e+\hr) -- (\r/2,\h/2+0.866*\e+\hr)
arc[start angle=450,end angle=270,x radius=\hr,y radius=\hr]
-- (-\r/2,\h/2-0.866*\e-\hr) -- cycle;

\draw[thick, fill=blue!12] (-\r/2,\h/2-0.866*\e) ellipse [x radius=\hr, y radius=\hr];
\draw[thick, dashed] (\r/2,\h/2+0.866*\e) ellipse [x radius=\hr, y radius=\hr];

\draw[brown, line width=3pt]
    (-\r/2,\h-\e*0.866) node[fill=brown, circle, inner sep=3pt] {}
    -- (-\r/2,\h/2-0.866*\e+\hr);
\draw[brown, line width=2.5pt, dashed]
    (-\r/2,\h/2-0.866*\e+\hr) -- (\r/2,\h/2+0.866*\e+\hr)
    -- (\r/2,\h+\e*0.866) node[fill=brown, circle, inner sep=3pt] {};
\draw[brown, line width=3pt]
    (-\r/2,-\e*0.866) node[fill=brown, circle, inner sep=3pt] {}
    -- (-\r/2,\h/2-0.866*\e-\hr);
\draw[brown, line width=2.5pt, dashed]
    (-\r/2,\h/2-0.866*\e-\hr) -- (\r/2,\h/2+0.866*\e-\hr)
    -- (\r/2,\e*0.866) node[fill=brown, circle, inner sep=3pt] {};

\begin{scope}[shift={(7,\h/2)},xscale=.95,yscale=.9]

\def\ne{0.95*\r}
\def\nfac{0.315}
\def\epsi{0.05}
\fill[blue!11, even odd rule]
    (0,0) ellipse [x radius=\r, y radius=\ne]
    (0,\nfac*\h) ellipse [x radius=\r, y radius=\ne];
\fill[blue!5]
    (-\r,0)
    arc[start angle=180,end angle=360,x radius=\r,y radius=\ne]
    -- (\r,\nfac*\h)
    arc[start angle=360,end angle=180,x radius=\r,y radius=\ne]
    -- cycle;
\draw[very thick] (0,0) ellipse [x radius=\r, y radius=\ne];
\fill[blue!8]
    (\hr,-\ne+\nfac*\h/2) arc[start angle=0, end angle=180, x radius=\hr, y radius=\nfac*\hr] -- (-\hr,\ne+\nfac*\h/2) arc[start angle=180, end angle=0, x radius=\hr, y radius=\nfac*\hr] -- cycle;
\draw[very thick] (0,\nfac*\h) ellipse [x radius=\r, y radius=\ne];

\draw[thick, fill=blue!14] (0,-\ne+\nfac*\h/2) ellipse [x radius=\hr, y radius=\nfac*\hr];
\draw[thick, dashed] (-\hr,-\ne+\nfac*\h/2) -- (-\hr,-\ne+\nfac*\h);
\draw[thick] (-\hr,-\ne+\nfac*\h) -- (-\hr,\ne+\nfac*\h/2);
\draw[thick, dashed] (\hr,-\ne+\nfac*\h/2) -- (\hr,-\ne+\nfac*\h);
\draw[thick] (\hr,-\ne+\nfac*\h) -- (\hr,\ne+\nfac*\h/2);
\draw[thick] (-\hr,\ne+\nfac*\h/2) arc[start angle=180, end angle=0, x radius=\hr, y radius=\nfac*\hr];
\draw[thick, dashed] (-\hr,\ne+\nfac*\h/2) arc[start angle=180, end angle=360, x radius=\hr, y radius=\nfac*\hr];
\draw[very thick, dashed] (-\r,0) arc[start angle=180, end angle=0, x radius=\r, y radius=\ne];
\draw[brown, line width=2pt]
    (3*\epsi,-\ne) node[fill=brown, circle, inner sep=2pt] {}
    -- (\epsi,-\ne+\nfac*\h/2-\nfac*\hr) -- (\epsi,-\ne+\nfac*\h/2+\nfac*\hr);
\draw[brown, line width=2pt, dashed]
    (\epsi,-\ne+\nfac*\h/2+\nfac*\hr) -- (\epsi,\ne+\nfac*\h/2-\nfac*\hr-0.05)
    -- (3*\epsi,\ne) node[fill=brown, circle, inner sep=2pt] {};
\draw[brown, line width=2pt]
    (-3*\epsi,-\ne+\nfac*\h) node[fill=brown, circle, inner sep=2pt] {}
    -- (-\epsi,-\ne+\nfac*\h/2+\nfac*\hr);
\draw[brown, line width=2pt, dashed]
    (-\epsi,-\ne+\nfac*\h/2+\nfac*\hr) -- (-\epsi,-\ne+\nfac*\h);
\draw[brown, line width=2pt]
    (-\epsi,-\ne+\nfac*\h) -- (-\epsi,\ne+\nfac*\h/2+\nfac*\hr) --
    (-3*\epsi,\ne+\nfac*\h) node[fill=brown, circle, inner sep=2pt] {};
\draw (-\r,0) -- (-\r,\nfac*\h);
\draw (\r,0) -- (\r,\nfac*\h);
\end{scope}
\end{tikzpicture}
\caption{Suggested bulk geometry dual to the free AS$^2$ saddle point with two heavy matter insertions computed by \eqref{eq:ASS_Z_2matter} and plotted in Figure~\ref{fig:ASSR_2matter_Plots}, shown from two different angles for clarity. The black circles are the $(0)$ and $(1)$ boundaries and the brown lines are the heavy matter worldlines supporting the wormhole.
\label{fig:ASSR_2matter_geometry}}
\end{figure}

The peak value of the blue corner point is equal to $e^{-\ell_{\rm wh}}$, with 
\bea
    \ell_\text{wh} = \frac{p}{2}\, \log\Big(\frac{p}{2\lambda \Delta}\Big).
\eea
We interpret $\ell_{\rm wh}$ as the distance  through the wormhole throat. In the ribbon phase with $\alpha>2$, this distance is much less than the ``long way around'' distance 
\bea
\ell_{\text{long}}\, = \, \mu\beta/2\, >\, p\log p
\eea
between the antipodal points connected by the wormhole.
The distance through the wormhole further decreases as we ramp up the mass of the heavy particles.\footnote{For our plots we take $\lambda\Delta$ to be a small fraction of $p$, since this makes our saddle point independent of $p$ (if we also specify $\nu\beta$). But our conclusions still hold if we say conservatively that $\lambda\Delta$ can only be parametrically large in $p$.} This suggests that the bulk geometry might look something like Figure~\ref{fig:ASSR_2matter_geometry}.


Note, however, that there are no analogous wormhole terms in the $B_{ab}$ piece of the two-point function \eqref{eq:ASS_2matter_Bab_body}. This is why the blue peaks in Figure~\ref{fig:ASSR_2matter_Plots} are ``bow-tie'' shaped. This discontinuity is actually exactly what we would expect if we imagined the wormhole throat as a microcosm of a full tube geometry, but supported by precisely 2 insertions of the MQ coupling as opposed to a constant $\mu$. Its interaction term would be
\begin{equation}\label{eq:delta_function_Hint}
    i\mu_0\Big(\delta(\tau-\beta/4)+\delta(\tau-3\beta/4)\Big)\sum_{i=1}^N\psi_i^{(0)}\psi_i^{(1)}.
\end{equation}
We analyze theories of this kind in Appendix \ref{app:D} and confirm this expectation.

Additionally, notice that the extra $\lambda\Delta$ in \eqref{eq:ASS_2matter_Bab_body} causes the green ridge in Figure~\ref{fig:ASSR_2matter_Plots} to appear to ``pinch'' at the locations of the heavy matter insertions. This is because the chords in these locations have to cross the heavy matter worldline, 
and are therefore suppressed by a crossing factor $e^{-\lambda\Delta}$. So the larger we make $\Delta$ to decrease the wormhole throat distance $\ell_\text{wh}$, the more we obstruct the communication between the regions on opposite sides of the heavy worldlines. 


In the limit where the heavy matter particle becomes infinitely heavy, its worldline effectively cuts the bulk geometry into two halves. Assuming that the rest of the bulk geometry maintains its topology, the dual spacetime takes the form of two thermal AdS$_2$ spacetimes each connected to a baby half-universe with an infinitely massive end-of-the-world brane. One would expect that the quantum state in this limit factorizes into a product state, where each factor describes an entangled state between the thermal AdS$_2$ and the baby half-universe. We will study the entanglement structure of the general AS$^2$ state in more detail in subsection \ref{subsect:ASSRHH}.

\subsection{AS$^2$ with one heavy operator}\label{subsect:ASSRmodified}
We can instead consider a modified AS$^2$ geometry with only heavy operator insertions $\mathbb{O}^{(0)}$ on one of the boundaries at $\tau=\beta/4$ and $\tau=3\beta/4$ but without any $\mathbb{O}^{(1)}$ insertions on the opposite boundary.\footnote{In higher dimensions, this amounts to an asymmetric smearing over the $S^{D-2}$ (possibly even just a single heavy particle) that leaves room for unimpeded entry and exit.} This saddle point loses the nice $0 \leftrightarrow 1$ symmetry (the analogue of rotation invariance in the higher dimensional AS$^2$ scenario) that helped us reduce the four independent $\overline{G_{ab}}$ to two, but makes up for this by keeping the two-point function on the empty side $|\overline{G_{11}}(\tau,\tau)|=1/2$ everywhere.  As explained above, our motivation for studying this case is that it should now be possible to go from one side of the heavy matter worldline to the other side without having to cross it. As we will see, this will also restore the missing wormhole terms in the $B_{ab}$ piece of the two-point function. 

As before, we start by analyzing the saddle point of the path integral computing the thermal expectation value
\bea\label{eq:ASS_Z_1matter}
    \tilde{Z}_{{\rm AS}^2}[\beta] = \Tr\Big[e^{-\frac{\beta}{2}H_{MQ}}\Big(\mathbb{O}^{(0)}\Big)^\dagger e^{-\frac{\beta}{2}H_{MQ}}\Big(\mathbb{O}^{(0)}\Big)\Big]
\eea
which we again write in the $G,\Sigma$ formalism with the effective action 
\bea
\int\mathcal{D}G\,\mathcal{D}\Sigma\,\Big(2G_{00}(3\beta/4,\beta/4)\Big)^{p\Delta}\,e^{-S^{\mathcal{J}=0}_\text{eff}[G,\Sigma]}.
\eea
The saddle point solution for the two-point function $G_{ab}(\tau,\tau')$ of \eqref{eq:ASS_Z_1matter} is written analytically in equations \eqref{eq:C.ASS_1matter_first}-\eqref{eq:C.ASS_1matter_z} and shown graphically in Figure~\ref{fig:ASSR_1matter_Realcolor_Plots}. 

\begin{figure}[t]
\centering
\hspace{-1cm}
\includegraphics[width = .5\textwidth]{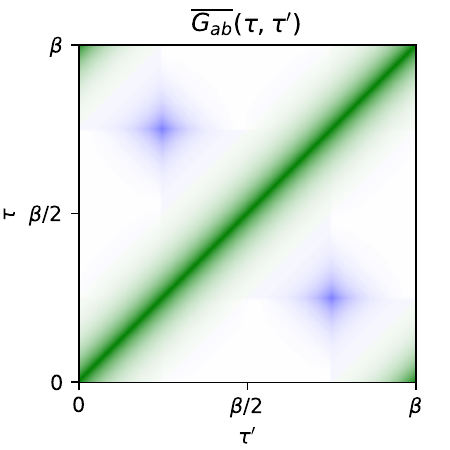}
\caption{Plot of the  chord factor $|\overline{G_{ab}}|$ for the AS$^2$ saddle point with one heavy matter insertion, colored to match the terms in equations \eqref{eq:ASS_2matter_Aab_body} and \eqref{eq:ASS_1matter_Bab_body}. The blue peaks reveal the presence of a wormhole connecting the diametrically opposite points $\beta/4$ and $3\beta/4$. The lack of discontinuities compared to Figure~\ref{fig:ASSR_2matter_Plots} indicates that chords can pass the locations $\beta/4$ and $3\beta/4$ without having to cross the heavy matter chord. This solution is computed with $\nu\beta=15$ and $\lambda\Delta=p/2$.
\label{fig:ASSR_1matter_Plots}}
\end{figure}

In the cylindrical regime \eqref{eq:ASS_regime}, the saddle point for this AS$^2$ geometry leads to the same chord factor $A_{ab}$ (corresponding to when the Hamiltonian chord does not cross $\beta/4$ or $3\beta/4$) to leading order in large $p$ as we found in the first scenario, given in equation \eqref{eq:ASS_2matter_Aab_body}. The leading order $B_{ab}$ chord factor (corresponding to when the Hamiltonian chord does cross $\beta/4$ or $3\beta/4$), on the other hand, now takes on the new form
\bea
\bigl(2B_{ab}(\tau,\tau')\bigr)^p \is \exp\left(-\min\begin{Bmatrix}\textcolor{green!30!black}{\mu|\tau-\tau'|}\\ \textcolor{green!30!black}{\mu(\beta-|\tau-\tau'|)}\\ \textcolor{blue}{\ell_\text{wh}+\mu\big|\beta-(\tau+\tau')\big|}\end{Bmatrix}\right) 
    \label{eq:ASS_1matter_Bab_body}
\eea
and the wormhole length (for both $A_{ab}$ and $B_{ab}$) is now 
\bea
    \ell_\text{wh} = \frac{p}{2}\, \log\Big(\frac{2p}{\lambda \Delta}\Big).\label{eq:ellwh_1matter}
\eea
The total chord factor is depicted graphically in Figure~\ref{fig:ASSR_1matter_Plots}. We see that it exhibits the same green ridge and blue peaks as in Figure \eqref{fig:ASSR_2matter_Plots} but now without the pinch or bow-tie shape. The quantitative reason for the former is that, compared with the previous case, the $B_{ab}$ chord factor no longer contains the $e^{-\lambda \Delta}$, as anticipated, since the Hamiltonian chords can now avoid crossing the heavy matter chord. The latter is again what we should expect for a wormhole throat supported by just one heavy chord as opposed to two, as explained in Appendix \ref{app:D}.


We again observe two blue off-diagonal peaks that highlight the presence of the wormhole. Notice that the blue term in \eqref{eq:ASS_2matter_Aab_body} decays exponentially with $\ell_\text{wh} + \mu(\beta/2-|\tau-\tau'|)$ while in \eqref{eq:ASS_1matter_Bab_body} it decays with $\ell_\text{wh} + \mu|\beta-(\tau+\tau')|$. This is exactly what one would expect geometrically: the $\ell_\text{wh}$ is the traversal distance through the wormhole throat, and the term depending on $\tau$ and $\tau'$ is the sum of the travel distances to and from the wormhole mouths.


\subsection{AS$^2$ in the tube phase}

In this subsection, we study the AS$^2$ geometry with one heavy matter insertion in the fine-tuned temperature range close the Hawking-Page transition. The geometry far from the wormhole mouths then enters the tube phase with a finite width. This regime provides a useful test on our geometric dictionary.


\begin{figure}[t]
\centering
\begin{tikzpicture}[scale=1.05]
\def\r{2.5}      
\def\h{3.75}      
\def\e{0.35}    
\def\hr{0.4}   

\fill[blue!5]
(-\r,0)
arc[start angle=180,end angle=360,x radius=\r,y radius=\e]
-- (\r,\h)
arc[start angle=360,end angle=180,x radius=\r,y radius=\e]
-- cycle;
\draw[very thick, fill=blue!8] (0,\h) ellipse [x radius=\r, y radius=\e];

\draw[very thick] (-\r,0) arc[start angle=180,end angle=360,x radius=\r,y radius=\e];
\draw[thick, dashed] (\r,0) arc[start angle=0,end angle=180,x radius=\r,y radius=\e];

\draw (-\r,0) -- (-\r,\h);
\draw (\r,0) -- (\r,\h);


\fill[blue!10]
(-\r/2,\h/2-0.866*\e+\hr) -- (\r/2,\h/2+0.866*\e+\hr)
arc[start angle=450,end angle=270,x radius=\hr,y radius=\hr]
-- (-\r/2,\h/2-0.866*\e-\hr);
\draw[dashed,thick,opacity=.0] (-\r/2,\h/2-0.866*\e-\hr) -- (\r/2,\h/2+0.866*\e-\hr);

\draw[violet!60, very thick] (-\r,\h/2) arc[start angle=180,end angle=360,x radius=\r,y radius=\e];
\draw[violet!40, very thick, dashed] (\r,\h/2) arc[start angle=0,end angle=180,x radius=\r,y radius=\e];

\draw[purple!40, thick, dashed] (-\r/2-\hr,\h/2-0.866*\e) -- (\r/2-\hr,\h/2+0.866*\e);

\draw[thick, fill=blue!12] (-\r/2,\h/2-0.866*\e) ellipse [x radius=\hr, y radius=\hr];
\draw[thick, dashed, opacity=.5, fill=blue!8] (\r/2,\h/2+0.866*\e) ellipse [x radius=\hr, y radius=\hr];

\draw[purple!40, thick, dashed] (-\r/2+\hr,\h/2-0.866*\e) -- (\r/2+\hr,\h/2+0.866*\e);
\draw[purple!40, thick] (-\r/2-\hr,\h/2-0.866*\e+0.02) -- (-\r/2+\hr-0.05,\h/2-0.866*\e+0.18);



\draw[brown, line width=2.5pt]
    (-\r/2,\h-\e*0.866) node[fill=brown, circle, inner sep=2.5pt] {}
    -- (-\r/2,\h/2-0.866*\e+\hr);
\draw[brown, opacity=.2,line width=2pt, dashed]
    (-\r/2,\h/2-0.866*\e+\hr) -- (\r/2,\h/2+0.866*\e+\hr)
    -- (\r/2,\h-\e*0.866);
\draw[brown, line width=2.5pt, dashed]
    (\r/2,\h-\e*0.866) -- (\r/2,\h+\e*0.866) node[fill=brown, circle, inner sep=2.5pt] {};

\node[fill=green!60!black, circle, inner sep=2pt](gb) at ({-\r*cos(70)}, {\h-\e*sin(70)}) {};
\node[fill=green!60!black, circle, inner sep=2pt](ga) at ({-\r*cos(50)}, {\h-\e*sin(50)}) {};
\draw[green!60!black, thick] (ga) to[out=270, in=90] ({-\r/2-2*\hr},{\h/2-0.866*\e}) arc[start angle=180, end angle=360, x radius=2*\hr, y radius=2*\hr] to[out=90, in=270] (gb);
\draw[<->,thick, violet] (3.5,0) -- (3.5,3.75) node [midway, right] {$\ell_{\rm tube}$};

\end{tikzpicture}
\medskip

\caption{Bulk geometry dual to the AS$^2$ saddle point with one heavy matter particle (indicated by the brown line) plotted in Figure~\ref{fig:ASSR_1matter_Plots}. 
The darker purple line wrapping the tube is the background chord generating $\ell_\text{tube}$ and the green line crossing it twice is a Hamiltonian chord representing the length computed in \eqref{eq:ellASS_2sigma}. The lighter purple lines threading the throat are the background chords generating $\ell_\text{throat}$.
\label{fig:ASSR_1matter_geometry}}
\end{figure}
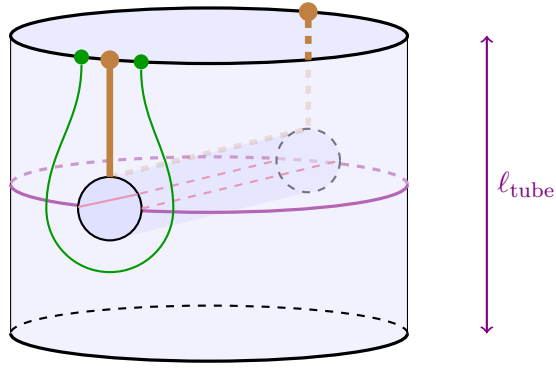


To probe this AS$^2$ tube geometry, we compute the leading order differences in the distances $\ell_{ab}$ far from, nearby, and through the wormhole. We again tune our temperature to the value \eqref{eq:beta_transition}, but now with $\alpha=2$. We let
\bea\label{eq:beta_ASS_transition}
    \beta\is\frac{2}{\mu}\,p\log\Big(\frac{2p}{\sigma}\Big)\,,\quad\sigma=\mathcal{O}(p^0).
\eea
We first consider the same calculation as in \eqref{eq:elltube_sigma}. Using the exact saddle solution in \eqref{eq:C.ASS_1matter_first}-\eqref{eq:C.ASS_1matter_z}, we find that when $|\tau-\tau'|=\mathcal{O}(1)$ and $\tau,\tau'$ sit on the same side of the geometry (i.e. when $G_{ab}=A_{ab}$),
\begin{align}
\ell_{00}^\text{same}(\tau,\tau')\big|_{\mathcal{J}=0}=\ell_{11}^\text{same}(\tau,\tau')\big|_{\mathcal{J}=0}&=\mu|\tau-\tau'|
\\[1.5mm]
    \ell_{01}^\text{same}(\tau,\tau')\big|_{\mathcal{J}=0}
    =\ell_{10}^\text{same}(\tau,\tau')\big|_{\mathcal{J}=0}
    &=\mu|\tau-\tau'|+ \ell_{\rm tube} 
    \label{eq:ellASS_sigma}
\end{align}
which suggests that we are looking at a tube-like bulk with an inherent length
\bea
\ell_{\rm tube} = \, \sqrt{2p \lambda \Delta }\,  e^{-\nu \beta/2}\, = \,  \sigma\sqrt{\frac{\lambda \Delta}{2p}}.
\eea 
This length goes to $0$ as we move from the tube regime into the AS$^2$ ribbon phase. When $|\tau-\tau'|=\mathcal{O}(1)$ but $\tau$ and $\tau'$ are on opposite sides of the geometry (i.e. when $G_{ab}=B_{ab}$) we instead find
\begin{align}
\ell_{11}^\text{opp}(\tau,\tau')\Big|_{\mathcal{J}=0}&=\mu|\tau-\tau'| 
\\*[1.5mm]
    \ell_{01}^\text{opp}(\tau,\tau')\Big|_{\mathcal{J}=0} 
    &=\mu|\tau-\tau'|+ \ell_{\rm tube}
    \\*[1.5mm]
    \ell_{00}^\text{opp}(\tau,\tau')\Big|_{\mathcal{J}=0}&=\mu|\tau-\tau'|+ 2\ell_{\rm tube} 
    \label{eq:ellASS_2sigma}
\end{align}
This is exactly what we would expect for a bulk geometry that looks like Figure~\ref{fig:ASSR_1matter_geometry}. The $\ell_{00}^\text{opp}$ calculation in \eqref{eq:ellASS_2sigma} especially supports this conclusion, as the heavy matter chord suppresses the naive bulk path, making the dominant contribution instead the path circumnavigating the wormhole mouth and crossing the tube waist \emph{twice}, giving an additional distance twice that of $\ell_{01}^\text{same}$ or $\ell_{01}^\text{opp}$. This path is shown by the green chord in Figure~\ref{fig:ASSR_1matter_geometry}. 

Repeating these calculations for chords going through the wormhole
%
leads to the same nice geometric structure, except that all lengths receive an additional contribution equal to the wormhole length \eqref{eq:ellwh_1matter}, and all of the $\ell_\text{tube}$'s are replaced with an analogous
\bea
\ell_\text{throat} = \,  \sigma\sqrt{\frac{2p}{\lambda \Delta}}.\label{eq:ellthroat}
\eea
So in the same way that the tube has an inherent length $\ell_\text{tube}$ in this phase, the wormhole throat has an inherent circumference $2\ell_\text{throat}$ even before turning on the coupling $\mathcal{J}$.



\subsection{Hartle-Hawking state}\label{subsect:ASSRHH}

\begin{figure}[t]
\centering
\begin{tikzpicture}[scale=1.2]

\def\h{3}
\def\r{0.3}

\fill[blue!5] (-4,0) -- (4,0) -- (4,\h) -- (-4,\h) -- cycle;
\draw[very thick] (-4,\h) -- (4,\h);
\draw[very thick] (-4,0) -- (4,0);
\draw[cyan!60!blue, thick] (0,\h) -- (0,0);
\draw[violet!50, thick] (-4,\h/2) -- (4,\h/2);
\draw[dashed] (4,\h) -- (4,0);
\draw[dashed] (-4,\h) -- (-4,0);

\foreach \thet in {-3.75, -3.25, -2.75, -2.25, -1.75, -1.25, -0.75, -0.25, 0.25, 0.75, 1.25, 1.75, 2.25, 2.75, 3.23, 3.75}{\draw[black!22] (\thet, 0) -- (\thet, \h);}

\draw[orange, thick]
    (-1,\h) node[fill=orange, circle, inner sep=2pt] {}
    -- (-1,0.9+\r) arc(180:270:\r) -- (2-\r,0.9) arc(270:360:\r) -- (2,\h)
    node[fill=orange, circle, inner sep=2pt] {};

\draw[orange, thick]
    (-3,0) node[fill=orange, circle, inner sep=2pt] {}
    -- (-3,1.75-\r) arc(180:90:\r) -- (-2-\r,1.75) arc(270:360:\r) -- (-2,\h)
    node[fill=orange, circle, inner sep=2pt] {};

\draw[orange, thick]
    (2.5,0) node[fill=orange, circle, inner sep=2pt] {}
    -- (2.5,1.75-\r) arc(0:90:\r) -- (1+\r,1.75) arc(270:180:\r) -- (1,\h)
    node[fill=orange, circle, inner sep=2pt] {};

\draw[orange, thick]
    (-0.5,0) node[fill=orange, circle, inner sep=2pt] {}
    -- (-0.5,0.65-\r) arc(0:90:\r) -- (-3.5+\r,0.65) arc(270:180:\r) -- (-3.5,\h)
    node[fill=orange, circle, inner sep=2pt] {};


\draw[orange, thick]
    (3,\h) node[fill=orange, circle, inner sep=2pt] {} -- (0,\h/2);

\draw[orange, thick]
    (3.5,\h) node[fill=orange, circle, inner sep=2pt] {}
    -- (3.5,2.25+\r) arc(180:270:\r) -- (4.5, 2.25);

\draw[orange, thick]
    (1.5,0) node[fill=orange, circle, inner sep=2pt] {}
    -- (1.5,0.65-\r) arc(180:90:\r) -- (4.5, 0.65);


\draw[orange, thick]
    (0.5,0) node[fill=orange, circle, inner sep=2pt] {}
    -- (0.5,0.4-\r) arc(0:90:\r) -- (-4.5, 0.4);

\draw[orange, thick]
    (-1.5,0) node[fill=orange, circle, inner sep=2pt] {} -- (0,\h/2);

\draw[orange, thick]
    (-2.5,0) node[fill=orange, circle, inner sep=2pt] {} -- (0,\h/2);

\draw[orange, thick]
    (-4.5,2.25) -- (-0.75-0.383*\r,2.25) arc(90:45:\r) -- (0,\h/2);

\draw[brown, line width=3.5pt]
    (0,\h) node[fill=brown, circle, inner sep=3pt] {} -- (0,\h/2+0.3);
\draw[thick, dashed, fill=white] (0,\h/2) ellipse [x radius=0.4, y radius=0.4];

\end{tikzpicture}
\medskip

\caption{One chord diagram contributing to the Hartle-Hawking state in the AS$^2$ phase of the DSSYK-MQ model. The top and bottom lines are half of the $(0)$ and 
$(1)$ boundaries, respectively. The dashed lines are $L$, $B$, and $R$ from left to right. The gray lines are the $\mu$-chords, and the brown line is the single heavy $\mathbb{O}^{(0)}$ worldline. The purple line is the $\ell_{\rm tube}$ background chord. The maximally entangled state $|\mathbb{1}_\text{AS$^2$}\rangle$ lives on the blue line and on $B$. The orange lines are Hamiltonian chords.
}
\label{fig:ASSR_HH_Term}
\end{figure}
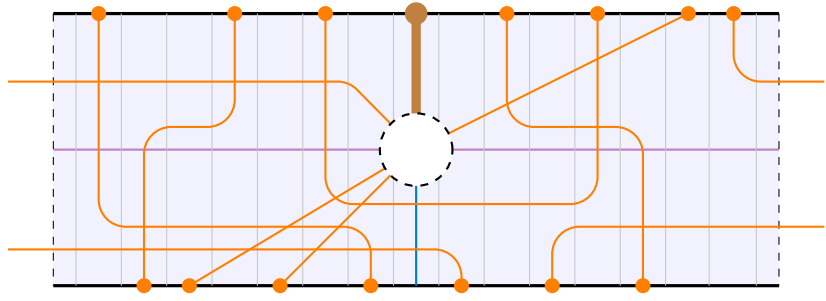

We take all of the above to be strong physical evidence that the chord diagram story in the AS$^2$ phase of the coupled SYK model is analogous to those in the disks and cylinder phases. The partition function is given by a sum over chord diagrams, where each diagram is obtained by placing pairs of nodes on the $(0)$ and $(1)$ boundaries and connecting each pair by the dominant chord with the largest chord factor. 
We can then define a Hartle-Hawking state whose norm gives this partition function. Cutting the geometry in the same way we cut the cylinder geometry in section \ref{sect:ChordStates} gives us a slice with the same $L$ and $R$ sections, but now also a closed circle $B$ section representing the cut through the waist of the wormhole. This is the baby universe. Figure~\ref{fig:ASSR_HH_Term} depicts an example of such a slicing. We will again perform the analysis first for the general AS$^2$ geometry (with one heavy matter insertion) including the AS$^2$ tube phase and then take $\sigma\rightarrow0$ to enter the AS$^2$ ribbon phase.

Chord states for this geometry live in a Hilbert space
\bea\label{eq:ASS_chord_Hilbert_space}
\mathcal{H}^\text{AS$^2$}_\text{chord}\is \mathcal{H}^\text{cylinder}_\text{chord} \otimes\mathcal{H}^\text{bu}_\text{chord}
\eea
with the same Hamiltonian \eqref{eq:tube_chord_Hamiltonian} as in the cylinder phase, but with $\sigma$ in \eqref{eq:trans_transfer_matrices} replaced with $\ell_{\rm tube}$. In the AS$^2$ tube phase, the baby universe is threaded by the heavy matter worldline, a number of Hamiltonian chords, and the background chords generating $\ell_\text{throat}$ from \eqref{eq:ellthroat}. Since Hamiltonian chords don't cross the heavy worldline, the state of the baby universe looks like a 2-particle state from standard DSSYK, where the background $\ell_\text{throat}$ chords act as the ``particle'' worldlines. We will denote the state of the baby universe with $b_i$ Hamiltonian chords by $|b_i\, ; \, \mathbb{O}\,\rangle_B$
so that the Hilbert space sector is
\bea
\label{hbuchord}
\mathcal{H}^\text{bu}_\text{chord}
\is    
\Bigl\{\,|b_i\, ; \, \mathbb{O}\,\rangle_B,\;\;b_i \in\mathbb{Z}_{\geq 0}\, \Bigr\}.
\eea
We prepare ket states as always by enforcing that no two open chords on the same section of the $L,R,B$ slice cross anywhere in our chord diagram. This makes the dynamics in the baby universe very simple, captured entirely by the standard chord inner product \cite{Lin:2023trc}. 

The Hartle-Hawking state is prepared by letting the Hamiltonian act on the infinite temperature PETS state $\big|\mathbb{1}_\text{AS$^2$}\,;\,\mathbb{O}\big\rangle$  prepared by the heavy operator $\mathbb{O}$
\bea
\label{eq:ASSR_Hartle_Hawking}
    \big|\text{HH}_\text{AS$^2$}\big\rangle \is e^{-(\beta/4)(H_L+H_R)}\big|\mathbb{1}_\text{AS$^2$}\,;\,\mathbb{O}\big\rangle.
\eea
See Figure~\ref{fig:ASSR_HH_Term} for a visualization of one term in this exponential. 

The maximally entangled state $|\mathbb{1}_\text{AS$^2$}\,;\,\mathbb{O}\rangle$ has some substructure in this phase. We know it looks like
\begin{align}\label{eq:ASSR_maxent_form}
    \big|\mathbb{1}_\text{AS$^2$}\,;\,\mathbb{O}\big\rangle&=\sum \Psi(m_i,n_i,b_i)\big|m_i\big\rangle_L\,\big|n_i\big\rangle_R\,\big|b_i\, ;\,\mathbb{O}\big\rangle_B\nonumber\\[3mm] &\equiv\sum_{b_i}e^{-\frac{1}{2}\ell_\text{wh}\sum_i b_i}\big|\mathbb{1}_\text{AS$^2$}(b_i)\big\rangle_{LR}\,\big|b_i\, ;\mathbb{O}\big\rangle_B
\end{align}
for some set of states $|\mathbb{1}_\text{AS$^2$}(b_i)\rangle_{LR}$. Here we have chosen to include the $e^{-\ell_\text{wh}}$ wormhole crossing factor in our definition of the state, rather than in the definition of the inner product on $B$. 


 We proceed by computing the matrix elements
\begin{equation}\label{eq:ASSR_maxent_elements}
    H(m_i,n_i\,;b_i)=\bigl\langle m_i\big|_L\,\bigl\langle n_i\big|_R\; \big|\mathbb{1}_\text{AS$^2$}(b_i)\big\rangle_{LR}
\end{equation}
via diagrammatic expansion. 
Our above restriction on the chords composing $b_i$ actually makes this rather straightforward, since we can treat those chords as an additional background for the rest of the calculation.

In the AS$^2$ tube phase, we need to sum over four components $\{b_{0L},b_{1L},b_{0R},b_{1R}\}$ in equation \eqref{eq:ASSR_maxent_form} specifying on which part of the tube boundary each $b$-chord originates.
The diagram shown in Figure~\ref{fig:ASSR_HH_Term} corresponds to $b_{0L}=1$, $b_{1L}=2$, $b_{0R}=1$, and $b_{1R}=0$, since these are the numbers of orange chords coming from the $B$ circle. It is a simple exercise to see that the answer to \eqref{eq:ASSR_maxent_elements} is then
\begin{align}\label{eq:ASSR_maxentH_answer}
H(m_i,n_i\,;b_i) &= \exp\Big(-
\bigg(\ell_{\rm tube}+
\lambda\,b_{1L} \Big)
(m_0-b_{0L})- 
\Big(\ell_{\rm tube}+
\lambda\,b_{1R} \Big)
(n_0-b_{0R})\bigg) \nonumber\\[2mm] &\begin{pmatrix}m_0\\b_{0L}\end{pmatrix}_{\! q}\begin{pmatrix}m_1\\b_{1L}\end{pmatrix}_{\! q}\begin{pmatrix}n_0\\b_{0R}\end{pmatrix}_{\! q}\begin{pmatrix}n_1\\b_{1R}\end{pmatrix}_{\! q}\;\big\langle m_0\nspc +m_1\nspc -b_{0L}\nspc -b_{1L}\spc\big|\,n_0\nspc +n_1\nspc -b_{0R}\nspc -b_{1R}\big\rangle
\end{align}
where $\Bigl(\!\begin{array}{c} x \\[-1.5mm] y \end{array}\!\Bigr)_q$ is the q-deformed binomial coefficient, and the inner product at the end is the 0-particle inner product defined in \eqref{eq:chord_inner_product}. This function vanishes if any of the binomial coefficients become ill-defined (e.g. if $b_{0L}>m_0$). To obtain the chord basis expansion of the state $ \big|\mathbb{1}_\text{AS$^2$}(b_i)\big\rangle_{LR}$ we  should then conjugate this matrix by the same inverse Gram matrix used in \eqref{eq:max_entangled_tube} to get
\begin{equation}
    \big|\mathbb{1}_\text{AS$^2$}(b_i)\big\rangle_{LR}=\sum_{m,n,r,s} G^{-1}(m_i,r_i)G^{-1}(n_i,s_i)H(r_i,s_i\,;b_i)\,\big|m_i\big\rangle_L\,\big|n_i\big\rangle_R.
\end{equation}

If we take $\sigma\rightarrow0$ to enter the AS$^2$ ribbon phase, we can reduce our sum to be over just two components $\{b_L,b_R\}$, and the state on $B$ becomes simply $|b_L+b_R\,;\,\mathbb{O}\rangle_B$. So our matrix elements $H$ become
\bea
H(m,n\,;b_L,b_R)=\begin{pmatrix}m\\b_{L}\end{pmatrix}_{\! q}\begin{pmatrix}n\\b_{R}\end{pmatrix}_{\! q}\;\bigl\langle m-b_L\,\bigl|\;n-b_R\bigr\rangle
\eea
which, when conjugated by the diagonal inverse Gram matrix in the 0-particle sector, leaves us with\footnote{Notice that at $b_L=b_R=0$ in the AS$^2$ ribbon phase, $H$ becomes equal to $G$, so $\big|\mathbb{1}_\text{AS$^2$} (0)\big\rangle_{LR}$ collapses to the tube maximally entangled state \eqref{eq:max_entangled_tube}.}
\begin{equation}
    \big|\mathbb{1}_\text{AS$^2$} (b_i)\big\rangle_{LR}\;\xrightarrow{\sigma\to 0}\;\sum_{m} \frac{1}{[b_L]_q!\;[b_R]_q!\;[m]_q!}\,\big|m+b_L\big\rangle_L\,\big|m+b_R\big\rangle_R
\end{equation}
and consequently
\begin{equation}
    \big|\mathbb{1}_\text{AS$^2$}\,;\,\mathbb{O} \big\rangle_{LR}\;\xrightarrow{\sigma\to 0}\;\sum_{m,b_L,b_R} \frac{e^{-\frac{1}{2}\ell_\text{wh}(b_L+b_R)}}{[b_L]_q!\;[b_R]_q!\;[m]_q!}\,\big|m+b_L\big\rangle_L\,\big|m+b_R\big\rangle_R\,\big|b_L+b_R\, ; \,\mathbb{O} \big\rangle_B.
\end{equation}

With this defined, we can see that the Hartle-Hawking state in this phase has genuine tripartite entanglement between the $L$, $R$, and $B$ sectors. This indicates that the chord Hilbert space of the baby universe is non-trivial.
Note, however, that states with nonzero $b_L+b_R$ are exponentially suppressed. This type of exponential suppression is to be expected, because the state on $L,R$ is a thermofield double state in the extreme low temperature limit. In fact, it is not hard to see that the chords between $L$ and $B$ are much less suppressed than those between $L$ and $R$. Since the norm of the state $\big|b_L+b_R\, ; \,\mathbb{O} \big\rangle_B$ is $\mathcal{O}(1)$ for $b_L+b_R\geq 0$, we are led to conclude that a non-negligible part (possibly even the majority) of the von Neumann entropy of the $L$ and $R$ sectors comes from entanglement with $B$.
We leave the computation of these various entanglement entropies as functions of $\beta$ and $\Delta$ to future work. 


\medskip


\section{Discussion}\label{sect:conc}

In this paper, we analyzed three different phases of the coupled DSSYK model \cite{Maldacena:2018lmt} corresponding to three topologically different spacetimes in $D=2$ quantum gravity. At high temperatures we have the ``disks'' phase, dual to a Euclidean black hole in AdS$_2$. At very low temperatures the model undergoes a transition into  the ``tube'' phase, dual to thermal AdS$_2$ without a black hole. And when we quench the low temperature state with a heavy matter operator, we get a new saddle point dual to the AS$^2$ geometry with a baby universe. For all three phases, we construct a chord Hilbert space and chord Hamiltonian such that the norm of the Hartle-Hawking state computes the partition function to leading order in large $N$ and $p$. The chord Hartle-Hawking state dual to the geometry with a baby universe exhibits nontrivial entanglement between the external spacetimes $L,R$ and the baby universe~$B$.

Our chord rules are derived from the free fermion theory with a Maldacena-Qi interaction but at zero SYK coupling $\mathcal{J}$. We then include the SYK interactions order by order, using the Wick rules of the free fermions and the disorder averaging over SYK couplings. Perhaps surprisingly, the $\mathcal{J}=0$ theory already encodes the baby universe topology. The intuitive reason is that the insertion of the two heavy matter operators dictates the presence of correlations between opposite points on the thermal circle. On the other hand, the MQ coupling impedes the propagation of these correlations along the thermal circle itself. The direct Wick contraction between the heavy operators opens up a new channel through which the Hamiltonian chords can travel. This provides the seed of the baby universe topology. 

The fact that we can already see the emergence of topology in the chord rules indicates that the main mechanism for creating the baby universe is entropic and combinatorial rather than dynamical. To fill in the complete geometry of the cylindrical phase and the baby universe, however, would require performing the full sum over the chord diagrams, or equivalently, developing a direct semiclassical saddle-point method for studying the double-scaled $G\Sigma$-theory at finite SYK coupling $\mathcal{J}$. This requires more work but looks very doable. See \cite{Sasieta:2025vck} for a mean field analysis of a closely related baby universe scenario based on a pure state in the DSSYK-MQ model.

In our description above, we constructed a basis for the two-sided SYK Hilbert space by means of Hamiltonian chords.  We could also imagine doing this with some very light matter field (composed of potentially as few as $\mathcal{O}(1)$ Majoranas) whose two-point function is given by some smaller power of $(2G_{ab})$ much less than $p$. These light matter chords are just as good of a probe of the bulk geometry, and the resulting chord basis would exhibit more entanglement between the $L$ and $R$ thermal AdS regions and the wormhole than does the Hamiltonian chord basis.

Our analysis allows us to get a better grasp on the AR puzzle in a concrete example \cite{Antonini:2024mci}. We hypothesize the following solution \cite{Liu:2025cml,Kudler-Flam:2025cki,liu2025filteringcftslargen}. The chord Hilbert space is obtained by a coarse-graining map acting on the boundary theory. This map is by construction insensitive to the detailed micro-structure of the couplings $J_{i_1...i_p}$ and $K_{i_1...i_{p'}}$, and remembers only that they are random and uncorrelated with each other. Microscopically, however, the intermediate states $|\spc b_i\spc ; \mathbb{O}\spc \rangle$ of the baby universe and the matrix elements of the heavy operator do depend on this micro-structure. To make this more concrete, it is useful to think about the random couplings $J_{i_1...i_p}$ and $K_{i_1...i_{p'}}$ as quantum numbers that are entangled with some external reference system $R$ that labels all the possible microscopic realizations of the model \cite{Almheiri:2021jwq}. It is then logical to identify the Hilbert space ${\mathcal{H}}_{\rm bu}$ of the baby universe with the $2^{p'/2} = 2^{p\Delta/2}$ dimensional Hilbert space of all possible choices of the operator $\mathbb{O}_\Delta$ of dimension~$\Delta$:
\bea
\label{hbuDelta}
{\mathcal H}^{\Delta}_{\rm bu} = \Bigl\{\mbox{span of all states} \ \mathbb{O}^{(i)}_\Delta|0\rangle\; ; \;  1\leq i \leq 2^{p'/2}\Bigr\} \qquad p'\equiv p \Delta
\eea

We can use this viewpoint to give a heuristic microscopic derivation of the wormhole crossing chord factor $e^{-\ell_{\rm wh}}$ as follows. Let's assume that the $b$-chords that traverse the wormhole describe the Wick contraction mediated by $\mathcal{H}_{\rm bu}$. The SYK Hamiltian $H$ acts on a given state in ${\cal H}_{\rm bu}$
\bea
\mathbb{O}_\Delta|0\rangle =\sum_{j_1,..,j_{p'}} K_{j_1\ldots j_{p'}} \psi_{j_1}\ldots \psi_{j_{p'}} |0\rangle 
\eea
via Wick contraction. The dominant matrix elements are those for which the normal ordered product contains $ s \equiv p/2$ Wick contractions
\bea
\label{HtwoO}
\textcolor{blue}{:} H \mathbb{O}_\Delta\textcolor{blue}{:} |0\rangle\;\; \supset \; \;
J_{i_1\ldots i_{2s}}  K_{j_1\ldots j_{p'}  }\textcolor{blue}{\psi_{i_1} .\spc .\spc .\spc
\psi_{i_{s}}}
\textcolor{blue}{\wick{\c3{\psi}{}_{i_{s +1}} \c2{.\spc}{ .\spc .\spc}\c1{\psi}{}_{i_{2s}}
\c1{\textcolor{green!30!black}{ \psi_{j_1}}}
\c2{\textcolor{green!30!black}{.\spc .\spc .\spc}}
\c3{\textcolor{green!30!black}{\psi_{j_{s}}}}}}
\textcolor{green!30!black}{\psi_{j_{s +1}}.\spc .\spc .\spc \psi_{j_{p'}}} |0\rangle\; \in {\mathcal{H}_{\rm bu}}
\eea
resulting in an operator with the maximal allowed number of Majorana oscillators contained in ${\mathcal{H}}_{\rm bu}$. Hence the Hamiltonian chords traverse the wormhole via a Wick contraction of the schematic form
$$
\begin{tikzpicture}[x=0.75pt,y=0.75pt,yscale=.5,xscale=.5,rotate=90]
\draw (395,333) node [anchor=north west][inner sep=0.75pt]   [align=left] {{\fontfamily{ptm}\selectfont \textit{{\large H}}}};
\draw (395,50) node [anchor=north west][inner sep=0.75pt]   [align=left] {{\fontfamily{ptm}\selectfont \textit{{\large H}}}};

\draw (335,365) node [anchor=north west][inner sep=0.75pt]   [align=left] {{{\large $\mathbb{O}^\dag_\Delta$}}};
\draw (325,34) node [anchor=north west][inner sep=0.75pt]   [align=left] {{{\large {$\mathbb{O}_\Delta$}}}};
\draw [color={rgb, 255:red, 65; green, 117; blue, 5 }  ,draw opacity=1 ][line width=1]    (291.91,43.37) -- (292.29,312.97) ;
\draw   (178.36,178.19) .. controls (178.36,102.97) and (237.81,42) .. (311.14,42) .. controls (384.47,42) and (443.91,102.97) .. (443.91,178.19) .. controls (443.91,253.4) and (384.47,314.37) .. (311.14,314.37) .. controls (237.81,314.37) and (178.36,253.4) .. (178.36,178.19) -- cycle ;
\draw  [color={rgb, 255:red, 74; green, 93; blue, 226 }  ,draw opacity=1 ][line width=1.5]  (359.72,52.14) .. controls (359.72,51.51) and (360.23,51) .. (360.86,51) .. controls (361.49,51) and (362,51.51) .. (362,52.14) .. controls (362,52.77) and (361.49,53.28) .. (360.86,53.28) .. controls (360.23,53.28) and (359.72,52.77) .. (359.72,52.14) -- cycle ;
\draw  [color={rgb, 255:red, 74; green, 93; blue, 226 }  ,draw opacity=1 ][line width=1.5]  (368.64,55.66) .. controls (368.64,55.03) and (369.15,54.52) .. (369.78,54.52) .. controls (370.41,54.52) and (370.92,55.03) .. (370.92,55.66) .. controls (370.92,56.29) and (370.41,56.8) .. (369.78,56.8) .. controls (369.15,56.8) and (368.64,56.29) .. (368.64,55.66) -- cycle ;
\draw  [color={rgb, 255:red, 74; green, 93; blue, 226 }  ,draw opacity=1 ][line width=1.5]  (375.64,60.66) .. controls (375.64,60.03) and (376.15,59.52) .. (376.78,59.52) .. controls (377.41,59.52) and (377.92,60.03) .. (377.92,60.66) .. controls (377.92,61.29) and (377.41,61.8) .. (376.78,61.8) .. controls (376.15,61.8) and (375.64,61.29) .. (375.64,60.66) -- cycle ;
\draw  [color={rgb, 255:red, 74; green, 93; blue, 226 }  ,draw opacity=1 ][line width=1.5]  (383.64,65.66) .. controls (383.64,65.03) and (384.15,64.52) .. (384.78,64.52) .. controls (385.41,64.52) and (385.92,65.03) .. (385.92,65.66) .. controls (385.92,66.29) and (385.41,66.8) .. (384.78,66.8) .. controls (384.15,66.8) and (383.64,66.29) .. (383.64,65.66) -- cycle ;
\draw [color={rgb, 255:red, 65; green, 117; blue, 5 }  ,draw opacity=1 ][line width=1]    (300,42) -- (300.91,315.37) ;
\draw [color={rgb, 255:red, 65; green, 117; blue, 5 }  ,draw opacity=1 ][line width=1]    (308.91,43.37) -- (309.91,314.37) ;
\draw [color={rgb, 255:red, 65; green, 117; blue, 5 }  ,draw opacity=1 ][line width=1]    (319,43) -- (318.91,313.37) ;
\draw [color={rgb, 255:red, 74; green, 95; blue, 226 }  ,draw opacity=1 ][line width=1]    (329.16,105.29) -- (328.3,254.29) ;
\draw [color={rgb, 255:red, 74; green, 90; blue, 226 }  ,draw opacity=1 ][line width=1]    (338.3,109.29) -- (336.91,251.37) -- (385.91,291.28) ;
\draw  [color={rgb, 255:red, 74; green, 93; blue, 226 }  ,draw opacity=1 ][line width=1.5]  (362.13,303.54) .. controls (361.62,303.18) and (361.5,302.46) .. (361.86,301.95) .. controls (362.23,301.44) and (362.94,301.32) .. (363.46,301.69) .. controls (363.97,302.05) and (364.09,302.77) .. (363.72,303.28) .. controls (363.35,303.79) and (362.64,303.91) .. (362.13,303.54) -- cycle ;
\draw  [color={rgb, 255:red, 74; green, 93; blue, 226 }  ,draw opacity=1 ][line width=1.5]  (370.04,300) .. controls (369.53,299.64) and (369.41,298.92) .. (369.78,298.41) .. controls (370.14,297.9) and (370.86,297.78) .. (371.37,298.15) .. controls (371.88,298.51) and (372,299.23) .. (371.63,299.74) .. controls (371.27,300.25) and (370.55,300.37) .. (370.04,300) -- cycle ;
\draw  [color={rgb, 255:red, 74; green, 93; blue, 226 }  ,draw opacity=1 ][line width=1.5]  (377.79,295.7) .. controls (377.36,295.24) and (377.4,294.52) .. (377.87,294.09) .. controls (378.33,293.67) and (379.06,293.7) .. (379.48,294.17) .. controls (379.9,294.64) and (379.87,295.36) .. (379.4,295.78) .. controls (378.93,296.21) and (378.21,296.17) .. (377.79,295.7) -- cycle ;
\draw  [color={rgb, 255:red, 74; green, 93; blue, 226 }  ,draw opacity=1 ][line width=1.5]  (384.32,291.54) .. controls (383.81,291.18) and (383.69,290.46) .. (384.05,289.95) .. controls (384.42,289.44) and (385.13,289.32) .. (385.65,289.69) .. controls (386.16,290.05) and (386.28,290.77) .. (385.91,291.28) .. controls (385.54,291.79) and (384.83,291.91) .. (384.32,291.54) -- cycle ;
\draw [color={rgb, 255:red, 74; green, 90; blue, 226 }  ,draw opacity=1 ][line width=1]    (338.3,109.29) -- (385.92,65.66) ;
\draw [color={rgb, 255:red, 74; green, 90; blue, 226 }  ,draw opacity=1 ][line width=1]    (329.16,105.29) -- (376.78,60.66) ;
\draw [color={rgb, 255:red, 74; green, 90; blue, 226 }  ,draw opacity=1 ][line width=1]    (327.38,252.68) -- (378.48,295.17) ;
\draw  [color={rgb, 255:red, 65; green, 117; blue, 5 }  ,draw opacity=1 ][line width=1.5]  (318.25,315.3) .. controls (317.73,314.93) and (317.61,314.22) .. (317.98,313.71) .. controls (318.35,313.19) and (319.06,313.08) .. (319.57,313.44) .. controls (320.09,313.81) and (320.21,314.52) .. (319.84,315.03) .. controls (319.47,315.55) and (318.76,315.67) .. (318.25,315.3) -- cycle ;
\draw  [color={rgb, 255:red, 65; green, 117; blue, 5 }  ,draw opacity=1 ][line width=1.5]  (309.13,315.54) .. controls (308.62,315.18) and (308.5,314.46) .. (308.86,313.95) .. controls (309.23,313.44) and (309.94,313.32) .. (310.46,313.69) .. controls (310.97,314.05) and (311.09,314.77) .. (310.72,315.28) .. controls (310.35,315.79) and (309.64,315.91) .. (309.13,315.54) -- cycle ;
\draw  [color={rgb, 255:red, 65; green, 117; blue, 5 }  ,draw opacity=1 ][line width=1.5]  (299.91,315.37) .. controls (299.4,315) and (299.28,314.29) .. (299.65,313.78) .. controls (300.01,313.27) and (300.72,313.15) .. (301.24,313.51) .. controls (301.75,313.88) and (301.87,314.59) .. (301.5,315.11) .. controls (301.14,315.62) and (300.42,315.74) .. (299.91,315.37) -- cycle ;
\draw  [color={rgb, 255:red, 65; green, 117; blue, 5 }  ,draw opacity=1 ][line width=1.5]  (318,43) .. controls (317.49,42.63) and (317.37,41.92) .. (317.74,41.41) .. controls (318.1,40.89) and (318.81,40.78) .. (319.33,41.14) .. controls (319.84,41.51) and (319.96,42.22) .. (319.59,42.74) .. controls (319.23,43.25) and (318.51,43.37) .. (318,43) -- cycle ;
\draw  [color={rgb, 255:red, 65; green, 117; blue, 5 }  ,draw opacity=1 ][line width=1.5]  (291.96,313.82) .. controls (291.45,313.46) and (291.33,312.74) .. (291.7,312.23) .. controls (292.07,311.72) and (292.78,311.6) .. (293.29,311.97) .. controls (293.8,312.33) and (293.92,313.05) .. (293.56,313.56) .. controls (293.19,314.07) and (292.48,314.19) .. (291.96,313.82) -- cycle ;
\draw  [color={rgb, 255:red, 65; green, 117; blue, 5 }  ,draw opacity=1 ][line width=1.5]  (308.32,42.83) .. controls (307.8,42.47) and (307.69,41.75) .. (308.05,41.24) .. controls (308.42,40.73) and (309.13,40.61) .. (309.65,40.98) .. controls (310.16,41.34) and (310.28,42.05) .. (309.91,42.57) .. controls (309.54,43.08) and (308.83,43.2) .. (308.32,42.83) -- cycle ;
\draw  [color={rgb, 255:red, 65; green, 117; blue, 5 }  ,draw opacity=1 ][line width=1.5]  (299.32,42.83) .. controls (298.8,42.47) and (298.69,41.75) .. (299.05,41.24) .. controls (299.42,40.73) and (300.13,40.61) .. (300.65,40.98) .. controls (301.16,41.34) and (301.28,42.05) .. (300.91,42.57) .. controls (300.54,43.08) and (299.83,43.2) .. (299.32,42.83) -- cycle ;
\draw  [color={rgb, 255:red, 65; green, 117; blue, 5 }  ,draw opacity=1 ][line width=1.5]  (327.32,43.83) .. controls (326.8,43.47) and (326.69,42.75) .. (327.05,42.24) .. controls (327.42,41.73) and (328.13,41.61) .. (328.65,41.98) .. controls (329.16,42.34) and (329.28,43.05) .. (328.91,43.57) .. controls (328.54,44.08) and (327.83,44.2) .. (327.32,43.83) -- cycle ;
\draw  [color={rgb, 255:red, 65; green, 117; blue, 5 }  ,draw opacity=1 ][line width=1.5]  (336.32,45.83) .. controls (335.8,45.47) and (335.69,44.75) .. (336.05,44.24) .. controls (336.42,43.73) and (337.13,43.61) .. (337.65,43.98) .. controls (338.16,44.34) and (338.28,45.05) .. (337.91,45.57) .. controls (337.54,46.08) and (336.83,46.2) .. (336.32,45.83) -- cycle ;
\draw  [color={rgb, 255:red, 65; green, 117; blue, 5 }  ,draw opacity=1 ][line width=1.5]  (291.32,43.83) .. controls (290.8,43.47) and (290.69,42.75) .. (291.05,42.24) .. controls (291.42,41.73) and (292.13,41.61) .. (292.65,41.98) .. controls (293.16,42.34) and (293.28,43.05) .. (292.91,43.57) .. controls (292.54,44.08) and (291.83,44.2) .. (291.32,43.83) -- cycle ;
\draw [color={rgb, 255:red, 74; green, 90; blue, 226 }  ,draw opacity=1 ][line width=1]    (329,93.33) -- (368.78,57.8) ;
\draw [color={rgb, 255:red, 74; green, 90; blue, 226 }  ,draw opacity=1 ][line width=1]    (337,75.33) -- (362,52.14) ;
\draw [color={rgb, 255:red, 65; green, 117; blue, 5 }  ,draw opacity=1 ][line width=1]    (328.91,43.57) -- (329,93.33) ;
\draw [color={rgb, 255:red, 65; green, 117; blue, 5 }  ,draw opacity=1 ][line width=1]    (337.91,45.57) -- (337,75.33) ;
\draw [color={rgb, 255:red, 74; green, 90; blue, 226 }  ,draw opacity=1 ][line width=1]    (328.31,263.33) -- (371,299.46) ;
\draw [color={rgb, 255:red, 74; green, 90; blue, 226 }  ,draw opacity=1 ][line width=1]    (336.63,281.06) -- (363.82,304.98) ;
\draw [color={rgb, 255:red, 65; green, 117; blue, 5 }  ,draw opacity=1 ][line width=1]    (327.92,313.22) -- (328.31,264.33) ;
\draw [color={rgb, 255:red, 65; green, 117; blue, 5 }  ,draw opacity=1 ][line width=1]    (337,311.33) -- (336.63,282.06) ;
\draw  [color={rgb, 255:red, 65; green, 117; blue, 5 }  ,draw opacity=1 ][line width=1.5]  (337.25,312.3) .. controls (336.73,311.93) and (336.61,311.22) .. (336.98,310.71) .. controls (337.35,310.19) and (338.06,310.08) .. (338.57,310.44) .. controls (339.09,310.81) and (339.21,311.52) .. (338.84,312.03) .. controls (338.47,312.55) and (337.76,312.67) .. (337.25,312.3) -- cycle ;
\draw  [color={rgb, 255:red, 65; green, 117; blue, 5 }  ,draw opacity=1 ][line width=1.5]  (327.25,314.3) .. controls (326.73,313.93) and (326.61,313.22) .. (326.98,312.71) .. controls (327.35,312.19) and (328.06,312.08) .. (328.57,312.44) .. controls (329.09,312.81) and (329.21,313.52) .. (328.84,314.03) .. controls (328.47,314.55) and (327.76,314.67) .. (327.25,314.3) -- cycle ;
\end{tikzpicture}
\vspace{-2mm}
$$
or in AdS/CFT terminology, $H$ travels through the wormhole via the OPE with the heavy operator ${\mathbb{O}}$, while placed at the threshold distance where the leading OPE preserves the scale dimension of ${\mathbb{O}}$.
The probability that a given set of $p/2$ Majorana's finds $p/2$ partners among $p' = p\Delta$ Majorana's equals
\bea
\biggl(\frac{p \Delta}{N}\biggr)^{p/2} = e^{-\frac p 2 \log\bigl(\frac{N} {p\Delta}\bigr)} = e^{-\ell_{\rm wh}}
\eea
which we recognize as the chord suppression factor for passing through the wormhole.
This combinatorial argument, supported by the more systematic analysis in Appendices \ref{app:A}-\ref{app:C}, indicates that we can indeed identify the baby universe Hilbert space ${\cal H}^{\rm bu}_{\rm chord}$ spanned by the chord basis \eqref{hbuchord} with the Hilbert space ${\cal H}^{\rm bu}_\Delta$ in \eqref{hbuDelta} spanned by all operators $\mathbb{O}_\Delta^{(i)}$ with scale dimension $\Delta$.

The baby universe states of the form \eqref{HtwoO} involve two mutually random couplings and thus are randomly oriented within ${\mathcal H}_{\rm bu}$. They would vanish upon averaging over all microscopic realizations of the model, or equivalently, upon tracing out the auxiliary reference quantum system $R$ that labels all of those microscopic realizations. The ``good'' semiclassical bulk observables are then those that are approximately unchanged by taking this trace: they are self-averaging.\footnote{They are analogous to pointer states, if we look at $R$ as a decohering environment.} The general lesson here is that any observable which can be formulated in the chord space is \emph{automatically} a good observable in this sense. The existence and size of the baby universe are among these.

We can also ask to what extent this coarse-graining map performs a large $N$ average, as discussed in \cite{Kudler-Flam:2025cki,Liu:2025cml,liu2025filteringcftslargen}. This requires a bit more care. Consider an incredibly capable experimentalist Alice who can very precisely measure chord state variables but is ignorant of $R$. If Alice repeatedly measures geodesic distances, she will find that they are discrete with increment $\lambda$. If she is also able to measure the critical temperature $\beta_{HP}\sim p\log(p)$ of the Hawking-Page transition, then she can also determine the value of $p$. With these two numbers she can compute $N$ and find it to be finite. So the map preparing the full set of chord Hilbert spaces for each topology does not appear to require performing a large $N$ average. The randomness of the $J$ and $K$ couplings appears sufficient for generating the non-trivial semiclassical notion of topology.



\section*{Acknowledgments}

 AS thanks Ahmed Almheiri, Juan Maldacena, and Suzanne Staggs for their willingness to sit on his pre-thesis committee and for their insightful comments. We thank Netta Engelhardt, Elliott Gesteau, Dan Harlow, Jonathan Karl,  Jonah Kudler-Flam, Hong Liu, Erik Verlinde, and Edward Witten for helpful conversations on the arguments surrounding closed universe quantum gravity. AS is supported by the National Science Foundation Graduate Research Fellowship Program under Grant No. DGE-2444107.


\pagebreak

\appendix

\section{Deriving the $\langle\delta G\,\delta G\rangle_0$ correlators}\label{app:A}

We begin as above with the effective Hubbard-Stratonovich action \eqref{eq:SeffGSig}. We set $\mathcal{J}=0$ and rephrase the integrals over $G_{ab}\Sigma_{ab}$ as traces, giving us
\begin{equation}\label{eq:A.SeffGSigTr}
    -\frac{S_{\text{eff}}^{\mathcal{J}=0}[G,\Sigma]}{N}=\frac{1}{2}\Tr\,\log \begin{pmatrix} \partial_\tau - \Sigma_{00} & i\nu+i\,\Sigma_{01} \\ -i\nu+i\,\Sigma_{10} & \partial_\tau - \Sigma_{11}\end{pmatrix} +\frac{1}{2}\Tr\Big(\Sigma_{00}G_{00}+\Sigma_{01}G_{10}+\Sigma_{10}G_{01}+\Sigma_{11}G_{11}\Big).
\end{equation}
We can then vary with respect to $\Sigma_{ab}$ to get the equation of motion
\begin{equation}\label{eq:A.MmatEOM}
    \begin{pmatrix} \partial_\tau - \Sigma_{00} & i\nu+i\,\Sigma_{01} \\ -i\nu+i\,\Sigma_{10} & \partial_\tau - \Sigma_{11}\end{pmatrix}^{-1}=\begin{pmatrix} G_{00} & iG_{01} \\ iG_{10} & G_{11}\end{pmatrix}
\end{equation}
where for shorthand we will define
\begin{equation}\label{eq:A.defAM}
    M\equiv\begin{pmatrix} G_{00} & iG_{01} \\ iG_{10} & G_{11}\end{pmatrix},\quad A\equiv\begin{pmatrix} \partial_\tau & i\nu \\ -i\nu & \partial_\tau \end{pmatrix}.
\end{equation}
The second trace in \eqref{eq:A.SeffGSigTr} can then be expressed as
\begin{equation}
    \frac{1}{2}\Tr\Bigg[\begin{pmatrix} \Sigma_{00} & -i\Sigma_{01} \\ -i\Sigma_{10} & \Sigma_{11}\end{pmatrix}\begin{pmatrix} G_{00} & iG_{01} \\ iG_{10} & G_{11}\end{pmatrix}\Bigg]=\frac{1}{2}\Tr\Big[(A-M^{-1})M\Big]=\frac{1}{2}\Tr\Big[AM-I\Big]
\end{equation}
which allows us to integrate out $\Sigma_{ab}$ by writing
\begin{equation}\label{eq:A.SeffAM}
    -\frac{S_{\text{eff}}^{\mathcal{J}=0}[G]}{N}=-\frac{1}{2}\Tr\Big[\log (M)\Big] +\frac{1}{2}\Tr\Big[AM-I\Big].
\end{equation}
This is equivalent to \eqref{eq:SeffGonly}, and is exact in finite $p$.\footnote{This is the action we should use to compute and compare saddle points where $\beta>\mathcal{O}(p^0)$.} We then define $\delta G_{ab}$ as in \eqref{eq:define_dGLR}-\eqref{eq:define_dGY} and analogously take
\begin{equation}\label{eq:A.defM0dM}
    M_0\equiv\begin{pmatrix} \overline{G_{00}} & i\overline{G_{01}} \\ -i\overline{G_{01}} & \overline{G_{00}}\end{pmatrix},\quad \delta M\equiv\frac{1}{2p}\begin{pmatrix} \text{sgn}(\tau-\tau')\,\delta G_{00} & i\,\delta G_{01} \\ -i\,\delta G_{10} & \text{sgn}(\tau-\tau')\,\delta G_{11} \end{pmatrix}
\end{equation}
such that $M=M_0+\delta M$ and $AM_0=I$.\footnote{This is easy to see in the Fourier basis, where $\tilde{\overline{G_{00}}}=\frac{i\omega}{\omega^2+\nu^2}$, $\tilde{\overline{G_{01}}}=\frac{\nu}{\omega^2+\nu^2}$, and $A=\begin{pmatrix} -i\omega & i\nu \\ -i\nu & -i\omega \end{pmatrix}$.} We can then simplify \eqref{eq:A.SeffAM} and expand the log as
\begin{align}
    -\frac{S_{\text{eff}}^{\mathcal{J}=0}[G]}{N}&=\frac{1}{2}\Tr\Big[-\log(M)+AM-I\Big]\nonumber\\
    &=\frac{1}{2}\Tr\Big[-\log\Big(A^{-1}A(M_0+\delta M)\Big)+A(M_0+\delta M)-I\Big]\nonumber\\
    &=\frac{1}{2}\Tr\Big[\log(A)-\log(I+A\,\delta M)+A\,\delta M\Big]\nonumber\\
    &=\frac{1}{2}\Tr\Big[\log(A)\Big]+\frac{1}{4}\Tr\Big[(A\,\delta M)^2\Big] + \mathcal{O}(1/p^3).
\end{align}
The first term is the topological piece that appears in \eqref{eq:SeffGSigJ0}-\eqref{eq:SeffGSigJ0sum}. A considerable amount of algebra (using the inherited (anti)symmetry and antiperiodicity properties of the $\delta G_{ab}$) can then produce
\begin{equation}\label{eq:A.Mabcd}
    \frac{1}{4}\Tr\Big[(A\,\delta M)^2\Big]=\frac{1}{16p^2}\iint\limits_{\;00}^{\,\beta\beta}d\tau\, d\tau'\;V^T(\tau,\tau')\begin{pmatrix} \partial_\tau\partial_{\tau'} & -\nu^2 & -2\nu\partial_\tau & 0 \\ -\nu^2 & \partial_\tau\partial_{\tau'} & 0 & -2\nu\partial_\tau \\ 2\nu\partial_\tau & 0 & -\partial_\tau\partial_{\tau'} & \nu^2 \\ 0 & 2\nu\partial_\tau & \nu^2 & -\partial_\tau\partial_{\tau'} \end{pmatrix}V(\tau,\tau')
\end{equation}
where
\begin{equation}\label{eq:A.defV}
    V=\begin{pmatrix} \text{sgn}(\tau-\tau')\,\delta G_{00} \\ \text{sgn}(\tau-\tau')\,\delta G_{11} \\ \delta G_{01} \\ \delta G_{10} \end{pmatrix}.
\end{equation}
Defining new fields $\delta G_\pm=\delta G_{00}\pm\delta G_{11}$ and $\delta H_\pm=\delta G_{01}\pm\delta G_{10}$ allows us to factorize and simplify the above system as
\begin{equation}\label{eq:A.MabcdOpm}
    \frac{1}{16p^2}\int_0^\beta d\tau\int_0^\tau d\tau'\Bigg[(\delta G_+\;\;\delta H_+)\;\mathbb{O}_+\begin{pmatrix}\delta G_+ \\ \delta H_+\end{pmatrix}\;+\;(\delta G_-\;\;\delta H_-)\;\mathbb{O}_-\begin{pmatrix}\delta G_- \\ \delta H_-\end{pmatrix}\Bigg]
\end{equation}
with
\begin{equation}\label{eq:A.defOpm}
    \mathbb{O}_\pm=\begin{pmatrix}\partial_\tau\partial_{\tau'}\mp\nu^2 & -\nu(\partial_\tau\mp\partial_{\tau'}) \\ +\nu(\partial_\tau\mp\partial_{\tau'}) & -\partial_\tau\partial_{\tau'}\pm\nu^2 \end{pmatrix}.
\end{equation}
Our effective action to leading order in $\lambda$ is then
\begin{equation}\label{eq:A.SeffFinal}
    -S_{\text{eff}}^{\mathcal{J}=0}[G]=\frac{N}{2}\Tr\Big[\log(A)\Big]+\frac{1}{8\lambda}\int_0^\beta d\tau\int_0^\tau d\tau'\Bigg[V_+^T\mathbb{O}_+V_+\;+\;V_-^T\mathbb{O}_-V_-\Bigg]
\end{equation}
for $V_\pm^T=(\delta G_\pm\,,\;\delta H_\pm)$. This is the explicit version of \eqref{eq:SeffdG}. 

We can now compute correlation functions in this $\mathcal{J}=0$ theory as
\begin{equation}\label{eq:A.CorrtoOpm}
    \langle V_\pm(\tau,\tau') V_\pm(\sigma,\sigma') \rangle_0 = -4\lambda \mathbb{O}^{-1}_\pm(\tau,\tau';\sigma,\sigma')
\end{equation}
for the Green's function $\mathbb{O}^{-1}$ that satisfies the relevant boundary conditions of $V_\pm$ inherited from the (anti)symmetry and antiperiodicity of $\delta G_{ab}$. Here we also need to impose physicality conditions on the propagators $G_{ab}$ (given in equation (5.80) in \cite{Maldacena:2018lmt}) as $\tau\rightarrow\tau'$:
\begin{equation}\label{eq:A.phyBC00}
    G_{00}(\tau,\tau')\rightarrow \frac{1}{2}\text{sgn}(\tau-\tau'),\quad G_{11}(\tau,\tau')\rightarrow \frac{1}{2}\text{sgn}(\tau-\tau')
\end{equation}
\begin{equation}\label{eq:A.phyBC01}
    \partial_\tau G_{01}(\tau,\tau')\rightarrow -\frac{\nu}{2}\text{sgn}(\tau-\tau'),\quad \partial_\tau G_{10}(\tau,\tau')\rightarrow \frac{\nu}{2}\text{sgn}(\tau-\tau')
\end{equation}

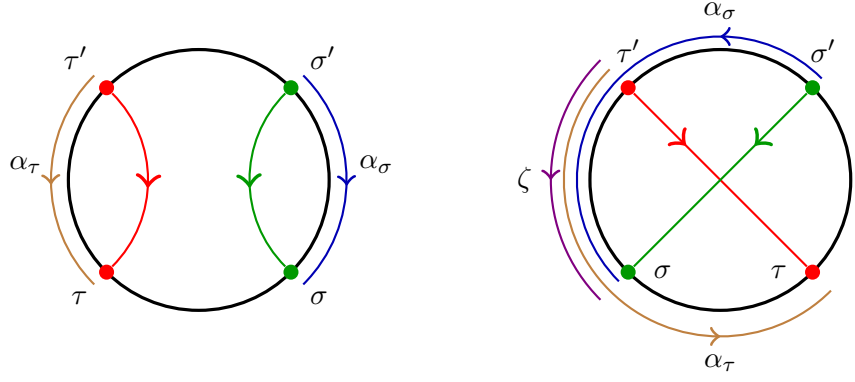
\begin{figure}[t]
\centering
\begin{tikzpicture}[scale = 1.15]

\def\r{1.5}

\draw[very thick] (0,0) circle (\r);

\node[fill=green!60!black, circle, inner sep=2pt, label={[label distance=0.6mm]45:$\sigma'$}](g1a) at ({\r*cos(45)}, {\r*sin(45)}) {};
\node[fill=red, circle, inner sep=2pt, label={[label distance=0.6mm]135:$\tau'$}](r1a) at ({\r*cos(135)}, {\r*sin(135)}) {};
\node[fill=red, circle, inner sep=2pt, label={[label distance=0.6mm]225:$\tau$}](r1b) at ({\r*cos(225)}, {\r*sin(225)}) {};
\node[fill=green!60!black, circle, inner sep=2pt, label={[label distance=0.6mm]315:$\sigma$}](g1b) at ({\r*cos(315)}, {\r*sin(315)}) {};
\draw[red, thick, postaction={decorate, decoration={markings,mark=at position 0.54 with {\arrow[scale=2]{>}}}}] (r1a) to[out=315, in=45] (r1b);
\draw[green!60!black, thick, postaction={decorate, decoration={markings,mark=at position 0.54 with {\arrow[scale=2]{>}}}}] (g1a) to[out=225, in=135] (g1b);

\draw[brown, thick, postaction={decorate, decoration={markings,mark=at position 0.52 with {\arrow[scale=1.5]{>}}}}]
  (135:{\r+0.2})
  arc[start angle=135, end angle=225, radius={\r+0.2}];
\node at (175:{\r+0.5}) {$\alpha_\tau$};
\draw[blue!70!black, thick, postaction={decorate, decoration={markings,mark=at position 0.52 with {\arrow[scale=1.5]{<}}}}]
  (315:{\r+0.2})
  arc[start angle=315, end angle=405, radius={\r+0.2}];
\node at (5:{\r+0.55}) {$\alpha_\sigma$};

\begin{scope}[shift={(6,0)}]
\draw[very thick] (0,0) circle (\r);

\node[fill=green!60!black, circle, inner sep=2pt, label={[label distance=1mm, xshift=4pt]above:$\sigma'$}](g1c) at ({\r*cos(45)}, {\r*sin(45)}) {};
\node[fill=red, circle, inner sep=2pt, label={[label distance=1mm]above:$\tau'$}](r1c) at ({\r*cos(135)}, {\r*sin(135)}) {};
\node[fill=green!60!black, circle, inner sep=2pt, label={[label distance=1mm]right:$\sigma$}](g1d) at ({\r*cos(225)}, {\r*sin(225)}) {};
\node[fill=red, circle, inner sep=2pt, label={[label distance=1mm]left:$\tau$}](r1d) at ({\r*cos(315)}, {\r*sin(315)}) {};
\draw[red, thick, postaction={decorate, decoration={markings,mark=at position 0.3 with {\arrow[scale=2]{>}}}}] (r1c) to[out=315, in=135] (r1d);
\draw[green!60!black, thick, postaction={decorate, decoration={markings,mark=at position 0.3 with {\arrow[scale=2]{>}}}}] (g1c) to[out=225, in=45] (g1d);

\draw[blue!70!black, thick, postaction={decorate, decoration={markings,mark=at position 0.25 with {\arrow[scale=1.5]{>}}}}]
  (45:{\r+0.15})
  arc[start angle=45, end angle=225, radius={\r+0.15}];
\node at (90:{\r+0.45}) {$\alpha_\sigma$};
\draw[brown, thick, postaction={decorate, decoration={markings,mark=at position 0.75 with {\arrow[scale=1.5]{>}}}}]
  (135:{\r+0.3})
  arc[start angle=135, end angle=315, radius={\r+0.3}];
\node at (270:{\r+0.6}) {$\alpha_\tau$};
\draw[violet, thick, postaction={decorate, decoration={markings,mark=at position 0.5 with {\arrow[scale=1.5]{>}}}}]
  (135:{\r+0.45})
  arc[start angle=135, end angle=225, radius={\r+0.45}];
\node at (180:{\r+0.75}) {$\zeta$};

\end{scope}

\end{tikzpicture}
\caption{Geometric parameters appearing in \eqref{eq:A.GpGp}-\eqref{eq:A.HmHm}. All arcs are directional. The red, green, orange ($\alpha_\tau$), and blue ($\alpha_\sigma$) arcs are always directed from the primed to the unprimed coordinate. The purple arc ($\zeta$) is always directed from a $\tau^{(\prime)}$ to a $\sigma^{(\prime)}$ coordinate. In the crossed diagram, choosing ``the other $\alpha_\tau$'' or ``the other $\alpha_\sigma$'' is fine, but the $\zeta$ arc must always be that which overlaps with both $\alpha_\tau$ and $\alpha_\sigma$. All arc lengths are {\bf nonnegative} values. An arc's orientation, written as $\text{sgn}(\cdot)$ in \eqref{eq:A.HmGm},\eqref{eq:A.HmHm} evaluates to $+1$ if  directed counterclockwise (same direction as $\tau=0\rightarrow\beta$) and $-1$ if clockwise.}
\label{fig:A.Circles}
\end{figure}

These become Dirichlet BCs for $\delta G_{00}$, $\delta G_{11}$ and Neumann BCs for $\delta G_{01}$, $\delta G_{10}$ at $\tau=\tau'$. It will turn out that the relevant quantities for computing chord diagrams are actually $\langle\delta G\,\delta G\rangle_0/(\overline{G}\,\overline{G})$, so we give these below in \eqref{eq:A.GpGp}-\eqref{eq:A.HmHm}. Each one is a piecewise function depending on whether or not the chord $(\tau\rightarrow\tau')$ crosses the chord $(\sigma\rightarrow\sigma')$ on the disk, as in Appendix H of \cite{Lin:2023trc}. We express these in terms of purely geometric features $\alpha_\tau$, $\alpha_\sigma$, and $\zeta$ to make manifest their connection to chord diagrams, see Figure~\ref{fig:A.Circles}. Observe also that each of the equations is symmetric under all possible geometrically-equivalent relabelings of these features (e.g. moving the point $\tau=0$, or swapping $\alpha_\tau$ for $\beta-\alpha_\tau$ in the crossed diagram).

\begin{align}
    \frac{\langle \delta G_+(\tau,\tau')\delta G_+(\sigma,\sigma')\rangle_0}{8\;\overline{G_{00}}(\tau,\tau')\overline{G_{00}}(\sigma,\sigma')}&=\lambda
    \begin{cases}
    -1: & \crossedB \\
    \displaystyle \frac{\sinh(\nu\alpha_\tau)\sinh(\nu\alpha_\sigma)}{\cosh(\nu(\beta/2-\alpha_\tau))\cosh(\nu(\beta/2-\alpha_\sigma))}: & \uncrossedB
    \end{cases}\label{eq:A.GpGp}\\[2ex]
    \frac{\langle \delta H_+(\tau,\tau')\delta G_+(\sigma,\sigma')\rangle_0}{8\;\overline{G_{01}}(\tau,\tau')\overline{G_{00}}(\sigma,\sigma')}&=\lambda
    \begin{cases}
    -1: & \crossedB \\
    \displaystyle \frac{-\cosh(\nu\alpha_\tau)\sinh(\nu\alpha_\sigma)}{\sinh(\nu(\beta/2-\alpha_\tau))\cosh(\nu(\beta/2-\alpha_\sigma))}: & \uncrossedB
    \end{cases}\label{eq:A.HpGp} \\[2ex]
    \frac{\langle \delta H_+(\tau,\tau')\delta H_+(\sigma,\sigma')\rangle_0}{8\;\overline{G_{01}}(\tau,\tau')\overline{G_{01}}(\sigma,\sigma')}&=\lambda
    \begin{cases}
    -1: & \crossedB \\
    \displaystyle \frac{\cosh(\nu\alpha_\tau)\cosh(\nu\alpha_\sigma)}{\sinh(\nu(\beta/2-\alpha_\tau))\sinh(\nu(\beta/2-\alpha_\sigma))}: & \uncrossedB
    \end{cases}\label{eq:A.HpHp} \\[2ex]
    \frac{\langle \delta G_-(\tau,\tau')\delta G_-(\sigma,\sigma')\rangle_0}{8\;\overline{G_{00}}(\tau,\tau')\overline{G_{00}}(\sigma,\sigma')}&=\lambda
    \begin{cases}
    \displaystyle \frac{-\cosh(\nu(\beta/2-\alpha_\tau-\alpha_\sigma+2\zeta))\cosh(\nu\beta/2)}{\cosh(\nu(\beta/2-\alpha_\tau))\cosh(\nu(\beta/2-\alpha_\sigma))}: & \crossedB \\
    0: & \uncrossedB
    \end{cases}\label{eq:A.GmGm} \\[2ex]
    \frac{\langle \delta H_-(\tau,\tau')\delta G_-(\sigma,\sigma')\rangle_0}{8\;\overline{G_{01}}(\tau,\tau')\overline{G_{00}}(\sigma,\sigma')}&=\lambda
    \begin{cases}
    \displaystyle \frac{\sinh(\nu(\beta/2-\alpha_\tau-\alpha_\sigma+2\zeta))\cosh(\nu\beta/2)}{\sinh(\nu(\beta/2-\alpha_\tau))\cosh(\nu(\beta/2-\alpha_\sigma))}\text{sgn}(\alpha_\tau)\text{sgn}(\zeta): & \crossedB \\
    0: & \uncrossedB
    \end{cases}\label{eq:A.HmGm} \\[2ex]
    \frac{\langle \delta H_-(\tau,\tau')\delta H_-(\sigma,\sigma')\rangle_0}{8\;\overline{G_{01}}(\tau,\tau')\overline{G_{01}}(\sigma,\sigma')}&=\lambda
    \begin{cases}
    \displaystyle \frac{\cosh(\nu(\beta/2-\alpha_\tau-\alpha_\sigma+2\zeta))\cosh(\nu\beta/2)}{\sinh(\nu(\beta/2-\alpha_\tau))\sinh(\nu(\beta/2-\alpha_\sigma))}\text{sgn}(\alpha_\tau)\text{sgn}(\alpha_\sigma): & \crossedB \\
    0: & \uncrossedB
    \end{cases}\label{eq:A.HmHm}
\end{align}

All correlators not given vanish identically. These equations are valid at any values of $\beta$ and $\lambda$,\footnote{While the assumption that $p\ll N$ never appeared explicitly in this derivation, we would be wary of exiting this regime.} to leading order in $1/N$ and $1/p$ as both go to $0$. Clearly, they generically depend on the values of the coordinates $\tau^{(\prime)},\sigma^{(\prime)}$ and not just their ordering on the circle. To recover the latter, we need to move into one of the phases discussed in section \ref{sect:MQPhases}. The simpler of the two is the disks phase, where $\nu\beta=\mathcal{O}(1/p)$ and there are no $01$-chords or $10$-chords (as discussed in section \ref{subsect:CRDisks}). Then the only surviving correlators are $\langle \delta G_\pm \delta G_\pm\rangle_0$, and these reduce to simply $-\lambda$ for crossed chords and $0$ otherwise. This reproduces two copies of the standard DSSYK correlators (to leading order in $p$)
\begin{equation}\label{eq:A.GGdisks}
    \langle \delta G_{00} \delta G_{00}\rangle_0^{\text{disks}} = \langle \delta G_{11} \delta G_{11}\rangle_0^{\text{disks}} = \begin{cases}-\lambda: & \crossedB \\ 0: & \uncrossedB\end{cases},\quad\quad \langle \delta G_{00} \delta G_{11}\rangle_0^{\text{disks}} = 0
\end{equation}
from which the series expansion in $\mathcal{J}$ reproduces two copies of the standard DSSYK chord diagram sum.

The cylinder phase is more subtle. We take $\nu\beta=\mathcal{O}(\log p)$ so that we have $
    \sigma=2p\,\exp(-\nu\beta)=\mathcal{O}(p^0)$. We also need to restrict ourselves to $N$ values sufficiently large such that\footnote{This is easily achieved in the ``triple scaled'' region of interest, where $\lambda\ll 1$, but still bears stating since the standard DSSYK does not carry this restriction.}
\begin{equation}\label{eq:A.large_enough_N}
    N\gg p\frac{e^{\nu\beta}}{\nu\beta} = \frac{p^2}{\sigma \log(p/\sigma)} \quad \implies \quad \lambda \ll \sigma \log p.
\end{equation}
This allows us to only consider diagrams where $\alpha_\tau$ and $\alpha_\sigma$ are each close to either $0$ or $\beta$. We can then distinguish between different classes of uncrossed diagrams: the two arcs are either ``nested'' $\uncrossedLLB$  or ``unnested'' $\uncrossedLRB$. The nested diagrams can further be distinguished by whether the $\tau^{(\prime)}$ chord is on the inside $\uncrossedLLRinG$  or on the outside $\uncrossedLLGinR$. 

With this in mind, the cylinder phase correlators (to leading order in $p$) are:
\begin{align}
    \frac{\langle \delta G_+(\tau,\tau')\delta G_+(\sigma,\sigma')\rangle_0}{8\;\overline{G_{00}}(\tau,\tau')\overline{G_{00}}(\sigma,\sigma')}&=\frac{\langle \delta G_-(\tau,\tau')\delta G_-(\sigma,\sigma')\rangle_0}{8\;\overline{G_{00}}(\tau,\tau')\overline{G_{00}}(\sigma,\sigma')}=\lambda
    \begin{cases}
    -1: & \crossedB \\
    0: & \uncrossedB
    \end{cases}\label{eq:A.GpGpTube}\\[2ex]
    \frac{\langle \delta H_+(\tau,\tau')\delta G_+(\sigma,\sigma')\rangle_0}{8\;\overline{G_{01}}(\tau,\tau')\overline{G_{00}}(\sigma,\sigma')}&=\lambda
    \begin{cases}
    -1: & \crossedB \\
    -2: & \uncrossedLLRinG \\
    0: & \uncrossedLLGinR,\uncrossedLRB
    \end{cases}\label{eq:A.HpGpTube} \\[2ex]
    \frac{\langle \delta H_+(\tau,\tau')\delta H_+(\sigma,\sigma')\rangle_0}{8\;\overline{G_{01}}(\tau,\tau')\overline{G_{01}}(\sigma,\sigma')}&=\lambda
    \begin{cases}
    -1: & \crossedB \\
    -2: & \uncrossedLLB \\
    0: & \uncrossedLRB
    \end{cases}\label{eq:A.HpHpTube} \\[2ex]
    \frac{\langle \delta H_-(\tau,\tau')\delta G_-(\sigma,\sigma')\rangle_0}{8\;\overline{G_{01}}(\tau,\tau')\overline{G_{00}}(\sigma,\sigma')}&=\lambda
    \begin{cases}
    \text{sgn}(\alpha_\tau)\text{sgn}(\zeta): & \crossedB \\
    0: & \uncrossedB
    \end{cases}\label{eq:A.HmGmTube} \\[2ex]
    \frac{\langle \delta H_-(\tau,\tau')\delta H_-(\sigma,\sigma')\rangle_0}{8\;\overline{G_{01}}(\tau,\tau')\overline{G_{01}}(\sigma,\sigma')}&=\lambda
    \begin{cases}
    \text{sgn}(\alpha_\tau)\text{sgn}(\alpha_\sigma): & \crossedB \\
    0: & \uncrossedB
    \end{cases}\label{eq:A.HmHmTube}
\end{align}
Using these to solve for the $\delta G_{ab}$ correlators, along with a brief exercise in spatial reasoning, shows that all of these can be summed up by the following rule: for each $\delta G_{ab}(\tau,\tau')$, place a node on circle $(a)$ at coordinate $\tau$ and another on circle $(b)$ at coordinate $\tau'$. Connect circles $0$ and $1$ by a cylinder. If the shortest chord\footnote{Our restriction on $\lambda$ in \eqref{eq:A.large_enough_N} means we never have to deal with pairs of nodes for which the ``shortest chord'' between them is ambiguous.} connecting $\tau$ and $\tau'$ must cross the shortest chord connecting $\sigma$ and $\sigma'$, the correlator gives $-\lambda$. Otherwise, it gives $0$.

\section{Deriving the chord diagram expansion}\label{app:B}

Once we have \eqref{eq:A.GpGpTube}-\eqref{eq:A.HmHmTube} in hand, we can proceed with Taylor expanding $Z_{MQ}[\beta]$ in $\mathcal{J}$ and distilling the sum over chord diagrams. Taking the effective action from \eqref{eq:A.SeffFinal} and restoring $\mathcal{J}$, we get 
\begin{equation}\label{eq:B.path_int}
    Z_{MQ}[\beta]=\int \mathcal{D}G\;\exp\Bigg[-S_{\text{eff}}^{\mathcal{J}=0}[G] + \frac{\mathcal{J}^2}{4p^2}\iint\limits_{\;00}^{\,\beta\beta}d\tau\, d\tau'\,\sum_{a,b} (2G_{ab}(\tau,\tau'))^p\Bigg]
\end{equation}
which can then be expressed as an expectation value in the $\mathcal{J}=0$ theory as in \eqref{eq:<expJ>}. The $\delta G$-independent piece is $S_0[\beta]=-\frac{N}{2}\Tr\log A$ as before, which we saw in section \ref{subsect:entropies} gave a topological term plus a ground state energy term. Following equations (384)-(389) in \cite{Lin:2023trc}, we write
\begin{equation}\label{eq:B.sum_n_sum_ab}
    Z_{MQ}[\beta]=e^{S_0[\beta]}\sum_{n=0}^\infty \frac{\mathcal{J}^{2n}}{\lambda^n n!}\Bigg\langle\int\prod_{j=1}^n \bigg( d\tau_j d\tau'_j \sum_{a,b} (2G_{ab}(\tau_j,\tau'_j))^p \bigg)\Bigg\rangle_0
\end{equation}
where the region being integrated over is $\tau_j\in[0,\beta)$, $\tau'_j\in[0,\tau_j)$. We then expand (schematically) to leading order in $p$\footnote{In the disks phase, this expansion is not valid for $G_{01}$ and $G_{10}$, since $\overline{G_{01}}$ is already suppressed in $p$. Instead, drop all such terms from expansion.}
\begin{equation}
    (2G)^p=\Big(2\overline{G}+\frac{\xi}{p}\delta G\Big)^p = (2\overline{G})^p \exp\Big(\frac{\xi \delta G}{2\overline{G}}\Big)
\end{equation}
where $\xi$ is a phase which we can safely set to $1$, since we have restricted ourselves to $\tau_j'<\tau_j$ and $p\equiv 0 \text{ mod } 4$. Expanding the product of sums over $a,b$ into a single sum over $(a_1,b_1,...,a_n,b_n)\in\{0,1\}^n$, we get
\begin{align}\label{eq:B.sum_n_sum_pi}
    Z_{MQ}[\beta]&=e^{S_0[\beta]}\sum_{n=0}^\infty \frac{\mathcal{J}^{2n}}{\lambda^n n!}\sum_{a_1...b_n}\int\Bigg\langle\prod_{j=1}^n d\tau_j d\tau'_j\; \Big(2\overline{G_{a_jb_j}}(\tau_j,\tau'_j)\Big)^p \exp\bigg(\frac{\delta G_{a_jb_j}(\tau_j,\tau'_j)}{2\overline{G_{a_jb_j}}(\tau_j,\tau'_j)}\bigg)\Bigg\rangle_0\nonumber\\
    &=e^{S_0[\beta]}\sum_{n=0}^\infty \frac{\mathcal{J}^{2n}}{\lambda^n n!}\sum_{a_1...b_n}\int\prod_{j=1}^n \bigg(d\tau_j d\tau'_j\; \Big(2\overline{G_{a_jb_j}}(\tau_j,\tau'_j)\Big)^p \bigg)\Bigg\langle\exp\bigg(\sum_{j=1}^n\frac{\delta G_{a_jb_j}(\tau_j,\tau'_j)}{2\overline{G_{a_jb_j}}(\tau_j,\tau'_j)}\bigg)\Bigg\rangle_0
\end{align}
where by $\overline{G_{10}}$ we mean $-\overline{G_{01}}$. Using the fact that the $\mathcal{J}=0$ theory is Gaussian to leading order in $p$, we can then write
\begin{equation}\label{eq:B.Gaussianity}
    \Bigg\langle\exp\bigg(\sum_{j=1}^n\frac{\delta G_{a_jb_j}}{2\overline{G_{a_jb_j}}}\bigg)\Bigg\rangle_0 = \exp\Bigg(\frac{1}{2}\sum_{i,j=1}^n \frac{\Big\langle\delta G_{a_ib_i} \delta G_{a_jb_j}\Big\rangle_0}{4\overline{G_{a_ib_i}}\,\overline{G_{a_jb_j}}}\Bigg) = \exp\Bigg(\sum_{i<j} \frac{\Big\langle\delta G_{a_ib_i} \delta G_{a_jb_j}\Big\rangle_0}{4\overline{G_{a_ib_i}}\,\overline{G_{a_jb_j}}}\Bigg)
\end{equation}
where $G_{a_jb_j}$ is always $G_{a_jb_j}(\tau_j,\tau_j')$. See that every set of $n$, $\{a_j,b_j\}$, $\{\tau_j,\tau_j'\}$ gives us a term that looks like a chord diagram with some penalty. We can use the generic temperature symmetries of the $\delta G$ correlators to reorganize this as a sum over $n_0$ nodes on circle $0$, $n_1$ nodes on circle $1$, and $c$ chords connecting the two. It is also convenient to rescale the integrals $x_j=\tau_j/\beta$ to pull out the unitful factors of $\beta$ and move around combinatorial factors to pave the way for a discrete sum over diagrams. This yields the explicit version of \eqref{eq:Z_sum_diagrams}:
\begin{align}
    Z_{MQ}[\beta]=&e^{S_0[\beta]}\sum_{\substack{n_{0},n_{1}=0 \\ n_{0}+n_{1} \text{ even}}}^\infty\frac{1}{n_{0}!\;n_{1}!}\Big(\frac{\beta\mathcal{J}}{\sqrt{\lambda}}\Big)^{n_{0}+n_{1}}\sum_{\substack{c=0 \\ n_{0}+c \text{ even}}}^{\min(n_{0},n_{1})}\nonumber\\
    &n_0!n_1!\int\limits_D dx_1...dx'_{(n_0+n_1)/2}\;\exp\bigg(p\sum_{j} \log|2\overline{G_{a_jb_j}}|+\sum_{j<k}\frac{\langle\delta G_{a_jb_j} \delta G_{a_kb_k}\rangle_0}{(2\overline{G_{a_jb_j}})(2\overline{G_{a_kb_k}})}\bigg)\label{eq:B.Z_sum_diagrams}
\end{align}
In this equation, the sums over $j$ run from $1$ to the total number of chords, $\frac{n_0+n_1}{2}$. The indices $(a_j,b_j)$ are $(0,1)$ for $1\leq j\leq c$, then $(0,0)$ for $c<j\leq \frac{n_0-c}{2}$, and finally $(1,1)$ for $j> \frac{n_0-c}{2}$. The $(x_j,x_j')$ for $1\leq j\leq c$ are from $0$ to $1$ unrestricted, while the rest are $x_j\in[0,1)$, $x_j'\in[0,x_j)$. The factor of $n_0!n_1!$ in front of the integral is there because if the integrand becomes dependent only on the order of the $(x_j,x_j')$ and not their values, $n_0!n_1!$ times the integral will become a sum over orderings. This is strictly true in the disks phase, and the sum is dominated by the $c=0$ term, so $Z_{MQ}[\beta]$ factorizes into the square of a single DSSYK partition function. In the tube phase, the $\delta G$ correlators only depend on ordering, but the $\log|2\overline{G}|$ term remains location dependent. See the above discussion on $\mu$-chords in section \ref{subsect:CRTube1}.

\section{Deriving the AS$^2$ saddle points}\label{app:C}

We start by putting the new effective action (with the $\Delta$ term included, defining $\Delta\equiv c/\lambda$) into a form analogous to \eqref{eq:A.SeffAM}:
\begin{equation}
\label{eq:C.SefflnAM}
    -\frac{S_{\text{eff},\Delta}^{\mathcal{J}=0}[G]}{N}=\frac{c}{2p}\Big(\log\big|2G_{00}(3\beta/4,\beta/4)\big|+\log\big|2G_{11}(3\beta/4,\beta/4)\big|\Big)-\frac{1}{2}\Tr\Big[\log (M)\Big] +\frac{1}{2}\Tr\Big[AM-I\Big]
\end{equation}
where $A,M$ are defined in \eqref{eq:A.defAM}. This system has sufficient symmetry to write $\overline{G_{11}}=\overline{G_{00}}$ and $\overline{G_{10}}=-\overline{G_{01}}$, so we can substitute that into \eqref{eq:C.SefflnAM} and vary to get the equation of motion
\begin{equation}\label{eq:C.2matter_EOM}
    \overline{M}=\begin{pmatrix}\partial_\tau -zK & i\nu\\-i\nu & \partial_\tau-zK\end{pmatrix}^{-1}
\end{equation}
where
\begin{equation}\label{eq:C.zKdef}
    z=\frac{c}{4p|\overline{G_{00}}(3\beta/4,\beta/4)|},\quad K(\tau,\tau')=2\Big(\delta(\tau-\beta/4)\delta(\tau'-3\beta/4)-\delta(\tau-3\beta/4)\delta(\tau'-\beta/4)\Big).
\end{equation}
We are going to treat $z$ as a free parameter in this differential equation, and then solve for $z$ as a function of $c$ later. Notice that when $c=0$, we necessarily have $z=0$, so this reduces to $A\overline{M}=I$ as before. We can invert the right side block matrix in \eqref{eq:C.2matter_EOM} by method of Schur complements since its off-diagonal blocks are proportional to the identity, yielding the pair of equations:
\begin{align}
    \Big(-\frac{1}{\nu}\big(\partial_\tau-zK\big)^2+\nu\Big)&\overline{G_{01}}=I \label{eq:C.2matter_EOM_block01}\\
    \overline{G_{00}}=-\frac{1}{\nu}\big(\partial_\tau-zK\big)&\overline{G_{01}} \label{eq:C.2matter_EOM_block00}
\end{align}
By requiring that $\overline{G_{00}}$ not be singular anywhere, we find that we can break $\overline{G_{01}}$ into four sub-blocks based on whether $\tau$ and $\tau'$ are each in $(-\frac{\beta}{4},\frac{\beta}{4})$ or $(\frac{\beta}{4},\frac{3\beta}{4})$. Take note that we are solving here for $\overline{G_{ab}}$ with domain $(-\frac{\beta}{4},\frac{3\beta}{4}\,]^2$ instead of the usual $[\,0,\beta\,)^2$. As always, we impose antiperiodicity on both $\tau\rightarrow\tau+\beta$ and $\tau'\rightarrow\tau'+\beta$, so this contains the same information as the usual formulation. The discontinuities at the $\frac{\beta}{4}$ and $\frac{3\beta}{4}$ boundaries produce $\delta$ functions in $\partial_\tau$, which must be canceled by the $\delta$ functions in $K$. This gives us a set of mixed boundary conditions on these blocks parameterized by $z$. Within the space of functions obeying these sub-block boundary conditions, the operator $\partial_\tau-zK$ acts like a simple derivative inside each sub-block and ignores the boundaries. This means that each sub-block $B$ of $\overline{G_{01}}$ satisfies
\begin{equation}
    \Big(-\frac{1}{\nu}\partial_\tau^2+\nu\Big)B(\tau,\tau')=\delta(\tau-\tau')
\end{equation}
so it must be that
\begin{equation}
    B(\tau,\tau')=\frac{1}{2}e^{-\nu|\tau-\tau'|}+b_1e^{(\tau-\tau')}+b_2e^{-(\tau-\tau')}+b_3e^{(\tau+\tau')}+b_4e^{-(\tau+\tau')}.
\end{equation}

\begin{figure}[h]
\centering
\includegraphics[width = .95\textwidth]{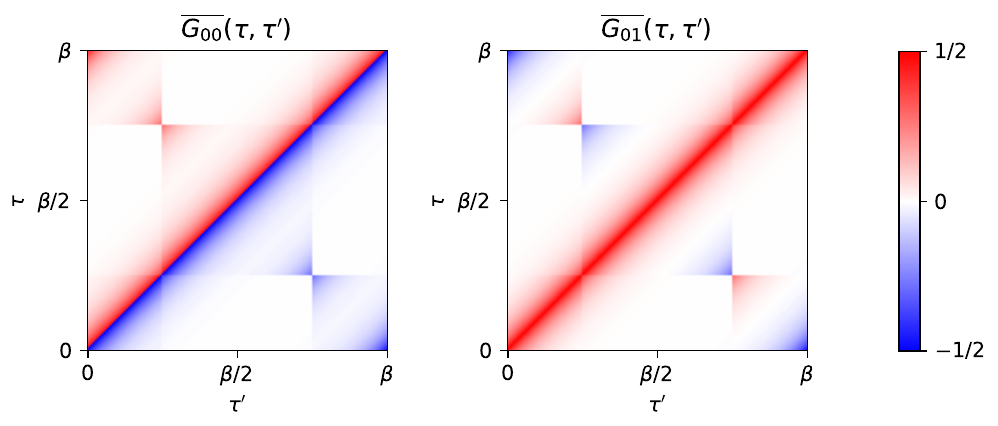}
\caption{Plots of $\langle G_{ab}\rangle$ in the free AS$^2$ saddle point with two heavy operators.  This solution is computed with $\nu\beta=15$ and $\lambda\Delta=p/5$. The peak values of $\overline{G_{ab}}$ near $(\beta/4,3\beta/4)$ and $(3\beta/4,\beta/4)$ indicate that this geometry contains a wormhole supported by the heavy particles' worldlines. The sharp edges reflect the presence of the two heavy matter chords.
\label{fig:ASSR_2matter_Realcolor_Plots}}
\end{figure}

\noindent Once we have used the mixed boundary conditions to solve for the coefficients in each sub-block, we can compute  $|\overline{G_{00}}(3\beta/4,\beta/4)|$ and then use \eqref{eq:C.zKdef} to solve for $z$. This yields:
\begin{align}
    \tau,\tau'&\in\Big(-\frac{\beta}{4},\frac{3\beta}{4}\,\Big]\\*
    \overline{G_{ab}}(\tau,\tau')&=\begin{cases}A_{ab}(\tau,\tau'):&\text{sgn}(\tau-\beta/4)=\text{sgn}(\tau'-\beta/4)\\B_{ab}(\tau,\tau'):&\text{sgn}(\tau-\beta/4)\neq\text{sgn}(\tau'-\beta/4)\end{cases}\label{eq:C.ASS_2matter_first}\\*
    A_{00}(\tau,\tau')&=\frac{1}{2}\text{sgn}(\tau-\tau')\Big[e^{-\nu|\tau-\tau'|}-r_A\Big(e^{-\nu(\beta/2+|\tau-\tau'|)}-e^{-\nu(\beta/2-|\tau-\tau'|)}\Big)\Big]\\*
    A_{01}(\tau,\tau')&=\frac{1}{2}\Big[e^{-\nu|\tau-\tau'|}-r_A\Big(e^{-\nu(\beta/2+|\tau-\tau'|)}+e^{-\nu(\beta/2-|\tau-\tau'|)}\Big)\Big]\\*
    B_{00}(\tau,\tau')&=\frac{1}{2}\text{sgn}(\tau-\tau')r_B\Big[e^{-\nu|\tau-\tau'|}+e^{-\nu(\beta-|\tau-\tau'|)}\Big]\\*
    B_{01}(\tau,\tau')&=\frac{1}{2}r_B\Big[e^{-\nu|\tau-\tau'|}-e^{-\nu(\beta-|\tau-\tau'|)}\Big]\\*
    r_A&=\frac{2z+(1+z^2)e^{-\nu\beta/2}}{(1+z^2)(1+e^{-\nu\beta})+4ze^{-\nu\beta/2}}\longrightarrow\begin{cases}\frac{2z}{1+z^2}:&z\gg e^{-\nu\beta/2}\\e^{-\nu\beta/2}+2z:&z\leq e^{-\nu\beta/2}\end{cases}\\*
    r_B&=\frac{1-z^2}{(1+z^2)(1+e^{-\nu\beta})+4ze^{-\nu\beta/2}}\longrightarrow\frac{1-z^2}{1+z^2}
\end{align}
To solve for $z$, we can then compute
\begin{equation}
    |\overline{G_{00}}(3\beta/4,\beta/4)|=\frac{1}{2}\Big(|A_{00}(3\beta/4,\beta/4)|+|B_{00}(3\beta/4,\beta/4)|\Big)
\end{equation}
and then use \eqref{eq:C.zKdef} find that
\begin{equation}
    z=\sqrt{\frac{c}{2p-c}}+\mathcal{O}(e^{-\nu\beta})\label{eq:C.ASS_2matter_z}
\end{equation}
for $0\leq c\leq p$. This is the solution plotted in Figure~\ref{fig:ASSR_2matter_Realcolor_Plots}. The recolored absolute value of this solution is plotted in Figure~\ref{fig:ASSR_2matter_Plots}. If we were to let $c=p$, we would get $z=1$, so $r_B=0$ and all $\overline{G_{ab}}$ would vanish whenever $\tau$ and $\tau'$ sit on opposite sides of the matter worldlines. Additionally, $r_A\rightarrow1$, so $|A_{00}(3\beta/4,\beta/4)|$ and $|A_{01}(3\beta/4,\beta/4)|$ would both approach $1/2$, meaning that $\tau=\beta/4$ and $\tau=3\beta/4$ would be directly adjacent points in the bulk. In other words, the geometry would fully separate.

\begin{figure}[t]
\centering
\includegraphics[width = .95\textwidth]{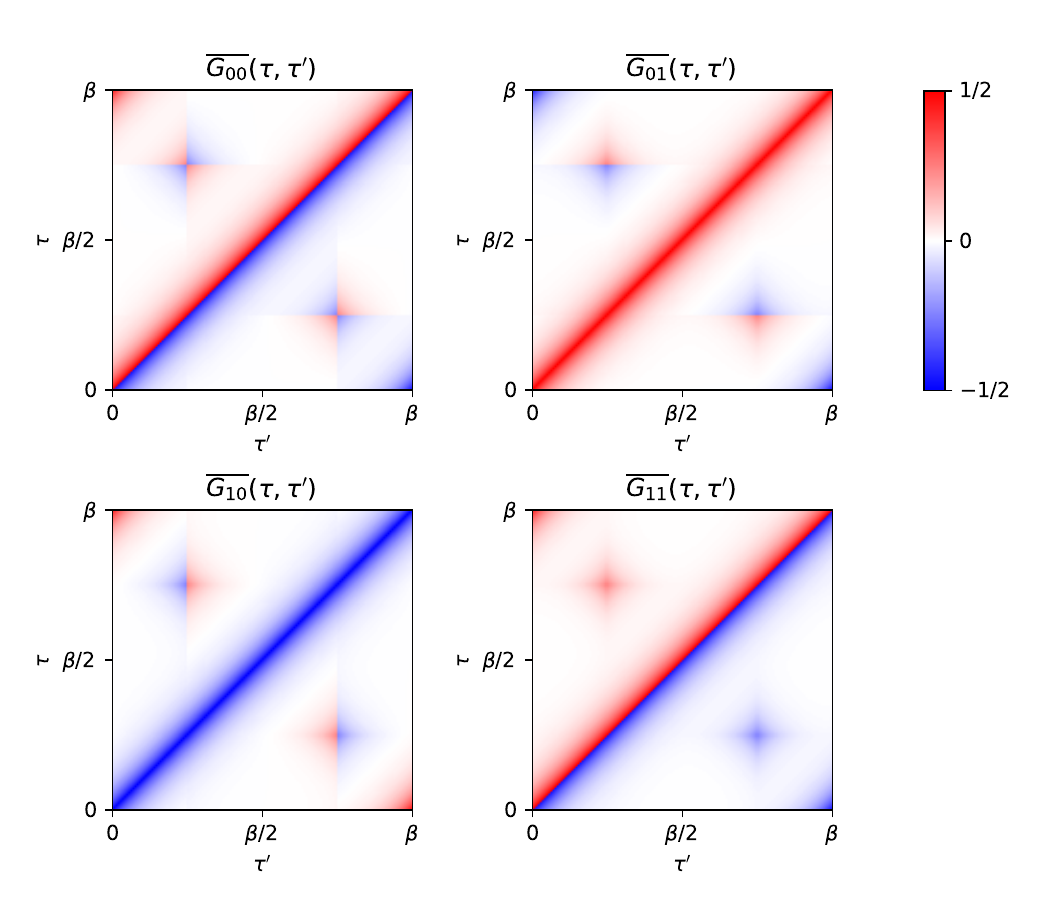}
\caption{Plots of $\langle G_{ab}\rangle$  in the free  AS$^2$ saddle point with a single heavy matter chord. This solution is computed with $\nu\beta=15$ and $\lambda\Delta=p/2$. We again clearly see the off-diagonal peak values that reveal the presence of the wormhole. Notice that $\overline{G_{11}}$ has no discontinuities away from $\tau=\tau'$ indicating that the dual geometry away from the heavy matter chord remains connected. 
\label{fig:ASSR_1matter_Realcolor_Plots}}
\end{figure}

For the AS$^2$ geometry with one $\mathbb{O}$ insertion, we simply remove one of the $\log$ terms in \eqref{eq:C.SefflnAM} to get 
\begin{equation}\label{eq:C.SefflnAM_1matter}
    -\frac{\tilde{S}_{\text{eff},\Delta}^{\mathcal{J}=0}[G]}{N}=\frac{c}{2p}\Big(\log\big|2G_{00}(3\beta/4,\beta/4)\big|\Big)-\frac{1}{2}\Tr\Big[\log (M)\Big] +\frac{1}{2}\Tr\Big[AM-I\Big]
\end{equation}
which gives an analogous equation of motion
\begin{equation}\label{eq:C.1matter_EOM}
    \overline{M}=\begin{pmatrix}\partial_\tau -zK & i\nu\\-i\nu & \partial_\tau\end{pmatrix}^{-1}
\end{equation}
 with $z,K$ defined as in \eqref{eq:C.zKdef}, and now a set of four block equations:
\begin{align}
    \Big(-\frac{1}{\nu}\partial_\tau\big(\partial_\tau-zK\big)+\nu\Big)&\overline{G_{01}}=I \label{eq:C.1matter_EOM_block01}\\
    \overline{G_{11}}=-\frac{1}{\nu}\big(\partial_\tau-zK\big)&\overline{G_{01}} \label{eq:C.1matter_EOM_block11}\\
    \Big(+\frac{1}{\nu}\big(\partial_\tau-zK\big)\partial_\tau-\nu\Big)&\overline{G_{10}}=I \label{eq:C.1matter_EOM_block10}\\
    \overline{G_{00}}=\frac{1}{\nu}\partial_\tau&\overline{G_{10}} \label{eq:C.1matter_EOM_block00}
\end{align}
Again requiring that neither $\overline{G_{00}}$ nor $\overline{G_{11}}$ be singular anywhere gives us mixed sub-block boundary conditions which can be solved to give the following:
\begin{align}
\tau,\tau'&\in\Big(-\frac{\beta}{4},\frac{3\beta}{4}\,\Big]\\
    \overline{G_{ab}}(\tau,\tau')&=\begin{cases}A_{ab}(\tau,\tau'):&\text{sgn}(\tau-\beta/4)=\text{sgn}(\tau'-\beta/4)\\B_{ab}(\tau,\tau'):&\text{sgn}(\tau-\beta/4)\neq\text{sgn}(\tau'-\beta/4)\end{cases}\label{eq:C.ASS_1matter_first}\\
    A_{00}(\tau,\tau')=A_{11}(\tau,\tau')&=\text{sgn}(\tau-\tau')\frac{\cosh\big(\nu(\frac{\beta}{2}-|\tau-\tau'|)\big)+z\cosh\big(\nu|\tau-\tau'|\big)}{2\cosh\big(\frac{\nu\beta}{2}\big)+2z}\\
    A_{01}(\tau,\tau')=-A_{10}(\tau,\tau')&=\frac{\sinh\big(\nu(\frac{\beta}{2}-|\tau-\tau'|)\big)-z\sinh\big(\nu|\tau-\tau'|\big)}{2\cosh\big(\frac{\nu\beta}{2}\big)+2z}\\
    B_{01}(\tau,\tau')=-B_{10}(\tau',\tau)&=\frac{\sinh\big(\nu(\frac{\beta}{2}-|\tau-\tau'|)\big)+z\,\text{sgn}(\tau-\tau')\sinh\big(\nu(\frac{\beta}{2}-(\tau+\tau'))\big)}{2\cosh\big(\frac{\nu\beta}{2}\big)+2z}\\
    B_{00}(\tau,\tau')&=\text{sgn}(\tau-\tau')\frac{\cosh\big(\nu(\frac{\beta}{2}-|\tau-\tau'|)\big)-z\cosh\big(\nu(\frac{\beta}{2}-(\tau+\tau'))\big)}{2\cosh\big(\frac{\nu\beta}{2}\big)+2z}\\
    B_{11}(\tau,\tau')&=\text{sgn}(\tau-\tau')\frac{\cosh\big(\nu(\frac{\beta}{2}-|\tau-\tau'|)\big)+z\cosh\big(\nu(\frac{\beta}{2}-(\tau+\tau'))\big)}{2\cosh\big(\frac{\nu\beta}{2}\big)+2z}
\end{align}
for
\begin{equation}\label{eq:C.ASS_1matter_z}
    z=\frac{c+\sqrt{c^2+8p\cosh^2\big(\frac{\nu\beta}{2}\big)c}}{4p\cosh\big(\frac{\nu\beta}{2}\big)}=\sqrt{\frac{c}{2p}}+\mathcal{O}(e^{-\nu\beta/2})\;,\quad 0\leq c\leq 2p
\end{equation}
This is the solution plotted in Figure~\ref{fig:ASSR_1matter_Realcolor_Plots}. The recolored absolute value of this solution is plotted in Figure~\ref{fig:ASSR_1matter_Plots}.

At $c=2p$, we would again get $z=1$, at which point $|A_{00}(3\beta/4,\beta/4)|$ and $|A_{11}(3\beta/4,\beta/4)|$ would both be exactly $1/2$. Just like in the above two-matter case, this means the points $\tau=\beta/4$ and $\tau=3\beta/4$ would have become, in some sense, the same bulk point. In both versions of the AS$^2$ geometry, we see that the maximum total mass of heavy particles is $c_\text{max}=2p$ (since before $c_\text{tot}=c+c$). Plugging this into $\Delta=c/\lambda$ gives $\Delta_\text{max}=2p/\lambda=N/p$, so $p'_\text{max}=p\Delta_\text{max}=N$. For the one-matter case, this is all of the $(0)$-site Majoranas. For the two-matter case, each insertion is half of that site's total Majoranas. Our $G,\Sigma$ action almost certainly breaks down at $p'\sim N$. From the trace combinatorics description, we know that the crossing penalty for a matter chord with $p'=N-b$ is actually the same as that with $p'=b$. Our $G,\Sigma$ action, which does not know this, could therefore only possibly contend with matter having up to $p'=N/2$. This means we should not trust $c>p$ in the one-matter AS$^2$ picture, and we are hesitant to trust the top edge $c=p$ in the two-matter AS$^2$ picture. To be safe, we should keep our total heavy particle $p'$ at least $\mathcal{O}(p)$ below $N/2$. This can be achieved by setting $c$ equal to any fixed fraction of $p$. Even more conservatively, we could restrict $c$ to only be parametrically large in $p$.

\section{Discrete Coupling Sites}\label{app:D}

Here we will consider the trace combinatorics of coupled SYK models with $r$ discrete interaction sites of the form
\begin{equation}\label{eq:D.delta_functions_Hint}
    H_r(\tau)=H^{(0)}_{SYK}+H^{(1)}_{SYK}+i\frac{\mu_0}{p}\bigg(\sum_{j=1}^r\delta(\tau-\tau_r)\bigg)\bigg(\sum_{i=1}^N\psi_i^{(0)}\psi_i^{(1)}\bigg).
\end{equation}
The AS$^2$ wormholes with one or two heavy operator insertion  can be thought of as microcosms of tube geometries in this theory with $r=1$ and $r=2$, respectively. In these theories, it is tenable to directly compute the partition function as a sum over chord diagrams without using a $G,\Sigma$ action. Without loss of generality, let $\tau_1=0$ and define $\tau_{r+1}\equiv\beta$. Then we write
\begin{equation}
    H_\text{int}=-\frac{i}{p}\sum_{i=1}^N\psi_i^{(0)}\psi_i^{(1)}
\end{equation}
so that
\begin{align}
    Z_r[\beta]&=\Tr\bigg(\exp\Big(-\int_0^\beta H_r(\tau)\,d\tau\Big)\bigg) 
    =\Tr\Bigg(\prod_{j=1}^r\bigg(e^{\mu_0H_\text{int}}\;e^{-(\tau_{j+1}-\tau_j)\big(H^{(0)}_\text{SYK}+H^{(1)}_\text{SYK}\big)}\bigg)\Bigg)\nonumber\\[1.5mm]
    &=\sum_{n_1...n_r=0}^\infty \frac{(-|\tau_2-\tau_1|)^{n_1}}{n_1!}...\frac{(-|\tau_{r+1}-\tau_r|)^{n_r}}{n_r!} \Tr\Bigg(\prod_{j=1}^r\bigg(e^{\mu_0H_\text{int}}\;\Big(H^{(0)}_\text{SYK}+H^{(1)}_\text{SYK}\Big)^{n_j}\bigg)\Bigg)
\end{align}
But $H_\text{SYK}^{(0)}$ commutes with $H_\text{SYK}^{(1)}$, so this reduces to a sum over terms of the form
\begin{equation}
\Tr\Bigg(\prod_{j=1}^r\bigg(e^{\mu_0H_\text{int}}\;\Big(H^{(0)}_\text{SYK}\Big)^{n_{0j}}\Big(H^{(1)}_\text{SYK}\Big)^{n_{1j}}\bigg)\Bigg).
\end{equation}
To compute this trace, we need to sum over all possible full contractions of fermions. Due to the randomness of the $J_{i_1...i_p}$'s, we can reduce this (to leading order in $N$) to all possible full contractions of the $H$'s as in \cite{Berkooz:2024lgq}. Each such full contraction will give us a single chord diagram. But now we have a handful more valid contraction patterns than in standard SYK. Abbreviating $H_\text{SYK}^{(a)}$ as $H_a$, we define:
\begin{align}
    \marknode{A}{H}_a\marknode{B}{H}_a&=\sum_{i_1...i_p}\big(J_{i_1...i_p}\big)^2\Big(\psi_{i_1}^{(a)}...\psi_{i_p}^{(a)}\Big)^2=2^p\frac{\mathcal{J}^2}{\lambda}\mathbb{1}\contractaa[green!60!black, thick]{A}{B}{1ex}{}\label{eq:D.green_green_cont}\\*
    \marknode{C}{H}_\text{int}\marknode{D}{H}_\text{int}&=-\frac{1}{p^2}\sum_i \psi_i^{(0)}\psi_i^{(1)}\psi_i^{(0)}\psi_i^{(1)}=\frac{2}{\lambda}\mathbb{1}\contractaa[blue, thick]{C}{D}{1ex}{}\label{eq:D.Hintbubble}\\*
    \marknode{E}{H}_a\marknode{F}{H}_{\overline{a}}\;\marknode{G}{H}_\text{int}^p&=\frac{1}{p^p}\sum_{i_1...i_p}\big(J_{i_1...i_p}\big)^2\Big(\psi_{i_1}^{(a)}...\psi_{i_p}^{(a)}\Big)\Big(\psi_{i_1}^{(\overline{a})}...\psi_{i_p}^{(\overline{a})}\Big)\Big(\psi_{i_1}^{(0)}\psi_{i_1}^{(1)}\Big)...\Big(\psi_{i_p}^{(0)}\psi_{i_p}^{(1)}\Big)\contractaa[red, thick]{E}{F}{1ex}{M}\arctoR[red, thick]{M}{G}{2.5ex}\nonumber\\*
    &=\Big(\frac{2}{p}\Big)^p\,\frac{\mathcal{J}^2}{\lambda}\mathbb{1}=\marknode{J}{H}_\text{int}^p\marknode{H}{H}_a\marknode{I}{H}_{\overline{a}}\contractaa[red, thick]{H}{I}{1ex}{N}\arctoL[red, thick]{N}{J}{2.5ex}\label{eq:D.red_red_cont}
\end{align}
using our $\mathcal{J}$ convention from \eqref{eq:SYK_J_distribution}. 

The general crossing rules are a bit complicated, but can be solved with some thorough combinatorial reasoning:\footnote{Notice that all of these rules are invariant under taking any contraction between two $H$'s and swapping it out for the ``other way around'' contraction, as we should expect for a cyclic trace.}
\begin{align}
    \marknode{A1}{H}_a\;\marknode{B1}{H}_b\;\marknode{C1}{H}_a\;\marknode{D1}{H}_b&=e^{-\delta_{ab}\lambda}\;\marknode{E1}{H}_a\;\marknode{F1}{H}_a\;\marknode{G1}{H}_b\;\marknode{H1}{H}_b\contractaa[green!60!black, thick]{A1}{C1}{1ex}{}\contractaa[green!60!black, thick]{B1}{D1}{2ex}{}\contractaa[green!60!black, thick]{E1}{F1}{1ex}{}\contractaa[green!60!black, thick]{G1}{H1}{1ex}{}\label{eq:D.green_on_green}\\[15pt]
    \marknode{I1}{H}_a\;\marknode{J1}{H}_\text{int}\;\marknode{K1}{H}_a&=e^{-\lambda/p}\;\marknode{L1}{H}_a\;\marknode{M1}{H}_a\;\marknode{N1}{H}_\text{int}\contractaa[green!60!black, thick]{I1}{K1}{1ex}{}\contractaa[green!60!black, thick]{L1}{M1}{1ex}{}\upnode[black!70, thick]{J1}{2.5ex}\upnode[black!70, thick]{N1}{2.5ex}\label{eq:D.gray_on_green}\\[20pt]
    \marknode{O1}{H}_a\;\marknode{P1}{H}_\text{int}^{n}\;\;...\;\;\marknode{Q1}{H}_{\overline{a}}\;\marknode{R1}{H}_\text{int}^{p-n}&=(-1)^n\;\marknode{T1}{H}_a\marknode{U1}{H}_{\overline{a}}\;\marknode{V1}{H}_\text{int}^p\contractaa[red, thick]{O1}{Q1}{1ex}{S1}\arctoR[red, thick]{S1}{R1}{2.5ex}\arctoL[red, thick]{S1}{P1}{2.5ex}\contractaa[red, thick]{T1}{U1}{1ex}{W1}\arctoR[red, thick]{W1}{V1}{2.5ex}\label{eq:D.red_phase}\\[15pt]
    \marknode{X1}{H}_a\;\marknode{Y1}{H}_\text{int}\;\;...\;\;\marknode{Z1}{H}_{\overline{a}}\;\marknode{A2}{H}_\text{int}^p&=e^{-\lambda/p}\;\marknode{B2}{H}_a\marknode{C2}{H}_{\overline{a}}\;\marknode{D2}{H}_\text{int}^p\;\marknode{E2}{H}_\text{int}\contractaa[red, thick]{X1}{Z1}{1ex}{F2}\contractaa[red, thick]{B2}{C2}{1ex}{G2}\upnode[black!70, thick]{Y1}{2.5ex}\upnode[black!70, thick]{E2}{2.5ex}\arctoR[red, thick]{F2}{A2}{2.5ex}\arctoR[red, thick]{G2}{D2}{2.5ex}\label{eq:D.gray_on_red}\\[15pt]
    \marknode{H2}{H}_a\;\;...\;\;\marknode{I2}{H}_b\;\marknode{J2}{H}_{\overline{a}}\;\marknode{K2}{H}_\text{int}^{n}\;\marknode{L2}{H}_b\;\marknode{M2}{H}_\text{int}^{p-n}&=\left(\begin{cases}e^{-\lambda n/p}:&b=a\\e^{-\lambda(1-n/p)}:&b=\overline{a}\end{cases}\right)\;\marknode{O2}{H}_a\marknode{P2}{H}_{\overline{a}}\;\marknode{Q2}{H}_\text{int}^{p}\;\marknode{R2}{H}_b\marknode{S2}{H}_b\contractaa[red, thick]{H2}{J2}{1ex}{N2}\contractaa[green!60!black, thick]{I2}{L2}{2ex}{}\arctoR[red, thick]{N2}{K2}{3ex}\arctoR[red, thick]{N2}{M2}{3ex}\contractaa[red, thick]{O2}{P2}{1ex}{T2}\contractaa[green!60!black, thick]{R2}{S2}{1ex}{}\arctoR[red, thick]{T2}{Q2}{2.5ex}\label{eq:D.green_on_red}\\[20pt]
    \marknode{A3}{H}_\text{int}^{p-n_b}\;\marknode{B3}{H}_a\;\marknode{C3}{H}_\text{int}^{n_b}\;\marknode{D3}{H}_b\;\marknode{E3}{H}_{\overline{a}}\;\marknode{F3}{H}_\text{int}^{n_a}\;\;...\;\;\marknode{G3}{H}_{\overline{b}}\;\marknode{H3}{H}_\text{int}^{p-n_a}&=\left(\begin{cases}e^{-\lambda\Big((n_a+n_b)/p-2n_an_b/p^2\Big)}:&b=a\\e^{-\lambda\Big(1-(n_a+n_b)/p+2n_an_b/p^2\Big)}:&b=\overline{a}\end{cases}\right)\contractaa[red, thick]{B3}{E3}{1ex}{I3}\contractaa[red!60!black, thick]{D3}{G3}{2ex}{J3}\arctoR[red, thick]{I3}{F3}{3ex}\arctoR[red, thick]{I3}{H3}{3ex}\arctoL[red!60!black, thick]{J3}{A3}{4ex}\arctoL[red!60!black, thick]{J3}{C3}{4ex}\nonumber\\[10pt]&\;\;\qquad \ \cdot\bigg(\marknode{A4}{H}_a\marknode{B4}{H}_{\overline{a}}\;\marknode{C4}{H}_\text{int}^p\bigg)\bigg(\marknode{D4}{H}_\text{int}^p\marknode{E4}{H}_b\marknode{F4}{H}_{\overline{b}}\bigg)\contractaa[red, thick]{A4}{B4}{1ex}{G4}\contractaa[red!60!black, thick]{E4}{F4}{1ex}{H4}\arctoR[red, thick]{G4}{C4}{2.5ex}\arctoL[red!60!black, thick]{H4}{D4}{2.5ex}\label{eq:D.red_on_red}
\end{align}
Using equations \eqref{eq:D.gray_on_green}, \eqref{eq:D.red_phase}, and \eqref{eq:D.gray_on_red}, we can figure out what the background geometry of $Z_r[\beta]$ looks like. To do this, we need to look at the ``empty'' geometry (including $H_\text{int}$'s but no $H_a$'s), insert a single Hamiltonian chord, and compute the expected diagram weight.

First, we need to account for all of the $H_\text{int}$ ``bubbles'' arising from \eqref{eq:D.Hintbubble}. If we take any fully contracted chord diagram in this theory, we can always generate another fully contracted diagram by inserting a pair of $H_\text{int}$'s right next to each other and contracting them together. We will call such a contraction a bubble. All of the $H_\text{int}$'s are indistinguishable,\footnote{They commute with each other, even when contracted to different partners.} so doing this $n$ times always generates a factor of $(\mu_0^2/\lambda)^n/n!$, regardless of the initial diagram. We can add up all of these bubbles separately to get an overall factor of $\exp(\mu_0^2/\lambda)$ for a single $H_\text{int}$ site. Since we have $r$ sites, we will divide out by
\begin{equation}
Z_0\equiv\Tr\Big(e^{r\mu_0H_\text{int}}\Big)=\exp(r^2\mu_0^2/\lambda)\Tr(\mathbb{1}).
\end{equation}

Let's compute the expected diagram weight for an $aa$-chord that stretches across $s$ of the $r$ $H_\text{int}$ sites. This means we want to look at
\begin{align}
    \frac{1}{Z_0}\Tr\bigg(e^{(r-s)\mu_0H_\text{int}}\;H_a\;e^{s\mu_0H_\text{int}}\;H_a\bigg)
    =\,\frac{1}{Z_0}\sum_{m,n=0}^\infty \frac{\big((r-s)\mu_0\big)^m}{m!}\frac{\big(s\mu_0\big)^n}{n!}\Tr\bigg(H_\text{int}^m\,H_a\,H_\text{int}^n\,H_a\bigg).
\end{align}
Of these $m,n$ insertions of $H_\text{int}$, $x$ will contract from one site to the other. The remaining insertions will form bubbles. Let the number of bubbles in each site be $y,z$. Changing our sum to be over $x,y,z$ and recalling that there are
\begin{equation}
    (2y-1)!!=\frac{(2y)!}{2^y\,y!}
\end{equation}
different ways to perfectly pair $2y$ objects, this reduces to
\begin{equation}
    \frac{1}{Z_0}\sum_{x,y,z=0}^\infty\frac{\big((r-s)s\mu_0^2\big)^x}{x!}\frac{\big((r-s)^2\mu_0^2\big)^y}{2^y\;y!}\frac{\big(s^2\mu_0^2\big)^z}{2^z\;z!}\Tr\bigg(\Big(\marknode{A5}{H}_\text{int}\marknode{B5}{H}_\text{int}\Big)^{y+z}\marknode{C5}{H}_\text{int}...\marknode{D5}{H}_\text{int}\marknode{E5}{H}_a\marknode{F5}{H}_\text{int}...\marknode{G5}{H}_\text{int}\marknode{J5}{H}_a\bigg)\contractaa[blue, thick]{A5}{B5}{1ex}{}\contractaa[blue, thick]{C5}{G5}{1.5ex}{}\contractaa[blue, thick]{D5}{F5}{1ex}{}\contractaa[green!60!black, thick]{E5}{J5}{2.5ex}{}
\end{equation}
where the ``$H_\text{int}...H_\text{int}$'' has $x$ insertions of $H_\text{int}$ in it. Using equations \eqref{eq:D.green_green_cont}, \eqref{eq:D.Hintbubble}, and \eqref{eq:D.gray_on_green} allows us to reduce this to
\begin{equation}
    \frac{1}{Z_0}\Big(2^p\frac{\mathcal{J}^2}{\lambda}\Tr\mathbb{1}\Big)\exp\bigg(\frac{\mu_0^2}{\lambda}\Big(2(r-s)s\,e^{-\frac{\lambda}{p}}+(r-s)^2+s^2\Big)\bigg)
\end{equation}
and then finally plugging in for $Z_0$ and simplifying gives
\begin{equation}
    \Big(2^p\frac{\mathcal{J}^2}{\lambda}\Big)\exp\bigg(-\frac{2\mu_0^2}{\lambda}(r-s)s\,\Big(1-e^{-\frac{\lambda}{p}}\Big)\bigg).
\end{equation}
The leading factor is precisely what we would expect in the uncoupled theory.\footnote{given our convention \eqref{eq:SYK_J_distribution} for $\mathcal{J}$, following \cite{Maldacena:2016hyu}, which makes $G,\Sigma$ simpler but leaves a $2^p$ in the trace combinatorics.} 
In the double scaling limit, this simplifies to
\begin{equation}\label{eq:D.green_rs_amp}
    2^p\,\frac{\mathcal{J}^2}{\lambda}\,\exp\bigg(-2\mu_0^2\frac{(r-s)s}{p}\bigg).
\end{equation}

Let's repeat this calculation for an $a\overline{a}$-chord. Now we are looking at
\begin{align}
    &\frac{1}{Z_0}\Tr\bigg(e^{(r-s)\mu_0H_\text{int}}\;H_a\;e^{s\mu_0H_\text{int}}\;H_{\overline{a}}\bigg)\, 
    =\,\frac{1}{Z_0}\sum_{m,n=0}^\infty \frac{\big((r-s)\mu_0\big)^m}{m!}\frac{\big(s\mu_0\big)^n}{n!}\Tr\bigg(H_\text{int}^m\,H_a\,H_\text{int}^n\,H_{\overline{a}}\bigg).
\end{align}
For this trace, we additionally have to select $p$ of the $m,n$ insertions of $H_\text{int}$ to contract with the $H_aH_{\overline{a}}$ as in \eqref{eq:D.red_red_cont}. Let $c$ of these come from $m$, and $p-c$ from $n$. Of the remaining insertions of $H_\text{int}$, define $x,y,z$ the same as before. We are then left with
\begin{align}
    \frac{1}{Z_0}&\sum_{c=0}^p \mu_0^p(r-s)^cs^{p-c}\begin{pmatrix}p\\c\end{pmatrix}\sum_{x,y,z=0}^\infty\frac{\big((r-s)s\mu_0^2\big)^x}{x!}\frac{\big((r-s)^2\mu_0^2\big)^y}{2^y\;y!}\frac{\big(s^2\mu_0^2\big)^z}{2^z\;z!}\nonumber\\[12pt]&\cdot\Tr\bigg(\Big(\marknode{A6}{H}_\text{int}\marknode{B6}{H}_\text{int}\Big)^{y+z}\marknode{Z6}{H}_\text{int}^{c}\marknode{C6}{H}_\text{int}...\marknode{D6}{H}_\text{int}\marknode{E6}{H}_a\marknode{F6}{H}_\text{int}...\marknode{G6}{H}_\text{int}\marknode{Y6}{H}_\text{int}^{p-c}\marknode{J6}{H}_{\overline{a}}\bigg)\contractaa[blue, thick]{A6}{B6}{1ex}{}\contractaa[blue, thick]{C6}{G6}{1.5ex}{}\contractaa[blue, thick]{D6}{F6}{1ex}{}\contractaa[red, thick]{E6}{J6}{2.5ex}{X6}\arctoR[red, thick]{X6}{Y6}{3ex}\arctoL[red, thick]{X6}{Z6}{2ex}\\[10pt]
    =\frac{\mu_0^p}{Z_0}&\Bigg(\sum_{c=0}^p\begin{pmatrix}p\\c\end{pmatrix} (r-s)^c(-s)^{p-c}\Bigg)\exp\bigg(\frac{\mu_0^2}{\lambda}\Big(2(r-s)s\,e^{-\frac{\lambda}{p}}+(r-s)^2+s^2\Big)\bigg)\bigg(\Big(\frac{2}{p}\Big)^p\frac{\mathcal{J}^2}{\lambda}\Tr\mathbb{1}\bigg).
\end{align}
The sum over $c$ simplifies into a binomial, leaving
\begin{equation}
    \bigg(\frac{2\mu_0}{p}\bigg)^p\Big(r-2s\Big)^p\,\frac{\mathcal{J}^2}{\lambda}\,\exp\bigg(-\frac{2\mu_0^2}{\lambda}(r-s)s\,\Big(1-e^{-\frac{\lambda}{p}}\Big)\bigg)
\end{equation}
which again simplifies in the double scaling limit to
\begin{equation}\label{eq:D.red_rs_amp}
    2^p\,\frac{\mathcal{J}^2}{\lambda}\,\bigg(\frac{\mu_0(r-2s)}{p}\bigg)^p\exp\bigg(-2\mu_0^2\frac{(r-s)s}{p}\bigg).
\end{equation}

There are a few key insights to be gleaned from \eqref{eq:D.green_rs_amp} and \eqref{eq:D.red_rs_amp}. Observe that when $r=2$ (relevant for the AS$^2$ geometry with two heavy operators), the amplitude for an $a\overline{a}$-chord to cross one of the $H_\text{int}$ sites (i.e. plugging in $s=1$) is exactly $0$. If we set $\mu_0=p/2$, then we also see that the amplitude for an $a\overline{a}$-chord to \emph{not} cross the $H_\text{int}$ sites (i.e. plugging in $s=0$) is the same as that for an $aa$-chord.\footnote{In this $\mu_0=\mathcal{O}(p)$ limit the exponential in the $s=1$ term of \eqref{eq:D.green_rs_amp} scales as $e^{-p/2}$, but this is okay because this is actually $(2G_{aa})^p$, so $G_{aa}$ is $\mathcal{O}(1)$.} This matches the ``bow-tie'' vanishing pattern shown in Figures~\ref{fig:ASSR_2matter_Plots} and~\ref{fig:ASSR_2matter_Realcolor_Plots}.

When $r=1$ (relevant for AS$^2$ with a single heavy operator), we find that \eqref{eq:D.green_rs_amp} is independent of $\mu_0$, and \eqref{eq:D.red_rs_amp} is equal to \eqref{eq:D.green_rs_amp} times $(\mu_0/p)^p$. Now exact equality between $aa$ and $a\overline{a}$ happens at $\mu_0=p$ instead of $p/2$, which matches the analysis of Appendix \ref{app:C} where $\Delta_\text{max}$ in the AS$^2$ with one heavy operator is equal to $\Delta_\text{max}^\text{tot}=2\Delta_\text{max}$ in standard AS$^2$. For finite $\mu_0$, we recover a wormhole length $\ell_\text{wh}=p\log(p/\mu_0)$. 

If we take $r=\mathcal{O}(p)$ and $\mu_0=\mathcal{O}(1)$, we find something that begins to look like the smooth tube geometry in Figure~\ref{fig:Tube_Chord_Diagram}. For $s=\mathcal{O}(1)$, the exponential shared by \eqref{eq:D.green_rs_amp} and \eqref{eq:D.red_rs_amp} becomes $\exp(-2\mu_0^2s)$, which looks like a ``traveling penalty'' $e^{-\mu\,\Delta\tau}$. We would then hope that the extra term in \eqref{eq:D.red_rs_amp} looks like the $\sigma$-chord penalty. This is not strictly true in this limit, as we get some additional exponential $s$ dependence, which makes it look like the $aa$-chords and $a\overline{a}$-chords see smooth tube geometries with different $\mu$'s. We expect this is indicative of combinatorial terms omitted from the exponential arguments of \eqref{eq:D.gray_on_green} and \eqref{eq:D.gray_on_red} that go to $0$ faster than $\mathcal{O}(1/p)$, but which become $\mathcal{O}(1)$ relevant when $r=\mathcal{O}(p)$. It could also be a real finite-$\lambda$ effect that the $G,\Sigma$ approach misses. We leave further analysis to future work.


\printbibliography

\end{document}